\documentclass[rmp, superscriptaddress, reprint]{revtex4-2}
\usepackage[T1]{fontenc}
\usepackage[utf8]{inputenc}

\usepackage{amsmath}
\usepackage{amsfonts}
\usepackage{natbib}
\usepackage{hyperref}
\usepackage{braket}
\usepackage{graphicx}
\usepackage{cleveref}
\usepackage{nicefrac}
\usepackage{dsfont}
\usepackage[squaren]{SIunits}
\usepackage[dvipsnames]{xcolor}
\usepackage{wasysym} 
\usepackage{array} 
\usepackage{multirow} 
\usepackage{booktabs}
\usepackage[normalem]{ulem}

\usepackage{supertabular}

\graphicspath{{Figures/}{Figures/APQRChapter}}

\newcommand{\id}{\mathds{1}}

\newcommand{\ignore}[1]{}

\begin{document}

\title{Quantum repeaters: From quantum networks to the quantum internet}

\author{Koji Azuma}
\email{koji.azuma@ntt.com}
\affiliation{NTT Basic Research Laboratories, NTT Corporation, 3-1 Morinosato Wakamiya, Atsugi, Kanagawa 243-0198, Japan}
\affiliation{NTT Research Center for Theoretical Quantum Physics, NTT Corporation, 3-1 Morinosato-Wakamiya, Atsugi, Kanagawa 243-0198, Japan}
\author{Sophia E. Economou}
\email{economou@vt.edu}
\affiliation{Department of Physics, Virginia Tech, Blacksburg, Virginia 24061, USA}
\author{David Elkouss}
\email{d.elkousscoronas@tudelft.nl}
\affiliation{
QuTech, Delft University of Technology, Lorentzweg 1, 2628 CJ Delft, The Netherlands
}
\affiliation{Networked Quantum Devices Unit, Okinawa Institute of Science and Technology Graduate University, Okinawa, Japan}
\author{Paul Hilaire}
\email{paul.hilaire@quandela.com}
\affiliation{Department of Physics, Virginia Tech, Blacksburg, Virginia 24061, USA}
\affiliation{Quandela SAS, 10 Boulevard Thomas Gobert, 91120, Palaiseau, France}
\author{Liang Jiang}
\email{liang.jiang@uchicago.edu}
\affiliation{Pritzker School of Molecular Engineering, The University of Chicago, Chicago, Illinois 60637, USA
}
\author{Hoi-Kwong Lo}
\email{hklo@ece.utoronto.ca}
\affiliation{Quantum Bridge Technologies\char`,{} Inc., 100 College Street, Toronto, ON M5G 1L5, Canada}
\affiliation{Department of Physics, University of Hong Kong, Pokfulam, Hong Kong}
\affiliation{Center for Quantum Information and Quantum Control, Department of Physics and Department of Electrical \& Computer Engineering, University of Toronto, M5S 3G4 Toronto Canada}
\author{Ilan Tzitrin}
\email{itzitrin@physics.utoronto.ca}
\affiliation{
Department of Physics, University of Toronto, Toronto, Canada}

\date{\today{}}

\begin{abstract}
A quantum internet is the holy grail of quantum information processing, enabling the deployment of a broad range of quantum technologies and protocols on a global scale. However, numerous challenges exist before the quantum internet can become a reality. Perhaps the most crucial of these is the realization of a quantum repeater, an essential component in the long-distance transmission of quantum information. As the analog of a classical repeater, extender, or booster, the quantum repeater works to overcome loss and noise in the quantum channels comprising a quantum network. Here, we review the conceptual frameworks and architectures for quantum repeaters, as well as the experimental progress towards their realization. We also discuss the various near-term proposals to overcome the limits to the communication rates set by point-to-point quantum communication. Finally, we overview how quantum repeaters fit within the broader challenge of designing and implementing a quantum internet.
\end{abstract}

\maketitle

\tableofcontents{}

\section{Introduction}\label{sec:into}
Following its rapid growth this century, the Internet has become an invaluable socioeconomic fixture, inextricable from almost all facets of day-to-day life. Access to a high-speed internet---the ability to send and receive digital information across the globe at almost the speed of light---has transformed from a luxury to a utility. 
However, the current Internet is not sustainable and scalable without
future innovation  \cite{LeonGarcia2021}. It has
been estimated in 2022 there are currently 7 billion connected
IoT (Internet of Things) devices on-line. This number is projected to increase to 25.4 billion by 2030 \cite{Howarth2021}.
As the number of devices increases exponentially over time,
the energy consumption in optical communication also grows
exponentially, thereby contributing to climate change.
The amount of local computing power needed to monitor
and control network traffic also grows exponentially.
The task of service
and network management is thus becoming more and more complex.
To move things forward, new concepts such as distributed intelligence and
distributed trust (e.g. blockchain) are probably needed. On the other hand, on the longer term,
it is widely recognized that a quantum internet and
distributed quantum computing will complement the classical internet.
The quantum internet will be provably secure and could provide exponentially
more computational power and sensing capability to specific tasks.

Indeed, analogously to this Internet, a new system is steadily emerging in theoretical literature and early experiment: the quantum internet~\cite{Kimble2008}, a means of transmitting quantum information globally. While serving a different purpose from the classical Internet, this new paradigm may prove disruptive in its own way. We dedicate this review to the progress that has been made in designing and building the quantum internet, focusing largely on its main building block, quantum repeaters. In addition to the basic theoretical concepts required to understand the components of the quantum internet, we survey its more technical architectural requirements as well as the experimental advances towards its implementation.

While classical information is often encoded digitally---as sequences of $0$s and $1$s, usually represented in electronic signals---it can also be housed in quantum mechanical states, which abide by different rules. The quantum states encoding the bits 0 and 1, mathematically represented by vectors and denoted by $\ket{0}$ and $\ket{1}$ (the \emph{computational-basis} states), can correspond to a variety of physical systems. Among the most popular and useful quantum information carriers is light---the state of the electromagnetic field associated with one or multiple photons.

Unlike the analogous classical states, quantum states can be superposed like waves. For instance, equal combinations of $\Ket{0}$ and $\Ket{1}$ include $\ket{+} \equiv \frac{1}{\sqrt{2}} (\ket{0} + \ket{1} )$ and
$ \ket{-} \equiv \frac{1}{\sqrt{2}} ( \ket{0} - \ket{1} )$, the \emph{conjugate-basis} states. Measuring a conjugate-basis state in the computational basis collapses the superposition, resulting in $\Ket{0}$ or $\Ket{1}$ at random with equal probability, a manifestation of a more general postulate of quantum mechanics known as Born's rule. The fact that the outcome of this measurement is probabilistic rather than deterministic is predicted by Heisenberg's uncertainty principle.

In addition to quantum superposition, Born's rule, and Heisenberg's uncertainty principle, the formalism of quantum mechanics
allows for subtle quantum correlations---dubbed \emph{entanglement}---to exist between remote physical systems.
For instance, two distant photons that are entangled may be in a so-called singlet state
$ \frac{1}{\sqrt{2}} (\ket{01} - \ket{10} )$, which can exhibit
stronger-than-classical correlations upon measurement. Not only is it impossible to describe independently the state of each photon in the singlet, but when measured along any common axis, the two photons always show
opposite results.
According to Schr\"{o}dinger, entanglement is the essence of quantum theory, but it is far from a theoretical curiosity. The existence of non-classical correlations associated with entangled states has been proven in several experiments via Bell tests~\cite{Brunner2014, PhysicsTodayBellTest},
which has led three experimental physicists---Alain Aspect, John Clauser, and Anton Zeilinger---to be awarded the Nobel Prize in Physics in 2022. 
Furthermore, in the last few decades, researchers have shown that entanglement is a powerful resource in quantum information processing, enabling many unusual applications that are impossible or impractical with only classical resources.

The quantum technologies enabled by our continuously evolving ability to understand, generate, manipulate, and entangle delicate quantum systems are the premise behind what is commonly referred to the Second Quantum Revolution~\cite{Dowling2003QuantumRevolution,lo1998introduction}. In the First Quantum Revolution, which occurred in the last century,
lasers and transistors---devices built upon the underlying principles of quantum mechanics---played
a crucial part in global economic growth. Now, we are already able to demonstrate primitives or complete protocols for the quintessential applications of quantum information: \emph{quantum cryptography}---unconditionally secure communication between parties---and \emph{quantum computation}---a method for exceeding the best-known scaling of certain classes of classical algorithms. 

These and other quantum information tasks can be accessed remotely if embedded within a quantum internet---a global network of quantum information processors, namely sources of quantum states, executors of quantum gates, and devices for quantum measurements~\cite{Wehner2018, VanDam2020}. Such a network can also provide secure access and enhance the performance of these applications of quantum information.

The security underlying the classical Internet is based on computational conjectures, which makes it vulnerable to hacking and eavesdropping. A quantum computer poses a threat to the contemporary cryptosystem because Shor's factoring algorithm~\cite{Shor1997} offers a way to break standard public-key encryption schemes, including RSA, Diffie-Hellman, and
elliptic curve crypto-systems within short timescales.
Owing to the extensive experimental progress in quantum computing in the
last few decades, its threat is now widely
acknowledged by many governments and organizations~\cite{NISTPostQuantum}. While certain classical solutions have been proposed to counter the threat, such as post-quantum cryptographic systems, these are still only conjectured to be secure against quantum attacks. 
Indeed, three candidate post-quantum crypto-systems in the NIST competition have  already been cracked easily by a PC \cite{Townsend2022}.
In reality, quantum key distribution (QKD) is the only known way to allow the unconditionally secure transmission of information---that is, a security founded in tested laws of physics and mathematical proofs~\cite{Bennett2014, Ekert1991, Xu2020,curty2021quantumleap}. However, commercialized fiber-based point-to-point QKD is limited to a distance of less than 400\,km, whereas satellite-to-ground QKD, intended to extend the communication distances, requires expensive components such as satellites and large telescopes. The quantum internet promises to significantly extend the range of QKD and other cryptographic protocols, thereby securing global communication and transactions.

In particular, a quantum internet will permit secure access to cloud-based quantum computing.
Major IT firms such as Google, IBM, Intel, Microsoft, Amazon, and Alibaba are actively
constructing their own quantum processors on the way to universal, scalable, and fault-tolerant quantum computers. These companies are working towards this goal alongside dedicated quantum-computing startup companies, which belong to a newly-forming ecosystem of quantum startups.
Companies such as IBM\footnote{This was followed by other companies---including ionQ, Quantinuum, Quandela, and Xanadu---proposing cloud accessible platforms based on either ion traps or photonics, potentially more promising platforms for remote access using quantum channels.} have already put small-scale quantum processors online for external access~\cite{IBMQuantumCloud}. The history of conventional computers suggests that the first few years in quantum computing epoch will see only a few large-scale quantum computers in the world. This means that users will have to engage with the devices through classical or quantum networks. With the help of innovative protocols for blind quantum computing~\cite{broadbent2009universal}, a future quantum internet will allow users to submit their jobs anywhere in the world privately and securely.

Quantum networking is also a crucial ingredient in distributed quantum computing, which allows separate quantum computers to cooperate on an algorithm.
At their early stages, quantum processors will be limited in size and complexity;
to achieve greater computing power, they will likely need to be networked through quantum channels, with quantum information flowing between them.
In this way, quantum networking is important even for short-distance communication between quantum computers. Other protocols enabled or improved by the quantum internet include
quantum teleportation~\cite{Bennett93},
quantum fingerprinting~\cite{Buhrman2001}, quantum sensing,
clock synchronization \cite{Jozsa2000,komar2014quantum}, and the linking of distant optical telescopes for sharper images~\cite{gottesman2012longer}.

Conceptually, it is known that sending quantum information
(i.e., qubits) can lower the
amount of required communication in distributed
information processing tasks, in comparison to sending
classical information (bits). The study of
the amount of required quantum communication is called
quantum communication complexity \cite{brassard2003quantum}.
Incidentally, the classical communication cost required in
quantum information processing is also an important subject \cite{Lo2000Cl}.

Building a quantum internet requires harnessing quantum states of light. Even in the far future, the photon---or a state of multiple photons---will likely be the information carrier of choice in quantum communication, as it can function as a ``flying'' qubit (as opposed to matter-based qubits, which are fixed in space) while minimally interacting with its environment. By encoding information in photonic degrees of freedom, quantum information can be transmitted through optical fibers or in free space over long distances with little decoherence.

Despite the advantages of light, there is enough absorption and scattering of photons in the media where they propagate---processes that lead to optical attenuation---that make loss the key physical hurdle in the construction of a quantum internet.
In a standard single-mode optical fiber, close to the standard telecommunication wavelength of 1550\,nm, the attenuation is 0.2\,dB/km~\cite{FOAReferenceGuide}. This means 1 of every 100 photons survives a journey of 100\,km on average.
Recently, ultra-low-loss (ULL) optical fibers have been commercialized with a loss as low as 0.15\,dB/km \cite{Corning2022}. 
These sorts of losses in optical channels yield fundamental limits to the rate at which two parties can establish a secret key with a point-to-point QKD protocol, given by
TGW bound~\cite{Takeoka2014} and PLOB bound
\cite{Pirandola2015}, and discussed in more detail in Secs.~\ref{sec:milestones} and \ref{sec:internet}.

\begin{table*}[tb]
\footnotesize
\begin{tabular}{cc}
\toprule 
\textbf{Reference} & \textbf{Topic}\tabularnewline
\midrule
\midrule 
\cite{Sangouard2011} & Quantum repeaters based on atomic ensembles and linear optics\tabularnewline
\midrule 
\cite{reiserer2015cavity} & Cavity-based quantum networks with single atoms and optical photons\tabularnewline
\midrule 
\cite{Heshami2016} &
Quantum memories and applications \tabularnewline
\midrule 
\cite{atature2018material} & Material platforms for spin-based photonic quantum technologies \tabularnewline\midrule
\cite{awschalom2018quantum} &
Quantum technologies with optically interfaced solid-state spins
\tabularnewline
\midrule 
\cite{Ruf2021} &
Quantum networks based on color centers in diamond \tabularnewline
\midrule 
\cite{Munro2015} & Primitives of quantum repeaters\tabularnewline
\midrule 
\cite{Muralidharan2016} & Generations of quantum repeaters\tabularnewline
\midrule 
\cite{Kimble2008} & Introductory work to the quantum internet\tabularnewline
\midrule 
\cite{Wehner2018} & Developmental stages of the quantum internet\tabularnewline
\midrule 
\cite{Xu2014} & Measurement-device-independent quantum cryptography\tabularnewline
\midrule 
\cite{Xu2020} & Realistic QKD\tabularnewline
\midrule 
\cite{Broadbent2016} & Quantum cryptography beyond QKD\tabularnewline
\midrule 
\cite{Fitzsimons2017} & Blind quantum computing\tabularnewline
\midrule
\cite{pirandola2020advances} & Advances in quantum cryptography\tabularnewline
\midrule
\cite{azuma2020tools} & Tools for quantum network design\tabularnewline
\bottomrule
\end{tabular}
  \caption{Related review articles.}
  \label{table_review}
\end{table*}

Nevertheless, quantum networks based on such point-to-point QKD links have already been built all over the world. Examples of ground-based
fiber networks include the Tokyo QKD network in Japan \cite{Sasaki2011}, the SECOQC
network in Europe \cite{peev2009secoqc}, the 2000~km Shanghai-Beijing network in China \cite{chen2021integrated}, and the Euro QCI network by the 27 EU member states \cite{Eurocommission2022}. Additionally, ground-to-satellite quantum transmission has been performed over
thousands of kilometers of free space.
This line of research has demonstrated that long-distance quantum communication in a global length scale is feasible with current satellite technology (see \cite{chen2021integrated}). Several theoretical papers envisioned a satellite-based quantum repeater network~\cite{Boone2015, Gundougan2021, Khatri2021}. However, because their foundation is point-to-point QKD, existing quantum networks rely on trusted relay nodes to achieve information-theoretically secure communication.
In these nodes, optical signals are measured to yield a classical output, and then new optical signals are generated and sent out.
This classical output is vulnerable to hacking and eavesdropping, meaning security is only achieved if the nodes can be trusted.

The architectural challenge of a long-distance quantum network is therefore to overcome the fundamental limit of point-to-point quantum communication, achieving high-rate secure communication without using trusted relay nodes. Unfortunately, conventional signal boosters, repeaters, extenders or amplifiers do not work for quantum signals because of the famous quantum no-cloning theorem~\cite{Dieks1982, Wootters1982},
which states that an unknown quantum state cannot be copied reliably. However, it is still possible to combat loss and noise without cloning quantum states; this is achieved with the help of quantum repeaters.

In quantum repeater protocols, instead of sending quantum signals (photons) directly from one user to
another, a sequence of intermediate nodes are set up. There, certain strategies can be used to combat errors induced by losses and other forms of noise, including entanglement distillation and purification, and quantum error detection and correction.
While practical quantum repeaters are not possible with existing technology, research towards this goal is active and involves many different fields of inquiry. Several matter-based systems exist to facilitate their implementation, including atomic ensembles, which can function as quantum memories; quantum dots, which can be used as on-demand sources of a host of photonic states; and cavity QED, which can be used to enhance light-matter interactions.
Since photons are often used as flying qubits and quantum memories often involve matter, the
quantum interface between light and matter is regarded as a key ingredient in quantum repeaters.

In addition to the many subfields of physics involved in the effort to build quantum repeaters, the pursuit of a quantum internet more generally is an interdisciplinary theoretical and experimental endeavor involving mathematicians,
computer scientists, and engineers.
Classical tools from network topology, protocol design, information theory, and
error correction, in addition to topics within quantum information, e.g., state preparation, quantum channels and measurements, and quantum error correction, are all needed for investigations into the quantum internet.

Several of the topics discussed in this work have been the focus of---or at least have gotten a mention in---previous reviews. We build on this body of work while discussing newer theoretical and experimental developments to keep pace with the dynamic field of quantum communication.
For instance, the review~\cite{Sangouard2011} chiefly covers quantum repeaters whose memories are implemented with atomic ensembles, while Ref.~\cite{Munro2015} focuses on the primitives used in quantum repeaters.
Reference~\cite{Muralidharan2016} (and Ref.~\cite{Munro2015} also) categorizes quantum repeater protocols into relevant generations which differ in performance and technological requirements.
In our review, we revisit this categorization, sorting repeaters based on the associated mechanisms for suppressing losses and errors. This gives us a more natural structure to understand newly emerging classes of repeaters, notably memoryless, error-corrected, and all-photonic repeaters, which have not been extensively featured in reviews.
In addition to our discussion of full-fledged repeaters, we dedicate a portion of our review to simpler protocols believed to be sufficient to beat repeaterless bounds, an important milestone for long-distance quantum communication. Reference~\cite{Xu2020} already tackles some of these ideas with an approach centered around their security in realistic implementations; in our work, we focus on performance, chiefly in terms of key distribution metrics.
References~\cite{Kimble2008, Wehner2018} review the progress towards the realization of the quantum internet. Notably, Reference~\cite{Wehner2018} introduces stages of development for the quantum internet, aligning with applications that grow in technological complexity. Here, we continue this discussion but additionally introduce an information-theoretic framework to derive fundamental limits of quantum communication over a quantum network, with views different from Refs.~\cite{pirandola2020advances} and \cite{azuma2020tools}. In Table~\ref{table_review}, we provide a list of the reviews just mentioned together with other works on applications of quantum communication that are not covered here.

The rest of this review is organized as follows.
In Secs.~\ref{sec:pre} and \ref{sec:primitives}, we present the preliminaries required to understand quantum repeaters and the physics behind the quantum internet.
In Sec.~\ref{sec:gens}, we overview the conceptual frameworks of quantum repeaters and use them to organize the existing proposals.
In Sec.~\ref{sec:memoryless} we discuss an important class of memoryless repeaters that intersect with the latest generations of theoretical proposals.
In Sec.~\ref{sec:milestones}, we review various near-term protocols, such as an adaptive version of measurement-device-independent QKD \cite{Lo2012} and twin-field QKD \cite{Lucamarini2018}, which
are regarded as milestones in the path to outperforming the PLOB bound en route to quantum repeaters. In Sec.~\ref{sec:exps}, we describe experimental advances towards optical-fiber-based quantum communication schemes featuring quantum repeaters.
Section~\ref{sec:internet} is dedicated to a discussion on the quantum internet, including the quantum/private capacities of quantum internet protocols and upper bounds on the capacities. Some concluding remarks are provided in Sec.~\ref{sec:conclusion}.

\begin{table}[bt]
    \centering
    \begin{tabular*}{8.7cm}{p{2.2cm}p{6.3cm}}
    \toprule
    \textbf{Abbreviation} &                                     \textbf{Meaning} \\
    \midrule
    \midrule
    BBSM &                     Boosted Bell State Measurement \\
        BDCZ &                           Briegel-D\"ur-Cirac-Zoller \\
      BM/BSM &                           Bell (State) Measurement \\
        CPTP &           Completely Positive and Trace-Preserving \\
         CSS &                             Calderbank-Shor-Steane \\
        CTSL &                    Childress-Taylor-S\o rensen-Lukin \\
          CV &                                Continuous Variable \\
        DLCZ &                             Duan-Lukin-Cirac-Zoller \\
          DV &                                  Discrete Variable \\
          EG &                           Entanglement Generation \\
          ES &                              Entanglement Swapping \\
         GBS &                            Gaussian Boson Sampling \\
         GHZ &                        Greenberger-Horne-Zeilinger \\
         GKP &                          Gottesman-Kitaev-Preskill \\
        HEGP &          Heralded Entanglement Generation Protocol \\
         LHC &                              Large Hadron Collider \\
        LIGO & 
        Laser Interferometer Gravitational-Wave Observatory
        \\
        LOCC &       Local Operations and Classical Communication \\
        MBQC &                Measurement-Based Quantum Computing \\
         MDI &                     Measurement Device Independent \\
         MIT &              Massachussets Institute of Technology \\
        NISQ &                   Noisy Intermediate Scale Quantum \\
        NIST &     National Institute for Standards in Technology \\
          NV &                                   Nitrogen Vacancy \\
         PBS &                            Polarizing Beamsplitter \\
        PLOB &                Pirandola-Laurenza-Ottaviani-Banchi \\
         PNR &                   Photon-Number-Resolving Detector \\
        PRCS &                    Phase-Randomized Coherent State \\
     EuroQCI &      European Quantum Communication Infrastructure \\
          QD &                                        Quantum Dot \\
         QED &                            Quantum Electrodynamics \\
         QKD &                           Quantum Key Distribution \\
          QM &                                     Quantum Memory \\
         QND &                             Quantum Non-Demolition \\
          QR &                                   Quantum Repeater \\
         RGS &                               Repeater Graph State \\
         RSA &                              Rivest-Shamir-Adleman \\
      SECOQC & Secure Communication based on Quantum Cryptography \\
       SNSPD &    Superconducting Nanowire Single Photon Detector \\
         SPD &                             Single Photon Detector \\
        SPDC &              Spontaneous Parametric Downconversion \\
          SW &                                 (Optical) Switches \\
      TF-QKD &                Twin-Field Quantum Key Distribution \\
         TGW &                                 Takeoka-Guha-Wilde \\
         ULL &                                     Ultra Low Loss \\
         \bottomrule
    \end{tabular*}
    \caption{Abbreviations used in this review.}
    \label{tab:abbreviation}
\end{table}

For clarity, we present the list (Table~\ref{tab:abbreviation}) of abbreviations that are used throughout the review.

\section{Preliminaries} \label{sec:pre}
In this section, we summarize relevant background concepts, including qubits, entanglement, and possible photonic encodings. Repeater primitives---including teleportation and entanglement swapping---are left to Sec.~\ref{sec:primitives}. Standard references, including~\cite{NielsenChuang2010}, can be used to supplement this part of the review.

\subsection{Qubits}\label{sec:qubits}
\begin{figure}
    \includegraphics[width=6cm]{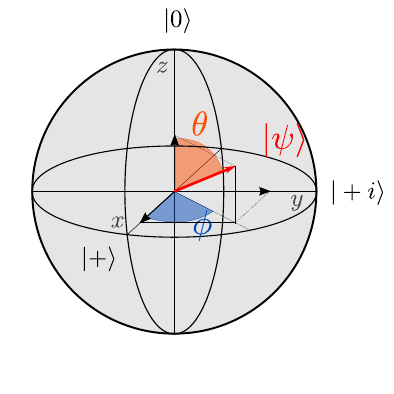}
    \caption{The Bloch sphere representation of a qubit. The $(x, y, z)$-components of a Bloch vector (displayed as an arrow) give the expectation values of the Pauli observable $X$, $Y$, and $Z$. 
    For instance, points $(0,0,1)$, $(1,0,0)$, and $(0,1,0)$ correspond to eigenstates $\ket{0}$, $\ket{+}=(\ket{0}+\ket{1})/\sqrt{2}$, and $\ket{+i}=(\ket{0}+i\ket{1})/\sqrt{2}$ of Pauli operators $Z$, $X$, and $Y$ with the eigenvalue of $+1$, respectively. 
    }
    \label{fig:bloch}
\end{figure}

A \emph{qubit}---the quantum mechanical analog of the classical bit and the fundamental unit of quantum information---is another name for a two-dimensional complex Hilbert space. A pure state $\ket{\psi}$ of any qubit can be written in the computational basis through
\begin{equation}
    \ket{\psi} = a\ket{0} + b\ket{1},
\end{equation}
where $a , b \in \mathbb{C}$ and $|a|^2 + |b|^2 = 1$. Setting $a = \frac{1}{\sqrt{2}}$ and $b = \pm \frac{1}{\sqrt{2}}$ gives states in the conjugate basis:
\begin{equation}
    \Ket{\pm} \equiv \frac{1}{\sqrt{2}} \Ket{0} \pm \Ket{1}.
\end{equation}
In quantum mechanics, the global phase of a state is irrelevant; thus, one can parametrize any pure qubit through two parameters, $a = \cos{\frac{\theta}{2}}$ and $b = e^{i\phi} \sin{\frac{\theta}{2}}$, revealing its Bloch sphere representation, illustrated in Fig.~\ref{fig:bloch}, where $\theta$ and $\phi$ are the polar and azimuthal angles, respectively. A qubit is realized experimentally by associating it with a two-dimensional space or subspace of a physical system. Although we will encounter matter (chiefly spin) qubits in this review, we are particularly interested in encodings into photonic systems, which we survey in Sec.~\ref{sec:photonic_encodings}.

Interactions with the environment or preparation errors can diminish the purity of a qubit---that is, introduce classical uncertainty. In this case, we must turn to a representation of the qubit as a statistical mixture of pure quantum states.
The general description of a state, which include \emph{mixed states}, is as a positive operator $\rho$ with unit trace, called a {\it density operator}.
The density operator of a pure state $\ket{\psi}$ is $\rho = \ket{\psi}\bra{\psi}$ with ${\rm Tr}[\rho^2]=1$, while a density operator $\rho$ with ${\rm Tr}[\rho^2]<1$ describes a mixed state.
In the case of a qubit, it can be written as
\begin{equation}
    \rho = \rho_{00} \Ket{0}\Bra{0} + \rho_{01} \Ket{0}\Bra{1} + \rho_{10} \Ket{1}\Bra{0} + \rho_{11} \Ket{1}\Bra{1},
\end{equation}
where the \emph{populations}, $\rho_{00}$ and $\rho_{11}$, are real and add to unity ($\rho_{00} + \rho_{11}$ = 1), the \emph{coherences}, $\rho_{01}$, and $\rho_{10}$, are complex conjugates ($\rho_{01}$ = $\rho^*_{10}$), and $\det [\rho]=\rho_{00}\rho_{11} - \rho_{01}\rho_{10} \ge 0$.

\emph{Unitary} transformations---operators $U$ with $U^{\dagger}U =UU^{\dagger}= \mathds{1}$---describe reversible, probability-preserving operations on qubits, i.e., \emph{quantum gates}. The \emph{Pauli gates} are defined through
\begin{align}
X &= \Ket{0}\Bra{1} + \Ket{1}\Bra{0}, \\
Y &= -i\Ket{0}\Bra{1} + i\Ket{1}\Bra{0}, \\
Z &= \Ket{0}\Bra{0} - \Ket{1}\Bra{1}.
\end{align}
$X$, $Z$, and $Y$ effect a phase flip, a bit flip, and a combination of the two on the qubit, respectively. A unitary $U$ is \emph{Clifford} if it maps any Pauli gate $P$ to a Pauli gate under conjugation, that is, $U P U^\dagger$ is also a Pauli gate. An example of a non-Pauli Clifford gate is the \emph{Hadamard gate}, defined by
\begin{equation}
    H = \frac{1}{\sqrt{2}} (\Ket{0}\Bra{0} + \Ket{0}\Bra{1} + \Ket{1}\Bra{0} - \Ket{1}\Bra{1}).
\end{equation}
An example of a non-Clifford gate is the $\frac{\pi}{8}$ or T gate, given through
\begin{equation}
    T = \Ket{0}\Bra{0} + e^{i\pi/4} \Ket{1}\Bra{1}.
\end{equation}
The above discussion is generalizable to systems of multiple qubits by taking tensor products; see Sec.~\ref{sec:entanglement}.

A \emph{measurement} process on a quantum system in a state $\rho$ is, in general, described by a set of Kraus (linear) operators $\{M_i\}_i$ satisfying $M_i^\dag M_i = \mathds{1}$. Performing the associated measurement results in an outcome $i$ 
with probability $p_i={\rm Tr}[M_i^\dag M_i \rho]$ and leaves the state in $M_i\rho M_i^\dag/p_i$. This is a formalization and generalization of Born's rule. For the particular case of a (destructive) Pauli measurement on a qubit, we may associate $M_0=\bra{v_0}$ and $M_1=\bra{v_1}$ with the eigenstates $\ket{v_0}$ and $\ket{v_1}$ of the corresponding operator; {\it $Z$-basis measurements} (corresponding to the Pauli $Z$) are specified by the Kraus operators $\{\bra{0},\bra{1}\}$, while {\it $X$-basis measurements} (corresponding to the Pauli $X$) are specified by the Kraus operators $\{\bra{+},\bra{-}\}$.

We also consider a \emph{quantum channel}, ${\cal N}$, which deterministically converts a given state $\rho$ into a state $\sigma$. This kind of transformation is useful to describe the actions of noise and transmission channels on quantum systems. 
Any quantum channel has an operator-sum representation, $\sigma={\cal N}(\rho)=\sum_i M_i \rho M_i^\dag$, specified by a set of Kraus operators $\{M_i\}_i$. Another representation is 
\begin{equation}
\sigma_{A'}={\cal N}_{A\to A'}(\rho_A)={\rm Tr}_{E'} [U_{AE} (\rho_A \otimes \ket{0}\bra{0}_E) U^\dag_{AE}], \label{eq:CPTP}
\end{equation} 
where $U_{AE}$ is a unitary operator acting on system ${\cal H}_{A}\otimes {\cal H}_{E}$ and $\ket{0}_E$ is a state of an auxiliary system (environment) $E$.
The map ${\cal N}$ must be completely positive and trace-preserving (CPTP). 

Three examples of common qubit errors described by channels are \emph{phase-flip}, \emph{bit-flip}, and \emph{depolarizing} noise, respectively written as
\begin{align}
    {\cal N} (\rho)=& (1-p)\rho +p Z\rho Z,\\
    {\cal N} (\rho)=&(1-p)\rho +p X\rho X,\\
    {\cal N} (\rho)=&(1-p)\rho +\frac{p}{3} (X\rho X+Y\rho Y+Z\rho Z),\label{eq:depolarizing}
\end{align}
where $0\le p \le 1$ corresponds to an error probability or channel strength. A \emph{pure-loss bosonic channel} is written by defining the action of $U_{AE}$ in Eq.~(\ref{eq:CPTP}) as
\begin{equation}
U_{AE}a_A U_{AE}^\dag = \sqrt{\eta} a_{A'} +\sqrt{1-\eta} a_{E'} \label{eq:lossch}
\end{equation}
in the Heisenberg representation. Here $a_X$ is the annihilation operator on bosonic system $X$, $0\le \eta \le 1$ is the transmittance of the channel, and $\ket{0}_E$ of Eq.~(\ref{eq:CPTP}) is the vacuum state of the bosonic system $E$. The pure-loss bosonic channel is used as a model for an optical fiber: in this case, the transmittance $\eta$ is related to the length $L$ of the fiber through $\eta=e^{-L/L_{\rm att}},$ with a constant $L_{\rm att}$ denoting the attenuation distance.

\subsection{Quantum no-cloning theorem}

\label{sec:no-cloning}

The quantum \emph{no-cloning theorem}~\cite{Wootters1982, Dieks1982} entails that it is impossible to create a copy of unknown quantum states. More precisely, given an unknown state $\ket{\psi}_A$, the theorem states that there exists no deterministic quantum operation that can copy $\ket{\psi}_A$ onto system $B$ to obtain $\ket{\psi}_A \otimes \ket{\psi}_B$. Originally demonstrated for pure states, the no-cloning theorem has later been extended to mixed states through the \emph{no-broadcasting theorem}~\cite{Barnum1996}. This no-go theorem has profound implications---helpful and unhelpful---for quantum information technologies. While it is at the core of the security of quantum key distribution \cite{Bennett1992,koashi1998,Koashi2002}, it also precludes building quantum repeaters analogously to classical signal extenders and furthermore creates challenges in the design and performance of quantum error-correcting codes. For example, the no-cloning theorem makes it impossible to use a classical-like repetition code to correct for errors acting on quantum states, and implies an upper bound of $50 \%$ on the loss that any quantum error-correcting code can tolerate. This directly impacts the performance of quantum repeater protocols based on quantum error correction, as addressed in Sec.~\ref{subsec:error_suppr}.

\subsection{Entanglement}
\label{sec:entanglement}

Here we present the formal definition of entanglement and introduce several important classes of entangled states.

\subsubsection{Definition and properties}
Entanglement---per Schr{\"o}dinger, a defining feature of quantum theory \cite{Schrodinger1935Probability}---refers to the impossibility of describing certain composite quantum states through independent specifications of their constituents. The existence of entanglement, as guaranteed by the formalism and postulates of quantum theory and confirmed by many experiments, has profound physical and metaphysical repercussions, as exemplified famously by Einstein, Podolsky and Rosen (EPR)~\cite{Einstein1935} and by Bell~\cite{Bell1964}, and since then by numerous physicists investigating its repercussions on increasingly rigorous footing. There are several equivalent formulations of entanglement---see, e.g.,~\cite{Horodecki2009}.
A useful one for our purpose is the view of entanglement as a resource for quantum information
tasks. Entanglement plays a central role in virtually every primitive and application of quantum information; for us, its most relevant uses are for the protocols we describe in Sec.~\ref{sec:primitives}: quantum teleportation and entanglement swapping, entanglement purification and distillation, and quantum error correction, all of which underlie quantum repeaters.
As a non-trivial resource with respect to local operations and classical communication
(LOCC), entanglement cannot be increased
by performing local operations (including local quantum gates and measurements),
classical communication (including adaptive schemes based on classical
outputs from other parties), or the combination of both. One may establish quantum entanglement by interacting systems via coupling Hamiltonians,
physically distributing entangled states between parties (such as by sending photons
over fiber channels), or performing collective measurements of observables
from different parties. The entanglement generation process depends
on the details of the physical system, as discussed in Sec.~\ref{sec:exps}.

\subsubsection{Entanglement in bipartite states}
The Hilbert space $\mathcal{H}$ of a bipartite system is the tensor product of the subsystem spaces $\mathcal{H}=\mathcal{H}_{A}\otimes\mathcal{H}_{B}$.
A \emph{separable} bipartite pure state is a tensor product of pure states
in $\mathcal{H}_{A}$ and $\mathcal{H}_{B}$,
\begin{align}
\label{eq:separable}
    \ket{\Psi}_{AB}&=\ket{\varphi}_A\otimes\ket{\phi}_B \\
    &=: \ket{\varphi}_A \ket{\phi}_B =:\ket{\varphi, \phi}_{AB}=:\ket{\varphi \phi}_{AB} \nonumber
\end{align}
with reduced density operators $\Psi_{A}:={\rm Tr}_B[\ket{\Psi}\bra{\Psi}_{AB}]=\ket{\varphi}\bra{\varphi}_{A}$ on subsystem $A$
and $\Psi_{B}:={\rm Tr}_A[\ket{\Psi}\bra{\Psi}_{AB}]=\ket{\phi}\bra{\phi}_{B}$ on subsystem $B$, obtained by tracing out the non-subscripted system. By contrast, an entangled
bipartite pure state cannot be described as a product of states from the
individual subsystems; that is, it cannot be written in the form~\eqref{eq:separable}.

Generally, we may write any bipartite pure state as
\begin{equation}
\ket{\Psi}_{AB}=\sum_{i,j} c_{ij}\ket{i}_{A}\otimes\ket{j}_{B},
\end{equation}
where $c_{ij}$ are complex numbers with $\sum_{i,j} |c_{ij}|^2=1$, $\{ \ket{i}_{A}\}$ and $\{\ket{j}_{B}\}$
are orthonormal bases of the two subsystems.
With the \emph{Schmidt decomposition},
we may find convenient orthogonal bases $\{\ket{v_i}_A\}$ and
$\{\ket{w_{j}}_B\}$ for the two subsystems, such that the bipartite
pure state can be expressed in a standard form with a single index:
\begin{equation}
\ket{\Psi}_{AB}=\sum_{j=1}^{r}\sqrt{p_{j}}\ket{v_j}_{A}\otimes\ket{w_{j}}_B,
\end{equation}
where $p_{j}>0$ for $j=1,\ldots,r$ and $\sum_{j=1}^{r}p_{j}=1$.
The integer $r$ is called the \emph{Schmidt rank}. The reduced density operators for the two subsystems
are $\Psi_{A}={\rm Tr}_B[\ket{\Psi}\bra{\Psi}_{AB}]=\sum_{j=1}^{r}p_{j}\ket{v_j}\bra{v_j}_A$
and $\Psi_{B}={\rm Tr}_A[\ket{\Psi}\bra{\Psi}_{AB}]=\sum_{j=1}^{r}p_{j}\ket{w_j}\bra{w_j}_B$. For $r=1$, the expression reduces to a separable bipartite pure
state. For $r\ge2$, the state $\ket{\Psi}_{AB}$ is entangled.

In the setting of mixed states, the definition of separability must be changed to include classical mixtures of tensor product states:
\begin{equation}
\label{eq:separable_mixed}
\rho_{AB}=\sum_{j}p_{j}\sigma_{A}^{(j)}\otimes\tau_{B}^{(j)},
\end{equation}
where $\{p_j\}$ is a probability distribution and $\sigma_A^{(j)}$ and $\tau_B^{(j)}$ are density operators. Since $\rho_{AB}$ can freely be generated by Alice and Bob with LOCC, the state must only include classical correlations and no entanglement.
This definition includes pure-state separability as a special case; therefore one can say that any state which cannot be written in the form ~\eqref{eq:separable_mixed} (that is, as a convex combination of product states) is entangled.

Quantifying the degree of entanglement in a mixed quantum state---finding an entanglement \emph{measure} or \emph{monotone} that does not confuse entanglement for classical correlations and does not increase over arbitrary LOCC operations---is a difficult problem. For this purpose one has at one's disposal the Schmidt rank, concurrence, negativity, or various entropic funtions of the reduced density operators, such as the von Neumann entropy~\cite{Bennett96b}. For mixed states of two qubits, one can
umambiguously compute the entanglement using one of the above tools, the concurrence~\cite{Wootters98}. However,
characterizing entanglement for general mixed states of higher dimensions is still an important and active area of research; see Ref.~\cite{Plenio2005, Horodecki2009} for detailed discussions of the difficulties of quantifying entanglement and of existing entanglement measures.

The simplest example of useful entanglement for quantum networks
is that between two qubits associated with two parties, with $\mathcal{H}_{A}=\mathrm{span}\left\{ \ket{0}_{A},\ket{1}_{A}\right\} $
and $\mathcal{H}_{B}=\mathrm{span}\left\{ \ket{0}_{B},\ket{1}_{B}\right\} $. Then,
the space $\mathcal{H}$ is spanned by the four orthogonal \emph{Bell states} or \emph{EPR pairs}:
\begin{equation}
\begin{split}
\label{eq:bell_basis}
\ket{\Phi^{\pm}}_{AB} & =\frac{1}{\sqrt{2}}(\ket{0}_{A}\ket{0}_{B}\pm\ket{1}_{A}\ket{1}_{B}),\\
\ket{\Psi^{\pm}}_{AB} & =\frac{1}{\sqrt{2}}(\ket{0}_{A}\ket{1}_{B}\pm\ket{1}_{A}\ket{0}_{B}).
\end{split}
\end{equation}
These four Bell states are equivalent up to Pauli gates:
$\ket{\Phi^{+}}_{AB}=Z_{B}\ket{\Phi^{-}}_{AB}=iY_{B}\ket{\Psi^{-}}_{AB}=X_{B}\ket{\Psi^{+}}_{AB}$.
Tracing out one of the qubits from any Bell state leaves the remaining qubit in a maximally mixed state, which implies that the Bell states are maximally entangled. We often use the Bell state to calibrate the amount of entanglement
shared between two parties; each Bell state counts as one entangled bit or \emph{ebit}  of entanglement, which can be used to teleport one qubit of quantum
information \cite{Bennett93} (see Sec.~\ref{sec:primitives} for a description of quantum teleportation).

\subsection{Entanglement in multipartite states}
We can generalize the definitions of entanglement in the previous subsection to systems with more than two parties. In this setting, there are several notions of separability. For example, a \emph{fully} separable state defined over multiple subsystems ($A,B,C,\ldots$) can be written as a convex combination of product
states 
\begin{equation}
\rho_{ABC\cdots}=\sum_{j}p_{j}\sigma_{A}^{(j)}\otimes\tau_{B}^{(j)}\otimes\gamma_{C}^{(j)}\otimes\cdots \label{eq:mulisep} .
\end{equation}
Similarly to the bipartite case, a multipartite state is entangled when the state cannot be written in the form (\ref{eq:mulisep}).

Two well-known families of entangled states of $M>2$ parties are the \emph{Greenberger--Horne--Zeilinger (GHZ)
state}
\begin{align}
\label{eq:GHZ}
\ket{\text{GHZ}_M} &=\frac{1}{\sqrt{2}}(\ket{\overset{M}{\overbrace{00\ldots0}}}+\ket{\overset{M}{\overbrace{11\ldots1}}}) \nonumber \\
&=\frac{1}{\sqrt{2}}(\ket 0^{\otimes M}+\ket 1^{\otimes M}),
\end{align}
and the \emph{W state}
\begin{equation}
\ket{{\rm W}_M}=\frac{1}{\sqrt{M}}\left(\ket{100\ldots0}+\ket{010\ldots0}+\cdots+\ket{000\ldots1}\right).
\end{equation}
The GHZ and W states cannot be transformed into each other through LOCC, thereby representing two different kinds of entanglement for three or
more parties \cite{Dur2000} (see, e.g., Ref.~\cite{Horodecki2009}) 

A broad and useful class of multipartite entangled states are the \emph{cluster states} or, more generally, the graph states, which we now describe.

\subsubsection{Graph states}
\label{sec:graph_states}
\begin{figure}
    \includegraphics[width=8cm]{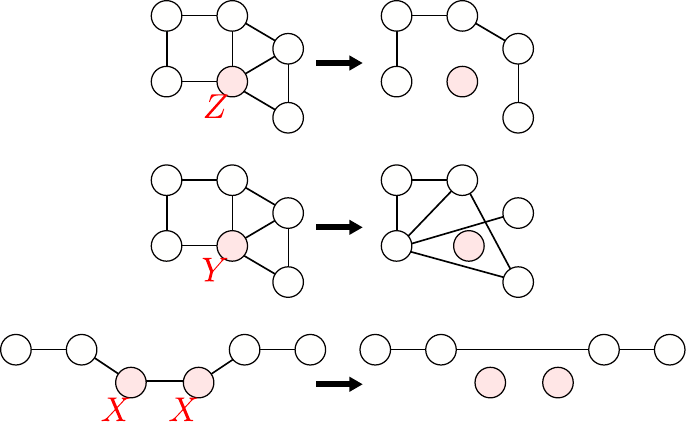}
    \caption{Graphical rules for operations on graph states. The effects of Pauli operations on the connections in the graph states are shown.}
    \label{fig:graph_state_rules}
\end{figure}

A \emph{graph state} $\Ket{G}$ is a multipartite entangled state associated with an undirected graph $G = (V, E)$, with $V$ a set of vertices and $E$ collection of undirected edges $\{ij\}=\{ji\}$, for $i, j \in V$ .
$\Ket{G}$ is then defined through
\begin{equation}
\Ket{G} \equiv \prod_{e \in E} C^Z_e \left( \bigotimes_{v\in V}\Ket{+}_v\right),
\end{equation}
where the controlled-$Z$ (CZ or controlled-phase) gate is a Clifford gate defined on qubits $i$ and $j$ through 
\begin{gather}
    C^Z_{ij} = \Ket{0}\Bra{0}_i \otimes \id_j + \Ket{1}\Bra{1}_i \otimes Z_j.
\end{gather}
$C^Z_{ij}$ is symmetric over $i\leftrightarrow j$, i.e., $C^Z_{ij}=C^Z_{ji}=C^Z_{\{ij \}}$, and $C^Z_{ij}$ and $C^Z_{i'j'}$ commute for any $i,j,i'$ and $j'$. 

A cluster state is a special kind of graph state whose underlying graph $G$ forms a lattice. Performing single-qubit adaptive measurements on a cluster state allow for the execution of a measurement-based quantum computation (MBQC)~\cite{Raussendorf2001}. Whereas one-dimensional (linear) cluster states allow for universal operations on a single qubit, a cluster state of a minimum of two dimensions is necessary to implement a universal set of multi-qubit gates, and additional dimensions are normally needed for error correction and fault tolerance \cite{Raussendorf2006,Raussendorf2007,Raussendorf2007a} (see, e.g., Ref.~\cite{Terhal2015}).

An alternative specification of the graph state is given by the {\it stabilizer formalism}: $\Ket{G}$ is the unique simultaneous eigenstate of all the (stabilizer generator) operators in $\mathcal{S} = \{X_a \otimes Z_{N_a} \vert a  \in V \}$ of commuting operators, where $Z_{N_a}:=\bigotimes_{v\in N_a} Z_v$ and $N_a$ is the set of all the vertices adjacent to vertex $a\in V$ in the graph $G$. 
We say $\Ket{G}$ is \emph{stabilized} by $\mathcal{S}$, making it a stabilizer state analyzable within the stabilizer formalism~\cite{Gottesman1997}.

A thorough and important review of discrete-variable qubit graph states is given in~\cite{Hein2004, Hein2006}. Let us distill their basic properties, illustrated in Fig.~\ref{fig:graph_state_rules}:
\begin{itemize}
    \item Application of local Clifford gates to a graph state is equivalent to that of a sequence of local complementations on the underlying graph (where a local complement of a graph $G$ at a node $a\in V$ is obtained by inverting the subgraph of $G$ induced by the neighborhood $N_a$).
    \item Pauli $Z$ measurement on a node decouples the node and breaks off its incident edges.
    \item Pauli $Y$ measurement on a node takes the local complementation at the node and decouples the node.
    \item Pauli $X$ measurement on two neighboring qubits in a linear cluster state decouples them but connects their other neighbors with an edge.
    \item The entanglement in a connected graph state is \emph{localizable}, meaning that it is possible to project any two qubits in the graph into a Bell pair by performing single-qubit (in particular, Pauli $Z$ or $X$) measurement on the other qubits.
\end{itemize}

The concept of the graph state can be generalized to continuous-variable (CV) bosonic systems, describable in the phase space formalism of the quantum harmonic oscillator with position operator $q$ and momentum operator $p$ such that $[q,p]=i$ ($\hbar =1$). In this case, there is a wealth of possible encodings to choose from. For example, for
a Gaussian graph state~\cite{Menicucci2006}, the plus state becomes the 0-momentum eigenstate of the momentum operator $p$,
\begin{equation}
    \Ket{+} \to \Ket{p=0},
\end{equation}
while for the Gottesman-Kitaev-Preskill (GKP) encoding \cite{GKP}, discussed in Sec.~\ref{sec:photonic_encodings}, the plus state becomes
\begin{equation}
        \Ket{+} \to \Ket{+_\text{GKP}} = \sum_{n=-\infty}^{\infty} \Ket{p=2n\sqrt{\pi}},
\end{equation}
where $\ket{p=2n\sqrt{\pi}}$ is the eigenstate corresponding to the eigenvalue $2n\sqrt{\pi}$ of the momentum operator $p$.
For both of these CV encodings, the CZ gate can be written as
 \begin{align}
    C^Z_{ij} \to e^{i ( q_i \otimes q_j)}
\end{align}
with the position operator $q_i$ for bosonic system $i$. Clifford operations on these encodings correspond to certain Gaussian operations in phase space, which are composed of squeezing, displacements, and rotations. In either case, one uses finitely-squeezed approximations to these states in practice. We give more details on these states in our discussion of photonic encodings in Sec.~\ref{sec:photonic_encodings} and as well as in Sec.~\ref{sec:bosonic_repeaters} on bosonic repeaters.

\subsection{Photonic encodings}

\label{sec:photonic_encodings}

\begin{table*}[bt]
\begin{tabular}{>{\centering}p{3cm}>{\centering}p{2.2cm}>{\centering}p{2.3cm}>{\centering}p{2.1cm}>{\centering}p{2.3cm}>{\centering}p{2cm}>{\centering}p{2cm}}
\toprule
 & \multicolumn{3}{c}{\emph{Single-rail encodings}} & \multicolumn{3}{c}{\emph{Dual-rail encodings}}\tabularnewline
\midrule
 & \textbf{Fock state} & \textbf{Coherent / cat} & \textbf{GKP}  & \textbf{Time-bin} & \textbf{Path} & \textbf{Polarization}\tabularnewline
\midrule
\midrule
\textbf{Cardinality} & DV & CV & CV & DV & DV & DV\tabularnewline
\midrule
\textbf{Physical basis} & Vacuum,

single photon & Coherent states: $\Ket{\pm\alpha}$ & GKP-0 and 1: (Eq.~(\ref{eq:GKP})) & Orthogonal

temporal modes & Orthogonal

spatial modes & Orthogonal polarizations\tabularnewline
\midrule
\textbf{Entanglement }

\textbf{w/ LO} & Deterministic & Deterministic & Deterministic & Probabilistic & Probabilistic & Probabilistic\tabularnewline
\midrule
\textbf{Single-mode }

\textbf{Clifford gates }

\textbf{w/ LO} & Probabilistic & Probabilistic & Deterministic (w/ squeezing) & Deterministic & Deterministic & Deterministic\tabularnewline
\midrule
\textbf{Single-mode}

\textbf{non-Clifford}

\textbf{gates w/ LO} & Probabilistic & Probabilistic & Probabilistic & Deterministic & Deterministic & Deterministic\tabularnewline
\bottomrule
\end{tabular}
    \caption{Descriptions of selected photonic encodings, including associated gate implementations.} 
    \label{tab:photonic_encodings}
\end{table*}

There are several degrees of freedom that one can exploit when encoding quantum information into light. Each one has own advantages and challenges. In this section we review some well-known photonic encodings, summarizing some of this information in Table~\ref{tab:photonic_encodings}.

A few ways exist for categorizing photonic encodings. One is through the cardinality of the Hilbert space. The state space of discrete-variable (DV) encodings is spanned by a finite number of orthogonal (more generally, linearly independent) states, whereas continuous-variable (CV) or bosonic encodings are spanned by infinitely (possibly countably) orthogonal (more generally, linearly independent) states. However, the line between the two kinds of encodings may not always be clear: DV systems can be viewed as finite subspaces of CV spaces, and our interest in CV systems may chiefly be to identify two-dimensional (qubit) subspaces. Furthermore, in practice, various imperfections and interactions with the environment increase the effective dimensionality of DV systems.

Another characterization of photonic encodings is in the number of ``rails.'' In the more restrictive definition, a \emph{single-rail} qubit is associated with the presence or absence of a single photon in an optical (spatial or temporal) mode. More broadly, however, one can view single-rail encodings as those where each state---including states of multiple photons---occupies a single optical mode. Conversely, a \emph{dual-rail} qubit is associated with the presence of a single photon in one of two orthogonal modes. For a single-rail encoding, it is possible to generate entanglement deterministically with linear optical resources, while linear-optical entangling operations are necessarily probabilistic in dual-rail encodings. Conversely, single-qubit rotations for certain single-rail encoded qubits may necessitate nonlinearlity (because the encoding can be based on a superposition of different photon number states, i.e., energy eigenstates), while there exist dual-rail encodings where arbitrary single-qubit rotations are possible only with linear optical elements. See, e.g.,~\cite{Kok2007}.

The following photonic encodings have been frequently considered within quantum information protocols:
\begin{figure}[bt]
    \includegraphics[width=6.7cm]{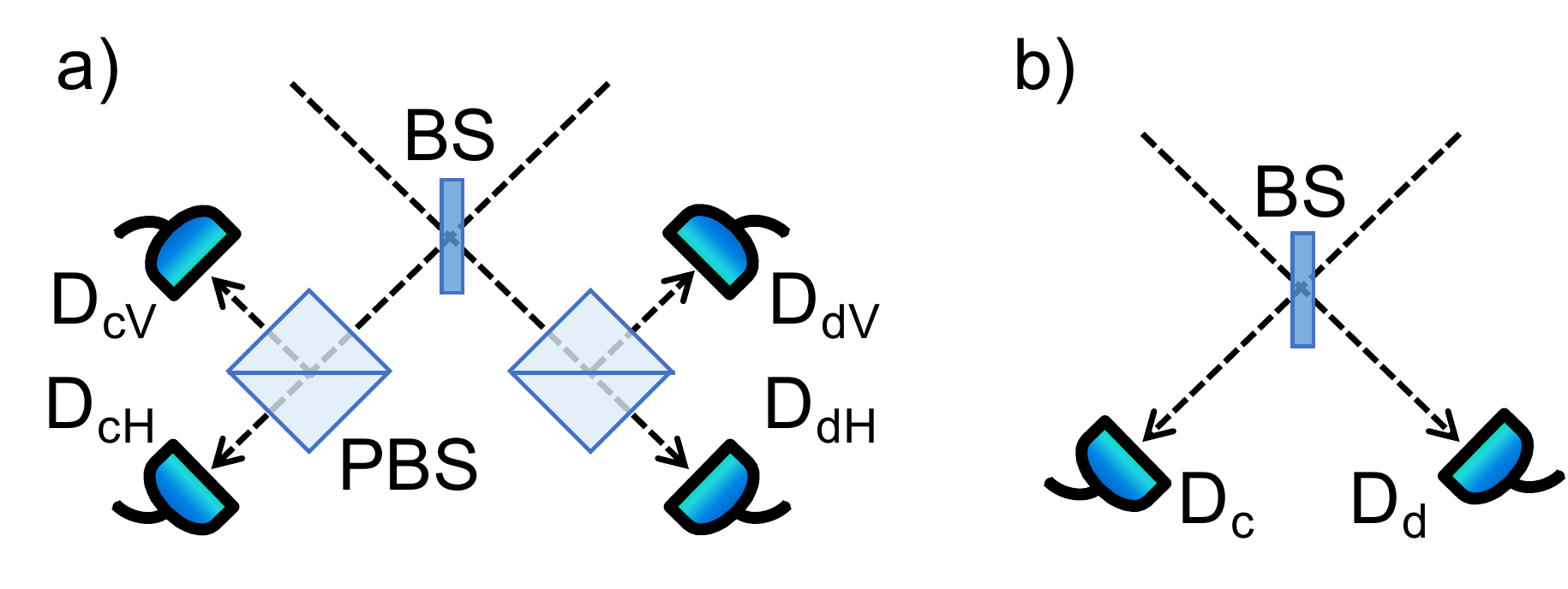}
    \caption{Examples of implementation of Bell measurement. a) Bell measurement for polarization-encoded qubits, spanned by horizontally and vertically polarized single-photon states $\ket{H}$ and $\ket{V}$. This is implemented by the application of a 50:50 beamsplitter (BS) on optical modes, followed by a polarization beamsplitter (PBS) on each of the two output modes and then by photon counting at all the output modes. Clicks in detectors ${\rm D_{cH}}$ and ${\rm D_{cV}}$, or in ${\rm D_{dH}}$ and ${\rm D_{dV}}$, project the received pair of the qubits into Bell state $\ket{\Psi^+}=(\ket{H}\ket{V}+\ket{V}\ket{H})/\sqrt{2}$, while clicks in detectors ${\rm D_{cH}}$ and ${\rm D_{dV}}$, or in ${\rm D_{cV}}$ and ${\rm D_{dH}}$, project the received pair of the qubits into Bell state $\ket{\Psi^-}=(\ket{H}\ket{V}-\ket{V}\ket{H})/\sqrt{2}$. Notice that this Bell measurement can succeed only when the input two optical pulses have 2 (or more) photons in total. b) Bell measurement for Fock-encoded qubits, spanned by the vacuum state $\ket{0}$ and the single-photon state $\ket{1}$. This is implemented by the application of a 50:50 beamsplitter (BS) on optical modes, followed by photon counting at the output modes. A click in the detector ${\rm D_{c}}$ (or ${\rm D_{d}}$) at the constructive-interference (destructive-interference) mode projects the received pair of the qubits into Bell state $\ket{\Psi^+}=(\ket{0}\ket{1}+\ket{1}\ket{0})/\sqrt{2}$ ($\ket{\Psi^-}=(\ket{0}\ket{1}-\ket{1}\ket{0})/\sqrt{2}$). Both implementations can distinguish $\ket{\Psi^\pm}$ from the other states only, and the success probabilities are thus $1/2$ even in the ideal cases.
    }
    \label{fig:BellM}
\end{figure}

\begin{itemize}
    \item \emph{Time-bin}: a photon takes one of two paths of an interferometer of different lengths. Then, $\Ket{0}$ is associated with one path, and $\Ket{1}$ with the other. This encoding is suited for fiber-based communication as it is unaffected by birefringence in optical fibers; however, it is difficult to interact two time-bin qubits, making the encoding preferred for quantum communication over computation.
    
    \item \emph{Polarization}: a kind-of dual rail encoding where a qubit is encoded into the polarization states of a single photon. Conventionally, $\Ket{0}$ is associated with a horizontally polarized photon, and $\Ket{1}$ with a vertically polarized photon. All single-qubit gates can be performed deterministically with waveplates and phase shifters, while linear-optical entangling gates are probabilistic, requiring beamsplitters, waveplates, measurements, and postselection. As an example of the two-qubit operations, an implementation of the Bell measurement is given in Fig.~\ref{fig:BellM}~(a).  
    This encoding prefers free-space over fiber-based communication, as it is vulnerable to birefringence within optical fibers.
    \item \emph{Path}: computational basis states are associated with orthogonal spatial modes. All single-qubit gates can be performed deterministically with beamsplitters and phase shifters; as with the polarization encoding, entangling gates with linear optical resources are probabilistic, requiring beamsplitters, phase shifters, measurements and postselection. As with the time-bin encoding, path-encoded photons prefer fiber-based over free-space communication.
    \item \emph{Fock}: a qubit is encoded into the Hilbert subspace of a single mode spanned by the vacuum state $\ket{0}$ and the single-photon state $\ket{1}$, corresponding to a single-rail qubit. With a phase shifter, we can rotate its Bloch vector along the $Z$-axis freely, but we cannot do so along the $X$-axis since $\ket{0}$ and $\ket{1}$ have different energy. On the other hand, a Bell state (such as $\ket{\Psi^\pm}=(\ket{01}\pm \ket{10})/\sqrt{2}$) can deterministically be obtained with single photon incident on a 50:50 beamsplitter. However, we can discriminate only Bell states $\ket{\Psi^\pm}$ from the others, with a 50:50 beamsplitter followed by two photon detectors (see Fig.~\ref{fig:BellM} (b)). This encoding is sensitive to phase drifts in a transmission channel, and thus, it is preferred in free-space over fiber-based communication.
    \item \emph{Coherent/Cat}: a qubit is encoded into the Hilbert subspace of a single mode spanned by coherent states $\ket{\alpha}$ and $\ket{-\alpha}$ with $\alpha>0$, corresponding to a single-rail qubit. The qubit basis states $\ket{\pm}$ are associated with cat states $\ket{\alpha}\pm \ket{-\alpha})/(2\sqrt{p_\pm})$ with $p_\pm:=(1\pm \langle -\alpha | \alpha\rangle)/2$. The states $\ket{\pm}$ are flipped by a $\pi$-phase shifter, and they are distinguished by a photon-number-resolving detector.
    This encoding is also sensitive to phase drifts in a transmission channel, and thus, it prefers free-space over fiber-based communication. 
    \item \emph{GKP}: the computational basis states are coherent superpositions of infinitely many regularly-spaced position eigenfuntions (i.e., infinitely squeezed states):
    \begin{align}
    \label{eq:GKP}
    \Ket{0_{\rm GKP}} &= \sum_{n=-\infty}^{\infty} \Ket{q=2n\sqrt{\pi}} ,\\
    \Ket{1_{\rm GKP}} &= \sum_{n=-\infty}^{\infty} \Ket{q=(2n+1)\sqrt{\pi}},
    \end{align}
    where $\ket{q=n\sqrt{\pi}}$ is the eigenstate corresponding to the eigenvalue $n\sqrt{\pi}$ of the position operator $q$.
     In realistic implementations, these unphysical infinite-energy states are replaced by their normalizable, finitely-squeezed counterparts. All single-qubit (many-qubit) Clifford gates---including entangling gates---are implementable deterministically through single-mode (multi-mode) Gaussian operations. Non-Clifford gates can be implemented with help of ancillary states and gate teleportation, i.e., they are only deterministic conditioned on the availability of the ancillae.
\end{itemize}

The above encoding schemes are ``digital,'' because they encode a discrete-variable (DV) quantum system with a finite dimensional Hilbert subspace of photonic modes. In contrast, we may also use the photonic modes for ``analog'' encoding, to store a continuous-variable (CV) ``analog'' quantum system with an infinite dimensional Hilbert space. For example, we can encode continuous-variable quantum information using squeezed states, which can be measured via homodyne and heterodyne detectors with a continuous-variable output. For quantum communication, the continuous-variable output can be used to generate secure secret keys \cite{Grosshans2004}. 

One major challenge of using CV encodings for quantum repeaters is the suppression of loss errors. Because of the theorem that Gaussian operations are of no use for protecting Gaussian states against Gaussian errors (including loss errors) \cite{Niset2009}, we have to introduce non-Gaussian operations (e.g., ``quantum scissors'' to truncate the number-state expansion \cite{Pegg1998}) or non-Gaussian ancillary resources (e.g., GKP stabilizer codes to encode an oscillator into many oscillators assisted by GKP ancilla \cite{noh2019}) to overcome loss errors.

\section{Quantum repeaters}
\label{sec:qrepeaters}

This section begins with a review of primitives for quantum repeaters. This is followed with an explanation of quantum repeater protocols through a conceptual classification based on methods to suppress loss and operation errors. We also review all-optical implementations of quantum repeaters.

\subsection{Repeater primitives} \label{sec:primitives}

Here we review quantum teleportation and entanglement swapping as primitives for quantum repeaters. We also summarize various tools for error suppression, which are necessary for quantum repeaters.

\subsubsection{Quantum teleportation}
\label{sec:teleportation}

\emph{Quantum teleportation} is a procedure for transferring quantum information from a sender to a distant receiver without transferring the physical system in which it is encoded \cite{Bennett93}. To accomplish this, the two parties must have established a classical communication link and pre-shared a maximally entangled state. Then, the teleportation consists of two steps. First, the sender locally performs a joint measurement between the state that she wants to transfer and her portion of the pre-shared entangled state. Then, she communicates the measurement outcome to the receiver via the classical channel, who must apply a local unitary operation to his quantum state to recover the original state. There exist quantum teleportation protocols for qudits\footnote{Recently, such high-dimensional teleportation is refocused in the context of quantum networks \cite{Bacco2021} thanks to experimental progress \cite{Hu2020tele,Luo2019}.}~\cite{Werner2001} and CV systems~\cite{Braunstein1998b}; here we focus on qubits to illustrate the concept. 

Suppose that Alice has a qubit in an arbitrary state $\ket{\psi}_{A_1}$ that she wants to send to Bob. Suppose also that she has already shared a Bell state with Bob, $\ket{\Phi^{+}}_{A_2B}$, from Eq.~\eqref{eq:bell_basis}. By performing a Bell-state measurement on her two qubits $A_1$ and $A_2$---that is, a projection onto one of Bell states of Eq.~\eqref{eq:bell_basis}---she will project Bob's qubit onto some state. This state of Bob's qubit $B$ is equal to the initial state $\ket{\psi}$, up to local rotations that are determined by the (random) outcome of Alice's measurement:
\begin{equation}
  \begin{split}
    \ket{\Phi^{+}}_{A_1A_2} & \rightarrow   \ket{\psi}_B, \\
    \ket{\Phi^{-}}_{A_1A_2} & \rightarrow   Z_B \ket{\psi}_B, \\
    \ket{\Psi^{+}}_{A_1A_2} & \rightarrow   X_B \ket{\psi}_B, \\
    \ket{\Psi^{-}}_{A_1A_2} & \rightarrow   Z_B X_B\ket{\psi}_B. \\
  \end{split}
\end{equation}
To conclude the teleportation, Alice must transfer the outcome of her measurement to Bob through a classical channel so that Bob can undo the Pauli byproduct and recover the original state $\ket{\psi}$. Even though Bob has a state locally equivalent to Alice's immediately after the Bell measurement, his ignorance at that point of the precise Pauli gate he has to apply means that Alice cannot transfer quantum information instantly to Bob. The quantum teleportation protocol therefore crucially needs classical communication, making it limited by the speed of light. This impossibility of faster-than-light communication assisted by quantum entanglement is known as the no-signaling principle~\cite{Eberhard1989}. 

Quantum teleportation allows a sender to send arbitrary quantum information encoded into a qubit by consuming an ebit (pre-shared with the receiver) and by sending two bits of classical information to the receiver. This implies that distributing ebits efficiently or quickly by using a quantum communication network is a fundamental question. 

\subsubsection{Entanglement swapping}
\label{sec:swapping}

Entanglement swapping \cite{Zukowski1993} can be thought of as an extension of quantum teleportation where Alice and Bob each share a two-qubit maximally entangled state with Charlie, $C$: $\ket{\Phi^{+}}_{AC_1}$ and $\ket{\Phi^{+}}_{C_2 B}$. After Charlie performs a Bell measurement on his systems $C_1$ and $C_2$, Alice's and Bob's qubits end up in one of the four Bell states, depending on the measurement outcome:
\begin{equation}
\begin{split}
    \ket{\Phi^{+}}_{C_1C_2} & \rightarrow   \ket{\Phi^{+}}_{AB}, \\
    \ket{\Phi^{-}}_{C_1C_2} & \rightarrow \ket{\Phi^-}_{AB} = Z_B \ket{\Phi^{+}}_{AB}, \\
    \ket{\Psi^{+}}_{C_1C_2} & \rightarrow   \ket{\Psi^+}_{AB}=X_B \ket{\Phi^{+}}_{AB}, \\
    \ket{\Psi^{-}}_{C_1C_2} & \rightarrow   \ket{\Psi^-}_{AB}=Z_B X_B \ket{\Phi^{+}}_{AB}.
\end{split} \label{eq:entswap}
\end{equation}

Although their qubits have not directly interacted, Alice and Bob have established a maximally entangled state. This is particularly useful in the context of quantum communication, as it means that entanglement can be propagated through a quantum network even between stationary nodes. Indeed, entanglement swapping is the crux of quantum repeater schemes based on heralded entanglement generation\footnote{The entanglement swapping operation can also be achieved using quantum Zeno effect, requiring no controlled gates~\cite{Bayrakci2022}.}; see Sec.~\ref{sec:IdealQR}, Sec.~\ref{sec:1GQR}, and Sec.~\ref{sec:2GQR}.

\subsubsection{Idealized quantum repeaters}\label{sec:IdealQR}

As shown in the quantum teleportation protocol of Sec.~\ref{sec:teleportation}, once Alice and Bob share a Bell pair (an ebit), Alice can send an unknown state of a qubit to Bob by LOCC, i.e., they can achieve quantum communication. Thus, the challenge of quantum communication reduces mainly to how to distribute a Bell pair between Alice and Bob in practice. Flying qubits---photons---appear to be the medium of choice for this. However, the transmittance $\eta$ of an optical fiber (and hence the ratio of photons sent to photons received) decreases exponentially with its length $L$, according to $\eta=e^{-L/L_{\rm att}}$ of Eq.~(\ref{eq:lossch}). In fact, the transmittance decreases as though it is multiplied by 0.1 every 50\,km in the case of typical optical fibers with attenuation length $L_{\rm att}=22$\,km (and even the quantum and private capacity of the pure-loss bosonic channel (\ref{eq:lossch}) is now known to be described by $-\log_2 (1-\eta)$ ($\approx \eta$ for $\eta \ll 1$) \cite{Pirandola2015} (see Sec.~\ref{sec:internet})).
Hence, simply linking Alice and Bob directly with an optical fiber is not enough to achieve efficient quantum communication, especially if they are far apart\footnote{Notice that the transmittance $\eta$ of the typical fiber with the length of 400\,km is about $10^{-8}$. Therefore, even if the system is designed to achieve the private capacity $-\log_2 (1-\eta)$ with the clock rate of 1\,GHz, the possible key rate is on the order of 10\,bits per second, which seems to be slow for practical applications. Hence, about 400\,km is sometimes said to be a practical distance limit of a fiber-based point-to-point quantum communication.}. 

Here we introduce a simple example to show how a quantum repeater protocol overcomes such an exponential increase of photon loss with the length of an optical fiber. The example is based on heralded entanglement generation and entanglement swapping; it is a simplified protocol designed to capture the main concept of the first-generation quantum repeater protocols \cite{Briegel98,Duan2001,Sangouard2011}, which will appear in Sec.~\ref{sec:1GQR}. The protocol is based on a concatenation allowed by the entanglement swapping of Sec.~\ref{sec:swapping}, dubbed a \emph{DLCZ}-like  protocol after the authors Duan, Lukin, Cirac and Zoller~\cite{Duan2001}. For simplicity, we assume that the fiber attenuation is the only error and all other operations are perfect. 

Suppose that we have a quantum memory $X$ which can establish a Bell state $\ket{\Phi^+}_{Xx}:=(\ket{0}_X\ket{H}_x+\ket{1}_X\ket{V}_x)/\sqrt{2}$ with an optical pulse $x$, where 
$\{\ket{0}_X,\ket{1}_X\}$ is the computational basis of the quantum memory while $\ket{H}_x$ and $\ket{V}_x$ are horizontally polarized and vertically polarized single-photon states of the pulse $x$, respectively. We also assume that an arbitrary state $a \ket{0} +b \ket{1}$ of the quantum memory can be converted into the state $a \ket{H} +b \ket{V}$ of a polarization qubit, if needed.
This kind of memory is an idealized version of a quantum memory, which can be realized by using two atomic ensembles \cite{Sangouard2011} (for example, we ignore any multi-photon excitations that arise in practice).
We also use a linear-optical Bell measurement for polarization-encoded qubits in Fig.~\ref{fig:BellM}~(a). This implementation works as a probabilistic Bell measurement. 

\begin{figure}[tb]
\includegraphics[width=8.6cm]{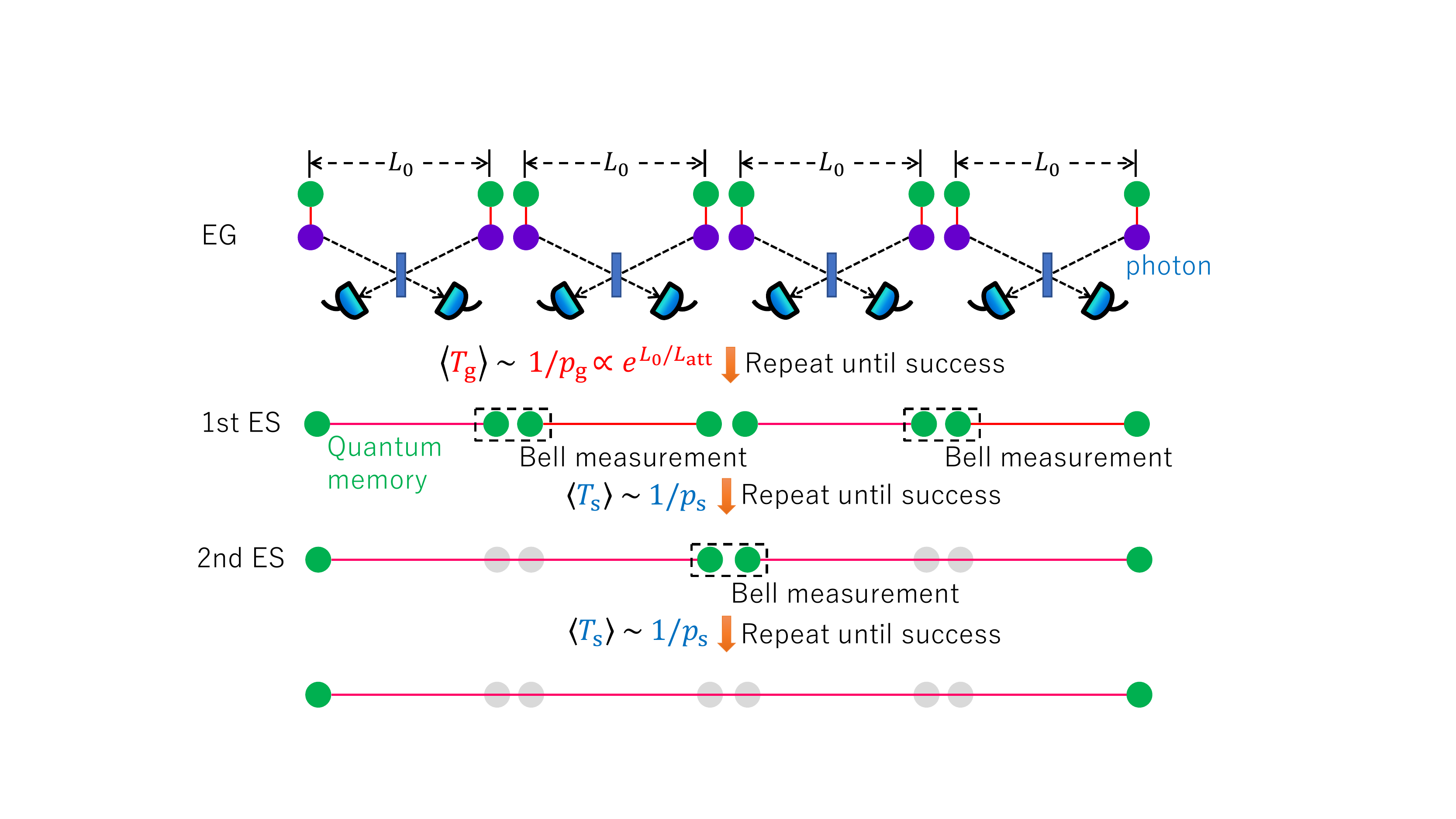}
\protect\caption{Idealized quantum repeater protocol. In this figure, three quantum repeater nodes (corresponding to the case of $N_{\rm QR}=3$) are located at regular intervals between Alice and Bob, who are separated by distance $L$, with $L_0=L/4$. The protocol starts by entanglement generation (EG) between adjacent repeater nodes with success probability $p_{\rm g}(L_0)=e^{-L_0/L_{\rm att}}/2$, followed by entanglement swapping (ES) with success probability $p_{\rm s}$. The EG protocol establishes a Bell pair between adjacent repeater nodes after the number of trials on the order of $\langle T_{\rm g}(L_0)\rangle=p_{\rm g}^{-1}(L_0)$. Given halves of a pair of Bell states, the ES protocol succeeds in swapping the entanglement after the number of trials on the order of $\langle T_{\rm s}\rangle=p^{-1}_{\rm s}$. If a trial of ES fails, we need to start again from EG to go back to the trial. 
Therefore, in the setting of this figure, the average of the total number of trials $T_{\rm tot}^{(3)}$ needed to establish a Bell pair between Alice and Bob is $\langle T_{\rm tot}^{(3)} \rangle \sim \langle T_{\rm g}(L_0) \rangle \langle T_{\rm s} \rangle^2 = p_{\rm s}^{-2} p_{\rm g}^{-1}(L_0) =2 p_{\rm s}^{-2} e^{L/(4 L_{\rm att})}$. This is of the order of the square root of $\langle T_{\rm tot}^{(1)} \rangle$, which is further of the order of the square root of $\langle T_{\rm tot}^{(0)} \rangle$.
}
\label{fig:IdealQR}
\end{figure}

We can generate a Bell state between stations $X$ and $Y$, separated by distance $l$, by combining such a quantum memory, the Bell measurement and optical fibers. To achieve this, the party $X$ (and the party $Y$) first establishes a Bell state $\ket{\Phi^+}_{Xx}$ ($\ket{\Phi^+}_{Yy}$) between her (his) own quantum memory $X$ ($Y$) and an optical pulse $x$ ($y$) locally, and then sends the single photon $x$ ($y$) to a measuring station in the middle of the parties over an optical fiber (modeled by Eq.~(\ref{eq:lossch})) (see a schematic for entanglement generation (EG) in Fig.~\ref{fig:IdealQR}). On receiving the pulses from the separated parties, the measuring station performs the linear-optical Bell measurement of Fig.~\ref{fig:BellM}~(a) on those pulses. This Bell measurement succeeds when both single photons $x$ and $y$ from the separated parties arrive at the measuring station without having been lost (during their travel over the lossy optical fiber), and the surviving photons are projected into a Bell state $\ket{\Psi^+}_{xy}$ or $\ket{\Psi^-}_{xy}$, which occurs with probability $p_{\rm g}(l)=e^{-l/L_{\rm att}}/2$. This success event entangles the quantum memories $XY$ of the separated parties into $\ket{\Psi^+}_{XY}$ or $\ket{\Psi^-}_{XY}$, according to Eq.~(\ref{eq:entswap}). This is called an (heralded) entanglement generation protocol.

If Alice and Bob, separated by distance $L$, run this entanglement generation protocol between them without any quantum repeater, the average of the number $T^{(0)}_{\rm tot}$ of trials needed to obtain a Bell pair will be 
\begin{equation}
    \langle T^{(0)}_{\rm tot}\rangle:=\langle T_{\rm g}(L) \rangle = p_{\rm g}^{-1} (L)= 2 e^{\frac{L}{L_{\rm att}}}, \label{eq:noQR}
\end{equation} 
which grows exponentially with the distance $L$.

Now, let us introduce an entanglement swapping protocol.
Suppose that a single quantum repeater node $C$ is located at the midpoint between Alice and Bob for simplicity, and it runs the above entanglement generation protocols in parallel with Alice and with Bob. 
Then, each of these entanglement generation protocols gives a Bell pair after trials on the order of $\langle T_{\rm g}(L/2) \rangle = 2 e^{L/(2L_{\rm att})} $.  Once it succeeds, the obtained Bell pair can be kept in quantum memories until both of the parallel protocols succeed. Thus, 
they can obtain Bell pairs not only between Alice and the node $C$, but also between the node $C$ and Bob, after trials on the order of $\langle T_{\rm g}(L/2) \rangle = 2 e^{L/(2L_{\rm att})} $, thanks to the parallelization.
Then, after receiving a classical signal to herald this successful sharing of Bell pairs, the node $C$ converts states of local quantum memories into polarization qubits, and then implements the linear-optical Bell measurement of Fig.~\ref{fig:BellM} (a) (corresponding to a schematic for entanglement swapping (ES) in Fig.~\ref{fig:IdealQR}).
This works as entanglement swapping to provide a Bell state between Alice and Bob with success probability $p_{\rm s}=1/2$ of the Bell measurement (in the ideal case).
Hence, the average of the number $T_{\rm s}$ of trials needed for the entanglement swapping to succeed (after the success of the entanglement generation protocols) is $\langle T_{\rm s} \rangle=  p_{\rm s}^{-1}$. On the other hand, if the Bell measurement fails, Alice and Bob start from scratch, i.e., from the parallel entanglement generation protocols. 
Thus, the average of the total number $T_{\rm tot}^{(1)}$ of trials needed to establish a Bell pair between Alice and Bob is 
\begin{equation}
    \langle T_{\rm tot}^{(1)} \rangle \sim \langle T_{\rm g}(L/2) \rangle  \langle T_{\rm s} \rangle =p_{\rm s}^{-1} p_{\rm g}^{-1}(L/2)  = 2 p^{-1}_{\rm s} e^{\frac{L}{2 L_{\rm att}}},    
\end{equation}
(see, e.g., Refs.~\cite{Sangouard2011,azuma2020tools} about the validity of this approximation). Therefore, by comparing this equation with Eq.~(\ref{eq:noQR}), we can conclude that, for a large distance $L$, the existence of a single quantum repeater node $C$ can provide the square-root improvement over the direct entanglement generation between Alice and Bob in the number of trials needed. 

The process for achieving this square-root improvement with entanglement swapping can be concatenated. If Alice and Bob use three equally-spaced quantum repeater nodes, they can achieve further square-root improvement (see Fig.~\ref{fig:IdealQR}); if they use seven, they can have further square-root improvement, and so forth. In particular, if Alice and Bob have $N_{\rm QR}=2^{n}-1$ quantum repeater nodes equally spaced between them, the average of the total number $T_{\rm tot}^{(N_{\rm QR})}$ of trials needed to have a Bell pair between Alice and Bob will be 
\begin{align}
\langle T_{\rm tot}^{(N_{\rm QR})} \rangle \sim& p_{\rm s}^{-n} p_{\rm g}^{-1}(L/2^n) =p_{\rm s}^{-n} e^{\frac{L}{2^n L_{\rm att}}}
\nonumber \\
=&2 p_{\rm s}^{-\log_2(N_{\rm QR}+1)} e^{\frac{L}{(N_{\rm QR}+1)L_{\rm att}} }.   
\end{align}
(again, see, e.g., Refs.~\cite{Sangouard2011,azuma2020tools} about this approximation). This shows the ultimate advantage of utilizing quantum repeaters: the exponential improvement in the number of trials needed to establish entanglement with the number $N_{\rm QR}$ of quantum repeater nodes. Since $p_{\rm s}$ is independent of distance $L$, this exponential improvement enables Alice and Bob to perform quantum communication efficiently over long distances.

This simple model does not include realistic imperfections such as memory errors and imperfect entanglement generation and swapping operations. In practice, these errors will accumulate and become non-negligible over longer distances. However, thanks to the existence of error suppression mechanisms explained in the next Sec.~\ref{subsec:error_suppr}, we can devise several kinds of quantum repeaters which work not only in the presence of loss but also other such imperfections.

\subsubsection{Tools for error suppression}
\label{subsec:error_suppr}

As shown in Sec.~\ref{sec:IdealQR} above, there exists a quantum repeater protocol which enables Alice and Bob to achieve efficient long-distance quantum communication, even with the use of optical fibers impacted by photon loss. However, this protocol was idealized; we assumed that the optical attenuation in fiber is the only source of error and that all other operations are perfect. In practice, there are many physical imperfections that compromise the quality of the resulting entanglement. Therefore, quantum repeater protocols need to be equipped with error suppression mechanisms, which we discuss in this section.

It is helpful to classify error suppression
techniques into two categories: those employing \emph{deterministic} error suppression (including quantum error correction \cite{Lidar13} and one-way
entanglement distillation \cite{Bennett1996}); and those leveraging  \emph{probabilistic} error suppression (including quantum error detection \cite{Lidar13}
and two-way entanglement purification \cite{Bennett96a,Deutsch1996,Briegel98}). 
The former class of techniques succeed deterministically, meaning they do not require users to share a heralding signal alerting each other of the success or failure of the error suppression; on the other hand, the probabilistic nature of the latter class necessitates users to alert each other of success or failure via classical communication and discard the failed instances. For networks with large spatial separation between
the nodes, the time delay associated with this classical heralding signaling
is highly relevant to the performance of the network---for reference, a photon takes roughly 0.5 ms to travel 100 km in an optical fiber. While deterministic error suppression has an advantage in this regard, probabilistic error suppression works even for states which are too noisy to be recovered through deterministic techniques. That is, the probabilistic techniques tend to have higher thresholds on tolerable error or loss probabilities~\cite{Bennett1996}. Let us now briefly summarize both of these types of approaches for suppressing errors.

\paragraph{Deterministic error suppression}
\label{sec:DetErrSupp}

\subparagraph{Quantum error correction}
\label{par:QEC}

The essence of \emph{quantum error correction} (QEC) is to use the redundancy in the entanglement of many physical qubits to encode a logical state and correct for errors. In particular, a qubit is encoded into a two-dimensional subspace of a large Hilbert space composed of many physical qubits rather than directly into a single physical system.
Quantum error correction is a deterministic
process; it is not impacted by the delays associated with classical heralding signals. 
For large-scale quantum networks, having this determinism favourably affects communication rates; on the other hand, physical implementations of QEC codes are demanding due their complexity, and exhibit lower thresholds (to work) on the errors affecting the physical qubits. These thresholds become more stringent as the variety and magnitude of errors increase.

\subparagraph{One-way entanglement distillation}

The purpose of one-way entanglement distillation (1-EDP) is to obtain an almost maximally entangled Bell pair from a set of noisier entangled pairs by applying (direct) one-way LOCC. Here ``one-way'' means that only one party has to communicate the results during the distillation process via
classical communication; there is no backward classical communication. 1-EDP is closely connected to quantum error correction \cite{Horodecki2009}; since there is a
direct mapping from a one-way hashing protocol \cite{Bennett1996} (or
one-way breeding protocol \cite{Bennett96a}) to a quantum error-correcting
code, we will treat them as equivalent at the protocol level.  
In practice, there may be subtle differences in the error accumulation and resource counts between one-way hashing protocols and quantum
error correction. 

\paragraph{Probabilistic error suppression}
\label{sec:ProbErrSupp}

\subparagraph{Quantum error detection}

QEC codes can also be used just to detect errors---that is, to herald the presence of error and discard the state without correcting the error. Quantum
error detection is a \emph{probabilistic} process; as a result, it takes time
to inform the relevant parties, through a classical signal, about the outcome of the error
detection, causing additional delay.

\subparagraph{Heralded entanglement generation protocol (HEGP)} \label{sec:HEGP}
A widely used error detection scheme is the heralded entanglement generation protocol (HEGP), which can generate entanglement on success and detect loss errors on failure. Since entanglement cannot be generated under LOCC, a party needs to generate an entangled state between a local qubit and a flying qubit locally and then to send the flying qubit over a quantum channel. A typical choice of flying qubit is a bosonic system, such as a photonic state; its quantum channel---a bosonic channel---has loss as the dominant noise process. The goal of HEGP, then, is to generate high-quality entanglement between separated parties in a heralded manner, notwithstanding losses in the channel.

Depending on how the quantum information is encoded in the optical modes, or how the local stationary qubits are entangled with the optical modes,
one ought to choose appropriate schemes to detect loss errors. For dual-rail (single-rail) discrete-variable encodings, one generates entanglement using two-photon (single-photon) interference of incoming optical modes from neighboring stations, while detecting loss errors according to click patterns of the photon detection \cite{Duan2001,Sangouard2011,childress2006fault,azuma2012quantum,azuma2012optimal,barrett2005efficient} after the interference.
For continuous-variable (e.g., GKP \cite{GKP}) encoding, one may generate entanglement by combining the two incoming optical modes from neighboring stations followed by homodyne measurements at the output ports.
The outcomes from the homodyne measurements provide information about the likelihood of loss errors, which may be used to determine whether the entanglement generation is successful or not \cite{Fukui2020}. 

If loss errors are detected, the heralded entanglement generation procedure is simply repeated until the two adjacent stations receive the confirmation of certain successful detection patterns via \textit{two-way} classical signaling. Instead of using this time multiplexing, we could also use spatial or frequency multiplexing to run the heralded entanglement generation protocol in parallel so that one of the multiplexed trials succeed with a high probability within a constant time \cite{Sinclair2014}.

\subparagraph{Two-way entanglement distillation protocol (2-EDP)}
\label{sec:2EP}

The purpose of two-way
entanglement distillation (2-EDP) is to produce an almost
maximally entangled pair from noisier entangled pairs
by applying two-way LOCC. 2-EDP allows both parties to communicate with each other using a classical channel, which enables them to compare measurement results or adaptively perform operations
conditioned on the outcomes from the other party. For example, if
the Bell states suffer from bit-flip errors, $\rho_{AB}=\left(1-\epsilon\right)\ket{\Phi^{+}}\bra{\Phi^{+}}_{AB}+\epsilon\ket{\Psi^{+}}\bra{\Psi^{+}}_{AB}$, 
separated parties may use two copies of the states to obtain one pair with a suppressed error of $O\left(\epsilon^{2}\right)$ by comparing measurement outcomes of a parity-check measurement on their own halves \cite{Bennett1996,Deutsch1996,Briegel98}. We may also extend the result
to suppress dephasing errors. For general depolarization errors, we
may use \emph{twirling} \cite{Bennett96a} or switching between phase
and bit errors \cite{Deutsch1996} to further suppress errors.

For ideal operations, we can quickly converge to perfect Bell pairs.
In principle, we can extract entanglement with a rate limited by the two-way distillable entanglement \cite{Bennett1996}. In practice, however, operation errors limit the ultimate fidelity of the
distilled Bell pairs. Various protocols have been proposed to distill entanglement
\cite{Bennett96a,Deutsch1996,JTSL07,Fujii09,Nickerson13,Krastanov19a,Riera-Sabat21}.
For example, one can use multiple copies of imperfect Bell pairs to purify a Bell
pair~\cite{Nickerson13, Fujii09}. One can also use a genetic algorithm to find
the optimal 2-EDP \cite{Krastanov19a}. Existing entanglement can also enhance the performance of 2-EDP~\cite{Riera-Sabat21}. Since there is a direct mapping from 2-EDP to quantum error detection \cite{Dur07}, we may treat them as equivalent
at the protocol level. In practice---just as in the relationship between QEC and 1-EDP---there may be subtle differences in error accumulation and resource counts
between quantum error detection and 2-EDP.

For CV encoding, due to the Gaussian entanglement distillation no-go theorem \cite{Eisert2002, Fiurasek2002, Giedke2002}, the CV repeaters use non-Gaussian operations at the entanglement distillation protocols \cite{Ralph2009, Fiurasek2010} to suppress loss errors.
Instead, we may distill entanglement using nondeterministic noiseless linear amplification (NLA) with quantum scissors \cite{Pegg1998, Ralph2009} or other non-Gaussian filtering with single-photon addition and subtraction operations \cite{Fiurasek2010}.

\begin{table}
\begin{tabular*}{8.7cm}{@{\extracolsep{\fill}}>{\centering}m{1.5cm}>{\centering}m{3.5cm}>{\centering}m{3.5cm}}
\toprule 
 & \emph{Deterministic} error suppression & \emph{Probabilistic} error suppression\tabularnewline
\midrule
\midrule 
\multirow{2}{1.5cm}{Schemes} & Quantum error correction  & Quantum error detection\tabularnewline
\cmidrule{2-3} \cmidrule{3-3} 
 & One-way entanglement distillation & Two-way entanglement purification\tabularnewline
\midrule 
Signaling & No delay & Delay\tabularnewline
\midrule 
\multirow{2}{1.5cm}{Threshold to work} & $\eta>1/2$ for loss of qubits or bosons  &  $\eta>0$ for loss of qubits or bosons \tabularnewline
\cmidrule{2-3} \cmidrule{3-3} 
 & $p< 1/4$ at least for depolarization of qubits &  $p< 1/2$ for depolarization of qubits\tabularnewline
\bottomrule
\end{tabular*}\caption{\label{tab:Deterministic-and-probabilistic} Comparison between deterministic and probabilistic
error suppression protocols.}

\end{table}

\paragraph{Comparison of deterministic and probabilistic quantum error suppression}

Deterministic error suppression has no corresponding
classical signaling delay. However, it imposes a threshold of 50\%
on the loss of qubits or bosonic systems (associated with the transmittance $\eta$ as $\eta>1/2$ if they are sent over a pure-loss channel, as in (\ref{eq:lossch})) \cite{Bennett1996,bennett1997capacities,giovannetti2003broadband,giovannetti2003entanglement}. Furthermore, this category of protocols will not work at all for qubits sent over a depolarizing channel (\ref{eq:depolarizing}) with strength $p> 1/4$ \cite{bennett1997capacities,Bennett1996,knill1997theory}, although they work for $p \lesssim 0.18929$ with the hashing protocol \cite{Bennett1996} and even for $p\lesssim 0.19130$ with a concatenated coding scheme \cite{fern2008lower}).
Probabilistic error suppression has an associated classical signaling delay, but it can tolerate larger errors. In principle, it works if the transmission probability of qubits or bosonic systems is nonzero \cite{bennett1997capacities,pirandola19} or if qubits are sent over a depolarizing channel with $p< 1/2$ \cite{Bennett1996,Deutsch1996}. 
We summarize and compare the properties of deterministic and probabilistic
error suppression protocols in Table \ref{tab:Deterministic-and-probabilistic}.

\subsection{Generations of quantum repeaters}
\label{sec:gens}

There are two major challenges for fiber-based quantum communication over long distances. First, as pointed out in Sec.~\ref{sec:IdealQR}, fiber attenuation during transmission leads to an exponential decrease in the entangled-pair generation rate. Second, several operational errors such as channel errors, gate errors, measurement errors, and quantum memory errors, severely degrade the quality of the obtained entanglement. Different from classical information, quantum information is encoded as quantum states that cannot be amplified or duplicated deterministically due to the quantum no-cloning theorem (see Sec.~\ref{sec:no-cloning}).

To overcome these challenges, quantum repeaters (QRs) have been proposed for the faithful realization of long-distance quantum communication \cite{Briegel98}.
As exemplified in Sec.~\ref{sec:IdealQR}, the essence of QRs is to divide the total distance of communication into shorter intermediate segments connected by QR stations, in which loss errors from fiber attenuation can be corrected. Active error suppression schemes are also employed at every repeater station to correct operation errors, i.e., imperfections induced by the channel, measurements, and gate operations. In the following, we will classify quantum repeaters according to how one suppresses loss and operation errors---using \textit{probabilistic} error suppression (Sec.~\ref{sec:ProbErrSupp})  or \textit{deterministic} error suppression (Sec.~\ref{sec:DetErrSupp})---which will lead to a different scaling of quantum communication rates.

\begin{table}
\begin{tabular*}{8.7cm}{@{\extracolsep{\fill}}>{\centering}b{1.3cm}>{\centering}b{1.7cm}>{\centering}b{1.7cm}>{\centering}b{1.7cm}>{\centering}b{1.7cm}}
\toprule
Errors & Error suppression & 1G & 2G & 3G\tabularnewline
\midrule
\midrule
\multirow{2}{1.3cm}{Loss error} & Probabilistic & $\checkmark$ & $\checkmark$ & \tabularnewline
\cmidrule{2-5} \cmidrule{3-5} \cmidrule{4-5} \cmidrule{5-5}
 & Deterministic &  &  & $\checkmark$\tabularnewline
\midrule
\multirow{2}{1.3cm}{Operation error} & Probabilistic & $\checkmark$ &  & \tabularnewline

\cmidrule{2-5} \cmidrule{3-5} \cmidrule{4-5} \cmidrule{5-5}
 & Deterministic &  & $\checkmark$ & $\checkmark$\tabularnewline
\midrule
\midrule
\multicolumn{2}{c}{Time scale} & $\max\left(\frac{L}{c},t_{0}\right)$ & $\max\left(\frac{L_{0}}{c},t_{0}\right)$ & $t_{0}$\tabularnewline
\midrule
\multicolumn{2}{c}{Cost coefficient} & $\textrm{poly}\left(L\right)$ & $\textrm{polylog}\left(L\right)$ & $\textrm{polylog}\left(L\right)$\tabularnewline
\bottomrule
\end{tabular*}\caption{\label{tab:QRGenerations}Three generations of quantum repeaters classified according to probabilistic or deterministic suppression of loss and operation errors. The time scale (key generation rate) and cost coefficient scale differently with the total distance $L$, repeater spacing $L_{0}$, and gate time $t_{0}$.}
\label{tab:generations}
\end{table}

For probabilistic error suppression protocols, we need \textit{two-way} classical signaling to inform relevant repeater nodes whether to proceed to the next step (if error suppression succeeds) or to make another attempt (if error suppression fails). 
A widely used error detection scheme to suppress loss errors is the heralded entanglement generation protocol (HEGP), as exemplified dual-rail photonic encoding in Sec.~\ref{sec:IdealQR}. 
For single-rail or CV encoding, photon click patterns may not immediately identify loss events, but we may use other non-Gaussian operations (e.g., non-deterministic noiseless linear amplification (NLA) with quantum scissors \cite{Pegg1998, Ralph2009} to suppress loss errors.
If loss errors are detected, one simply repeats the heralded entanglement generation procedure until the two adjacent stations receive the confirmation of certain successful detection patterns via \textit{two-way} classical signaling.
Similarly, to achieve \textit{probabilistic} suppression of operation errors, a popular error detection scheme is the two-way entanglement distillation protocol (2-EDP), which consumes several low-fidelity Bell pairs to probabilistically generate a smaller number of higher-fidelity Bell pairs \cite{Deutsch1996,Dur1998}. Like HEGP, to confirm the success of purification, \textit{two-way} classical signaling between repeater stations for exchanging measurement results is required.
The time delays from the two-way classical signaling may decrease the communication rates.

To achieve \textit{deterministic} error suppression of loss errors or operation errors, we may use quantum error correction \cite{Jiang2008,Munro2010,Fowler2010,Muralidharan2014,Azuma2015,li2013long} or one-way entanglement distillation \cite{Bennett1996, Zwerger2018}. The key idea is to encode a logical qubit into a block of physical qubits that are sent through the lossy channel, and then to use quantum error correction to restore the logical qubit. One may also include \textit{one-way} classical signaling to assist the deterministic one-way entanglement distillation protocols \cite{Bennett96a,Bennett1996}, but the additional one-way (forward) classical signaling from the sender does not affect the quantum channel capacity. Hence, all the deterministic error suppression (even when assisted by one-way signaling) can correct no more than $50\%$ loss, which is consistent with the no-cloning theorem \cite{Stace2009,Muralidharan2014}, and not more than $25\%$ depolarizing errors (see Table~\ref{tab:Deterministic-and-probabilistic}). The existence of these finite thresholds itself implies the need of quantum repeater nodes in the case of the use of deterministic error suppression, as such errors tend to depend on the communication distance \cite{Briegel98}.

Based on the methods adopted to suppress loss and operation errors, we can classify various QRs into three categories, as shown schematically in Table~\ref{tab:generations}.
We refer to these as first, second, and third generations of QRs \cite{Muralidharan2014,Munro2015,Muralidharan2016}, 
to imply the increasing difficulty in technology with improved performance\footnote{We may also classify QRs using other criteria, such as the physical platform, different operations, and so on. For instance, see Ref.~\cite{razavi2018introduction}. We will discuss various physical platform and implementation in Sec.~\ref{sec:exps}.}.
Note that the combination of \textit{deterministic} suppression of loss errors and \textit{probabilistic} suppression for operation errors, which only does not appear in Table~\ref{tab:generations}, is sub-optimal compared to the other three combinations. 

Each generation of QRs performs best for a specific regime of operational parameters such as local gate speed, gate fidelity, and coupling efficiency. We consider both the temporal and physical resources consumed by the three generations of QRs and identify the most efficient architecture for different parameter regimes. The results can guide the design of efficient long-distance quantum communication links that act as elementary building blocks for future quantum networks.

\subsubsection{First-generation repeaters}\label{sec:1GQR}

\begin{figure}[bt]
\includegraphics[width=8.5cm]{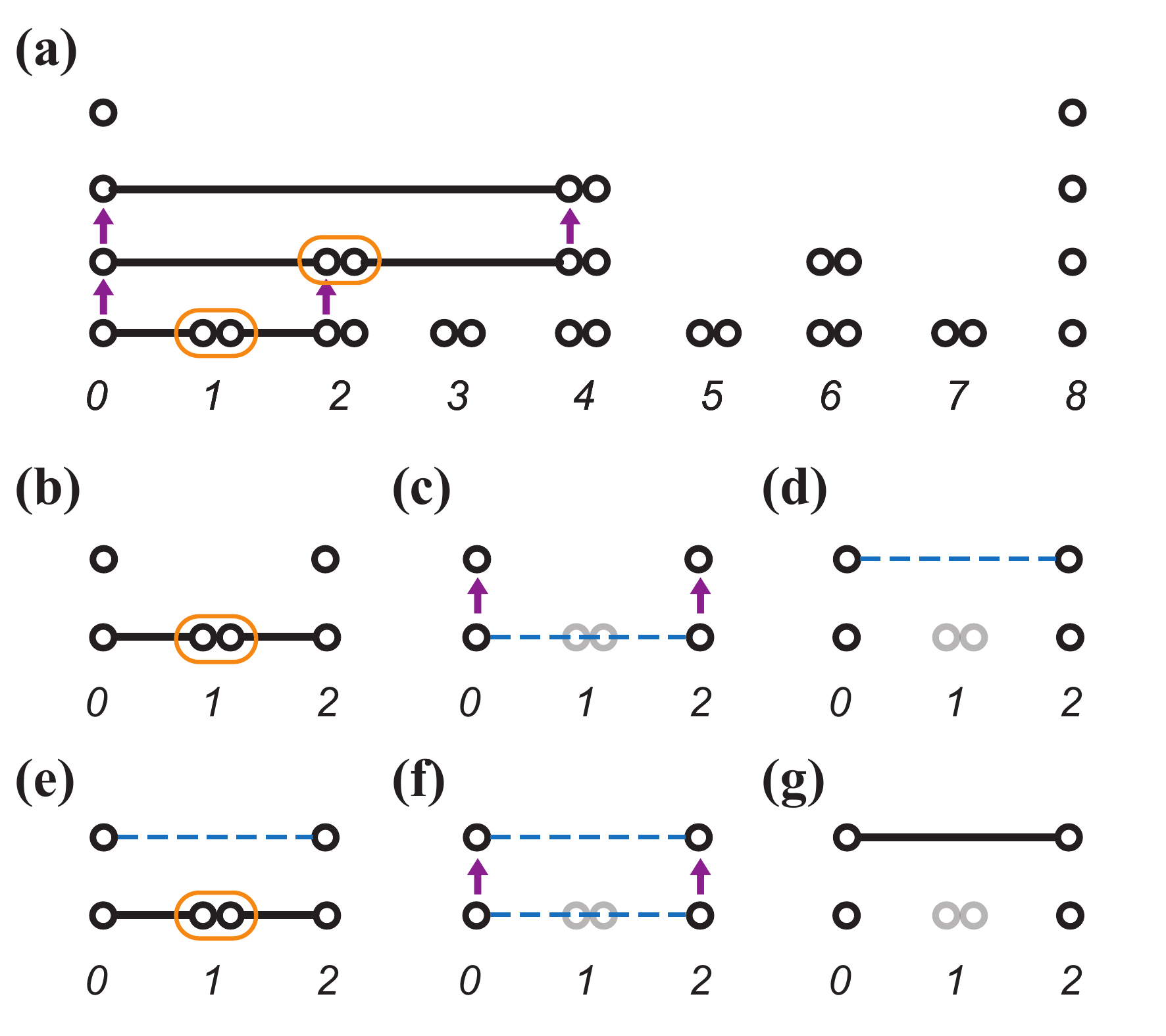}
\protect\caption{The first generation repeater protocol (BDCZ scheme \cite{Briegel98}). (a) In a realization based on the pumping protocol with $N + 1 = 9$ nodes, the number of qubits per node is bounded by $2 \log_2 2N = 8$. Each (orange) oval surrounding two vertices (or two qubits) describes application of Bell measurement to the two qubits for entanglement swapping.  
(b–d) Two entangled pairs with distance 1 are connected through entanglement swapping [(orange) oval)] at node 1 to produce an entangled state with distance $2$, which is stored in the qubits [as described by (purple) arrows] at higher level. (e–g) Another entangled state with distance $2$ is produced to purify the entangled state [as described by (purple) arrows] stored in qubits at higher level. Similarly, entangled states with distance $2^n$ can be connected to produce entangled state with distance $2^{n+1}$, which may be further purified, as indicated in (a). Figure from Ref.~\cite{Jiang2007}.
Copyright (2007) National Academy of Sciences.}
\label{fig:BDCZ}
\end{figure}

The first generation of QRs uses \textit{probabilistic} error suppression
to overcome practical imperfections---for example, HEGP can herald
the successful entanglement generation while overcoming loss errors
and 2-EDP can use two-way classical signaling to recognize successful
entanglement distillation to suppress operation errors~\cite{Briegel98,Sangouard2011,azuma2012quantum,PhysRevLett.96.240501,childress2006fault,munro2008high,zwerger2012measurement,kok2003construction}.
Since we have explained the principle of QRs by exemplifying a simplified first-generation QR protocol in Sec.~\ref{sec:IdealQR}, here we start by briefly summarizing how QRs from this generation can be used to
correct losses with a simple example in which we assume there are
no operation errors. Alice and Bob, separated by a distance $L_{{\rm tot}}$,
want to share a maximally-entangled qubit pair that they can use,
e.g., to teleport a quantum state or to distill a private key.
They are connected by a lossy medium such as a telecom fiber, having
the typical loss of $0.2$\,dB/km (that is, an attenuation length
$L_{{\rm att}}\approx22$\,km). Supposing that $L_{{\rm tot}}\gg L_{{\rm att}}$,
the direct transmission of a photon between Alice and Bob succeeds
with a vanishingly small probability on the order of $e^{-L_{{\rm tot}}/L_{{\rm att}}}\ll1$.

The solution provided by first-generation QRs is to divide the total
distance $L$ between Alice and Bob into smaller lengths with the
help of $N_{{\rm QR}}=L_{{\rm tot}}/L_{0}-1$ quantum repeater nodes. Here we assume
that the nodes are evenly separated by an internodal distance $L_{0}$ and $L_{0}=L_{{\rm tot}}/2^{n}$ for simplicity.
The role of each QR node is to share entanglement with its adjacent
nodes: we use an HEGP strategy to create high-quality entanglement
between a quantum memory and its counterpart to the immediate left,
and between another memory and its counterpart in the adjacent QR
node to the right. Each HEGP trial also takes a time $T_{{\rm trial}}=t_{{\rm op}}+t_{{\rm c}}$,
which depends on the total time $t_{{\rm op}}$ of operations and on the
time $t_{{\rm c}}=L_{0}/c$ for photons to arrive at the central measuring station and the classical signaling back to the QR node. 

A typical HEGP procedure has success probability $P_{{\rm ent}}$
depending on the photon collection efficiency, fiber transmission
efficiency, and photon detection efficiency. For the dual-rail encoding,
without ancillary photons, the success probability $P_{{\rm ent}}\le1/2$ even in the lossless
limit, limited by linear optics and by the photon loss probability \cite{calsamiglia2001maximum}.
However, we may use more advanced encoding to achieve a higher success
probability $P_{{\rm ent}}>1/2$ \cite{azuma2009optimal,azuma2012quantum,Martin2019}.
In any case, if it succeeds, the HEGP tends to present high-quality entanglement
between nearest neighbor nodes even under the existence of photon
loss. Due to the probabilistic nature of HEGP, for the first-generation
QR protocol to proceed, it is necessary to inform the adjacent nodes
whether the HEGP has succeeded or not. In the case of a failure, the
process is repeated until it succeeds. The entanglement generation
procedure therefore succeeds in an average time $\langle T_{{\rm ent}}\rangle=P_{{\rm ent}}^{-1}T_{{\rm trial}}=P_{{\rm ent}}^{-1}(L_{0}/c+t_{{\rm op}})$.
In the case of a success, the entanglement can be stored in the quantum
memories.

At each QR node, we can store entangled qubit pairs shared with an
adjacent node, say the node on the immediate left, during the time
required to produce an entangled pair with the adjacent QR node on
the right. Thanks to this functionality of quantum memories, we see
that not all the entanglement needs to be generated at the same time
throughout the network; this is the reason that this strategy can
outperform direct photon transmission and a quantum relay protocol \cite{waks2002security,jacobs2002quantum,de2004long} (which uses repeater nodes but only distributes photonic Bell pairs from sending repeater nodes to their adjacent receiver nodes, in which Bell measurement is performed soon after receiving halves of the Bell pairs). When a QR
node finally shares an entangled pair of qubits with each of its adjacent
nodes, it performs entanglement swapping between its two quantum memories,
such that if it succeeds, a maximally entangled pair is now shared
between its two adjacent nodes. After repeating these entanglement
swapping steps at each QR node, Alice and Bob end up with a maximally-entangled
pair at a rate much higher than what is achievable with direct fiber
transmission (see Sec.~\ref{sec:IdealQR} for detail).

So far, we have only considered loss errors and have thus assumed
that information can be manipulated, transferred and stored faithfully.
In practice, this is not the case; we ought to also handle operation
errors, which eventually reduce the fidelity of the two qubits shared
by Alice and Bob. This is achieved through an entanglement distillation
scheme, which can be incorporated in first-generation QRs, for example,
using a nested purification QR scheme, as introduced in the following
paragraph.

As illustrated in Fig.~\ref{fig:BDCZ}, we start with distilled high-fidelity
entangled pairs with separation $L_{0}=L_{{\rm tot}}/2^{n}$, created
and stored in adjacent stations. At the $k$-th nesting level, two
entangled pairs of distance $L_{k-1}=2^{k-1}L_{0}$ are connected
by entanglement swapping to extend entanglement to a distance $L_{k}=2^{k}L_{0}$
\cite{Zukowski1993}. As practical gate operations and entanglement
swapping (Fig.~\ref{fig:BDCZ}~(b-d)) inevitably cause the fidelity of entangled pairs to drop,
2-EDP may be incorporated at each level of entanglement extension
(Fig.~\ref{fig:BDCZ}~(e-g)) \cite{Deutsch1996,Dur1998}. With $n$
nesting levels of connection and distillation, a high-fidelity entangled
pair over distance $L_{n}=L_{{\rm tot}}$ can be obtained. Suppose
$T_{k-1}$ is the average time needed to prepare a distilled entangled
pair over distance $L_{k-1}$, average time to prepare a distilled
entangled pair over distance $L_{k}$ is
\begin{equation}
T_{k}=\alpha_k T_{k-1}+ \beta_k L_{k}/c=\alpha_k T_{k-1}+\beta_k 2^{k}t_{\rm c},
\end{equation}
where $t_{{\rm c}}=L_{0}/c$ is the communication time between neighboring repeater stations, 
$\alpha_k$ and $\beta_k$ are dimensionless numbers capturing the time overhead
associated with the entanglement swapping, distillation, and multiple rounds of classical communication. For simplicity, we assume each nesting level has similar overheads $\alpha_k \approx \alpha$ and $\beta_k \approx \beta$ for $k \ge 1$. The average time to generate distilled entangled pairs between neighboring repeaters is $T_0 = \beta_0 t_c$, with the time overhead $\beta_0$ associated with photon efficiency, entanglement generation and purification between neighboring repeater stations.
From
the recursive relation, we can obtain the average time to generate
a distilled entangled pair over distance $L_{n}=L_{{\rm tot}}$ is
\begin{equation}
T_{\rm tot}=T_{n}\sim\left(L_{\rm tot}/L_{0}\right)^{\log_{2}\left[\max\left(\alpha,2\right)\right]} \max\left(\beta,\beta_0 \right) t_{\rm c},
\end{equation}
which increases polynomially with $L_{\rm tot}$ depending on the value
of $\alpha$. 

For the simple mode of loss-only channel, $\alpha\approx\frac{3}{2}\frac{1}{P_{\rm swap}}$,
with prefactor $\frac{3}{2}$ for the time overhead associated with
the requirement that two entangled pairs on both sides should be ready
for entanglement swapping \cite{Jiang2007,azuma2020tools,Sangouard2011}, and $P_{\rm swap}$
for the success probability of entanglement swapping. For example,
$P_{\rm swap}\le1/2$ for the Duan-Lukin-Cirac-Zoller quantum repeater
protocol based on atomic ensembles and linear optics~\cite{Duan2001}.
To overcome operation errors, we need entanglement distillation from
at least two copies of entangled pairs, and hence $\alpha\ge2$ for all
entanglement distillation schemes (e.g., the Briegel-D\"{u}r-Cirac-Zoller
(BDCZ) protocol~\cite{Briegel98} and the Childress-Taylor-S\o rensen-Lukin
(CTSL) protocol \cite{childress2006fault}), unless we use multiplexing
in generating entangled pairs \cite{Dur1998}. 

The first generation of QRs reduces the exponential overhead in direct
state transfer to only polynomial overhead, which is limited by the
two-way classical signaling required by HEGP between non-adjacent
repeater stations. The communication rate still decreases polynomially
with distance and thus becomes very slow for long-distance quantum
communication. The communication rate of first-generation QRs can
be boosted using temporal, spatial, and/or frequency multiplexing
associated with the internal degrees of freedom for the quantum memory
\cite{Sangouard2011,Banarota2011}.

The first generation of QRs can also be very efficient in entanglement resources. As shown in Fig.~\ref{fig:BDCZ}, the BDCZ protocol~\cite{Briegel98,Dur1998} has a self-similar structure with $n=\log_2 \frac{L_{\rm tot}}{L_{0}}$ nesting levels. We start with the elementary entangled pairs with initial fidelity\footnote{A general definition of the fidelity between states $\rho$ and $\sigma$ is given by $F(\rho,\sigma):=\|\sqrt{\rho}\sqrt{\sigma}\|^2$, where $\|X\|:={\rm Tr} \sqrt{X^\dag X}$ is the trace norm \cite{Jozsa1994}. The ``initial fidelity'' here means the fidelity of an initial state $\rho$ to a Bell state $\ket{\Phi^+}$, i.e., $F=F(\rho, \ket{\Phi^+})=\bra{\Phi^+}\rho\ket{\Phi^+}$.} $F$ and distance $L_0$ between neighboring repeater nodes. In the $j$-th nesting level (with $j=1,2,\cdots,n$), a repeater node performs entanglement swapping to convert two initial entangled pairs with fidelity $F$ and the length $2^{j-1} L_{0}$ into an entangled pair with fidelity $F’(\le F$ in general) and length of $2^j L_0$. The extended entangled pairs with fidelity $F'$ are collected, and $M$ pairs of them are used to distill a purified entangled pair with the initial fidelity $F$ and the length of $2^j L_0$ through an entanglement distillation protocol. These imply that each purified entangled pair with fidelity $F$ and with the length of $2^j L_0$ can be regarded as having been made from $2  M$ entangled pairs with fidelity $F$ and with the length of $2^{j-1} L_0$. Therefore, an entangled pair with fidelity $F$ and with the length of $L_{\rm tot}=2^n L_0$ can be made up from $(2  M)^n=(L_{\rm tot}/L_0)^{1+\log_2 M}$ elementary entangled pairs.

In addition, the first generation of QRs can be highly efficient even in terms of quantum memory resources, if the purification of an unpurified entangled pair with the length of $2^jL_0$ $(j=n,n-1,\cdots,0)$ can be done by a sequential application of the pumping protocol which ``pumps'' entanglement to the entangled pair out of a fixed unpurified auxiliary entangled pair with the same length of $2^jL_0$ (see Fig.~\ref{fig:BDCZ}) \cite{Briegel98,Dur1998}. Here, how much entanglement is purified depends on both the initial fidelity and the shape of the fixed auxiliary pair.
During the purification, we just need two pairs of memories, one for storing the entangled pair to be pumped and the other for storing the auxiliary entangled pair for each round, and the purification is regarded as having started from two unpurified pairs with the length of $2^jL_0$. One of these two unpurified pairs, as the auxiliary entangled pair, should be prepared repeatedly during the pumping purification, and it can be regarded as having been obtained by connecting two purified entangled pairs with the length of $2^{j-1}L_0$ through entanglement swapping. As a result, a purified entangled pair with the length of $2^j L_0$ can be regarded as having been made from an unpurified entangled pair (to be pumped at the $j$-th nesting level) with the length of $2^jL_0$ and from two purified pairs with the length of $2^{j-1}L_0$. 
By considering this recursively from $j=n$ to $j=1$, a purified entangled pair with the length of $2^n L_0(=L_{\rm tot})$ is regarded as having been made from $1$ unpurified entangled pair (to be pumped at the $n$-th nesting level) with the length of $2^nL_0$, $2$ unpurified pairs (to be pumped at the $n-1$-th nesting level) with the length of $2^{n-1}L_0$, $\cdots$, $2^{n-1}$ unpurified pairs (to be be pumped at the $2$nd nesting level) with the length of $2 L_0$, and $2^n$ {\it purified} pairs with the length of $L_0$. Since each of these entangled pairs needs two quantum memories, the maximum number $N_{\rm tot}$ of memories required during the protocol is $N_{\rm tot}=2\sum_{j=0}^n 2^{j}=2(2^{n+1}-1)=2(2L_{\rm tot}/L_0-1)=4L_{\rm tot}/L_0-2$. For example, we have $n=3$ in Fig.~\ref{fig:BDCZ}~(a), where 30 quantum memories are written, corresponding to $N_{\rm tot}$ memories.

There are different variations of the BDCZ protocol.
Its measurement-based implementation using graph
states is given in Ref.~\cite{zwerger2012measurement}. The DLCZ
protocol simplifies it with the use of atomic ensembles and linear optics~\cite{Duan2001}.
Room-temperature quantum repeaters have also been proposed using nitrogen-vacancy defect centers in diamond \cite{childress2006fault,Ji2022}.
Reference~\cite{Sangouard2011} provides a detailed review on various
first-generation quantum repeaters based on atomic ensembles and linear
optics, where HEGPs are based on Fock-state encoding, polarization
encoding, and time-bin encoding. 
The concept of the nested purification in the BDCZ protocol, as well as the concatenation of quantum error-correcting codes \cite{knill1997theory}, is generalized to distribute entangled pairs with fixed error to clients in a quantum network with arbitrary topology, regardless of their distance \cite{azuma2023networking}.

We can further generalize the BDCZ protocol by introducing CV encoding. For example, we can take a hybrid CV-DV approach by interfering optical coherent-state signals to generate DV entanglement between repeater stations \cite{azuma2012quantum,PhysRevLett.96.240501,childress2006fault,munro2008high}. Moreover, we can design CV quantum repeaters to efficiently distribute CV entangled states with high fidelity over long distances \cite{Dias2017, Furrer2018, Seshadreesan2020}. Due to the Gaussian entanglement distillation no-go theorem \cite{Eisert2002, Fiurasek2002, Giedke2002}, the CV repeaters use non-Gaussian operations at the entanglement distillation protocols \cite{Ralph2009, Fiurasek2010} to suppress loss errors.

 \subsubsection{Second-generation repeaters}\label{sec:2GQR}
 The second generation of QRs uses \textit{probabilistic} error suppression (see Sec.~\ref{sec:ProbErrSupp}) for loss errors and \textit{deterministic} error suppression (see Sec.~\ref{sec:DetErrSupp}) for operation errors \cite{Jiang2008,Munro2010,li2013long,Mazurek2014Long-distanceMemory}.
 For example, we may first prepare the encoded states $|0\rangle_{L}$ and $|+\rangle_{L}$ using the Calderbank-Shor-Steane (CSS) codes and store them at two adjacent stations. CSS codes are considered because of their fault-tolerant implementation of preparation, measurement, and encoded CNOT gates \cite{Jiang2008, NielsenChuang2010}.
 Then, an encoded Bell pair $|\Phi^{+}\rangle_{L}=\frac{1}{\sqrt{2}}\left(|0,0\rangle_{L}+|1,1\rangle_{L}\right)$ between adjacent stations can be created via teleportation-based non-local CNOT gates \cite{Gottesman1999,Jiang2008} applied to each physical qubit in the encoded block using the entangled pairs generated through HEGP process. Finally, QEC is carried out when entanglement swapping at the encoded level is performed to extend the range of entanglement. Second-generation QRs use QEC to replace 2-EDP and therefore avoid the time-consuming two-way classical signaling between non-adjacent stations. The communication rate is then limited by the time delay associated with two-way classical signaling between adjacent stations and local gate operations.  If the probability of accumulated operation errors over all repeater stations is sufficiently small, we can simply use the second generation of QRs \textit{without} encoding. For instance, proposals based on single ion qubits, to which we can apply deterministic Bell measurement, fall into this category \cite{Asadi2020,Asadi2018,Sangouard2009}.

We can generate entangled pairs through the HEGP process adapted for different photonic encoding schemes (see Sec.~\ref{sec:photonic_encodings}). For dual-rail photonic encoding (time-bin, polarization, or path), we may use linear optics and photon detectors to herald the successful Bell measurement and also detect photon loss errors (e.g., Fig.~\ref{fig:BellM}~(a)). The potential limitation is that the success probability of the Bell measurement will be upper-bounded by 50\% for dual-rail encoding. Alternatively, we may use bosonic encodings, such as GKP states, for HEGP~\cite{Fukui2020}. Different from the dual-rail encoding schemes, the GKP encoding can achieve deterministic Bell measurement with linear optics and homodyne detection \cite{GKP}. In the presence of loss errors, there will be vacuum noise added to the system, which can be detected by the homodyne measurement. The GKP encoding can correct small added vacuum noise up to certain level, above which it is better to report the presence of large noise and restart the process.

Similar to the first-generation repeaters, we can also give bounds on the achievable communication rate for the second generation repeaters, which is limited by the HEGP and 2-EDP between neighboring repeater stations. For example, we have 
$R \leq {\langle T_{\rm ent} \rangle}^{-1} \leq [2(L_0/c + t_{\rm op})]^{-1}.$
By reducing the distance $L_0$ to zero, and neglecting $t_{\rm op}$, we see that this bound can, in principle, go to infinity. Yet, assuming $L_0 \to 0$ would require infinitely many QR nodes, $N_{\rm QR} \to \infty$, and thus an infinite amount of resources (quantum memories). 

The physical resources required for the second generation of QRs depend on the size of the CSS code, $n_{\rm code}$. At each repeater station, we need at least $2 n_{\rm code}$ qubits for storing the encoded states $|0\rangle_{L}$ and $|+\rangle_{L}$, and we also need additional memory qubits to store and purify entanglement between neighboring repeater stations \cite{Jiang2008}. Hence, the total number of quantum memory qubits is $N_{\rm tot} \sim n_{\rm code} \frac{L_{\rm tot}}{L_0}$.

The size of the encoding block, $n_{\rm code}$, only needs to increase poly-logarithmically with the total distance $L_{\rm tot}$. Asymptotically, there are CSS codes with $n_{\rm code} \le 19t$, which can correct up to $t$ (bit-flip and dephasing) errors [obtained from the Gilbert-Varsharov bound, see Eq. (30) in \cite{Canderbank96}]. This implies that we only need $n_{\rm code} \propto t \sim \ln \frac{L_{\rm tot}}{L_0}$ that increases logarithmically with $L_{\rm tot}$ \cite{Jiang2008}. In practice, however, it is might be challenging to initialize large CSS encoding block fault tolerantly with imperfect local operations. To avoid complicated initialization, we may construct larger CSS codes by concatenating smaller codes with $r$ nesting levels, and the code size scales polynomially with the code distance, $n_{\rm code} \propto t^r \sim (\ln \frac{L_{\rm tot}}{L_0})^r $. Alternatively, we may consider the Bacon-Shor code \cite{Bacon06}; the encoding block scales quadratically with the code distance $n_{\rm code}=(2t+1)^2 \sim (\ln \frac{L_{\rm tot}}{L_0})^2$, and the initialization can be reduced to the preparation of $(2t+1)$-qubit Greenberger-Horne-Zeilinger (GHZ) states.
For finite total distance $L_{\rm tot}$, a more useful performance metric for comparing the QR protocols should quantify both the amount of physical resources, as well as the communication rate (see Sec.~\ref{sec:CompareQRs}).

\subsubsection{Third-generation repeaters}\label{sec:3GQR}

The third generation of QRs relies on \emph{deterministic} error suppression, such as QEC and one-way hashing (see Sec.~\ref{sec:DetErrSupp}), to correct both loss and operation errors~\cite{Fowler2010,Munro2012,Muralidharan2014}. The quantum information can be directly encoded in a block of physical qubits that are sent through the lossy channel. If the loss and operation errors are sufficiently small, the received physical qubits can be used to restore the whole encoding block, which is re-transmitted to the next repeater station. The third generation of QRs only needs \textit{one-way} signaling and thus can achieve very high communication rates, just like classical repeaters only limited by local operation delays.

Various choices of quantum error-correcting codes can be used for the third generation of QRs \cite{knill1996concatenated}.
For qubit-based quantum error correction, we may use quantum parity codes \cite{Ralph2005} with moderate coding blocks ($\sim 200$ qubits) to efficiently overcome both loss and operation errors \cite{Munro2012,Muralidharan2014}. The surface code \cite{Raussendorf2007,Raussendorf2007a} or the tree-cluster code \cite{Varnava2006} can suppress more loss errors---up to 50\%---with larger encoding blocks. For quantum codes based on $d$-level quantum systems (e.g., based on time-bin encoding), we can implement quantum polynomial \cite{cleve1999share} codes to approach loss tolerances up to 50\% \cite{Muralidharan17} and quantum Reed-Solomon codes \cite{LiZ08} to further improve the key generate rate \cite{Muralidharan18}. If we treat each optical mode as a continuous variable system, we may use bosonic quantum error-correcting codes (e.g., cat codes \cite{Leghtas13b, Mirrahimi14}, binomial codes \cite{Michael16}, and GKP codes \cite{GKP, Albert2018, Noh2018quantum}) to correct loss errors. The advantage of bosonic codes is that they can efficiently use the large Hilbert space of bosonic systems and reduce the number of bosonic modes, which might be advantageous to maximize the usage of our optical quantum channel bandwidth \cite{LiL17}. To further suppress the residual errors from the first-level bosonic codes, we may concatenate it with a second-level DV encoding, which leads to a concatenated CV-DV encoding scheme. To reduce the
resource cost with respect to an architecture for which all repeaters are the same, we may introduce two different types of repeaters, correcting errors at two different levels, respectively~\cite{Rozpedek2021}.

Note that the second and third generations of QRs can achieve communication rates much faster than the first generation over long distances, but they are technologically more demanding. For example, they require high-fidelity quantum gates, as QEC only works well when operation errors are below the fault-tolerance threshold. The repeater spacing for the third generation of QRs is smaller compared to the first two generations of QRs because error correction can only correct a finite amount of loss errors deterministically (only up to 50$\%$ loss error rates deterministically~\cite{Muralidharan2014, Stace2009}).

Similar to the second generation of QRs, the physical resources required for the third generation of QRs depend on the size of the quantum error-correcting code. We may use $n_{\rm code}$ to characterize the size of the encoding blocks based on qubits or bosonic modes. At each repeater station, we need $O( n_{\rm code})$ quantum memories to perform error correction suppressing not only operation errors, but loss errors as well. The total number of quantum memories (in terms of qubits or bosonic modes) needed is $N_{\rm tot} \sim n_{\rm code} \frac{L_{\rm tot}}{L_0}$.
In principle, we may use QEC over optical modes to fully replace the need of the traditional atomic or solid-state quantum memory, which inspires the design of all-photonic quantum repeaters as discussed in. Sec~\ref{sec:memoryless}.

For the specific application of quantum key distribution, we may use QRs to generate random secret classical bits shared by remote parties. Since the ultimate goal is to generate secret keys, rather than the entangled states, we might slightly relax the requirement of quantum memories. In particular, in this case, even for first and second generation repeaters, there is no need of long-lived quantum memories to store the entangled states at the {\it end} stations, because they can be measured simultaneously with all intermediate repeater stations \cite{Jiang2008} (see Sec.~\ref{sec:stages-of-QI}). However, notice that the first and second generation QRs still need quantum memories at repeater nodes, whose required memory time is longer than that of third generation QRs.

\subsubsection{Comparison of three generation of QRs}\label{sec:CompareQRs}
To present a systematic comparison of different QRs in terms of efficiency, we need to consider both temporal and physical resources. The temporal resource depends on the rate, which is limited by the time delay from the two-way classical signaling (in first- and second-generation repeaters) and the local gate operation (in the second and third generations) \cite{Jiang2007}. The physical resources depend on the total number of qubits needed for HEGP (first and second generations) and QEC (second and third generations) \cite{Muralidharan2014,Bratzik2014}. One may quantitatively compare the three generations of QRs using a cost function  \cite{Muralidharan2014} related to the required number of qubit memories to achieve a given transmission rate.
If a total of $N_{\rm tot}$ qubits are needed to generate secure keys at $R$ bits/second, a cost function is defined as
\begin{equation}
C(L_{\rm tot})=\frac{N_{\rm tot}}{R}=\frac{N_{\rm s}}{R}\times \frac{L_{\rm tot}}{L_{0}} ,
\end{equation}
where $N_{\rm s}$ is the number of qubits needed per repeater station, $L_{\rm tot}$ is the total communication distance, and $L_{0}$ is the spacing between neighboring stations.
Since the cost function scales at least linearly with $L_{\rm tot}$, to demonstrate the additional overhead associated with $L_{\rm tot}$, a \textit{cost coefficient} can be introduced as
\begin{equation}
C'(L_{\rm tot})=\frac{C(L_{\rm tot})}{L_{\rm tot}},
\end{equation}
which can be interpreted as the resource overhead (qubits $\times$ time) for the creation of one secret bit over 1~km (with target distance $L_{\rm tot}$). Besides the fiber attenuation (with $L_{\rm att}=20$\,km for telecom wavelengths), the cost coefficient also depends on other experimental parameters, in particular the coupling efficiency $\eta_{\rm c}$, the gate error probability $\epsilon_{\rm G}$, and the gate time $t_{0}$.

\begin{figure}[bt]
\includegraphics[width=\linewidth]{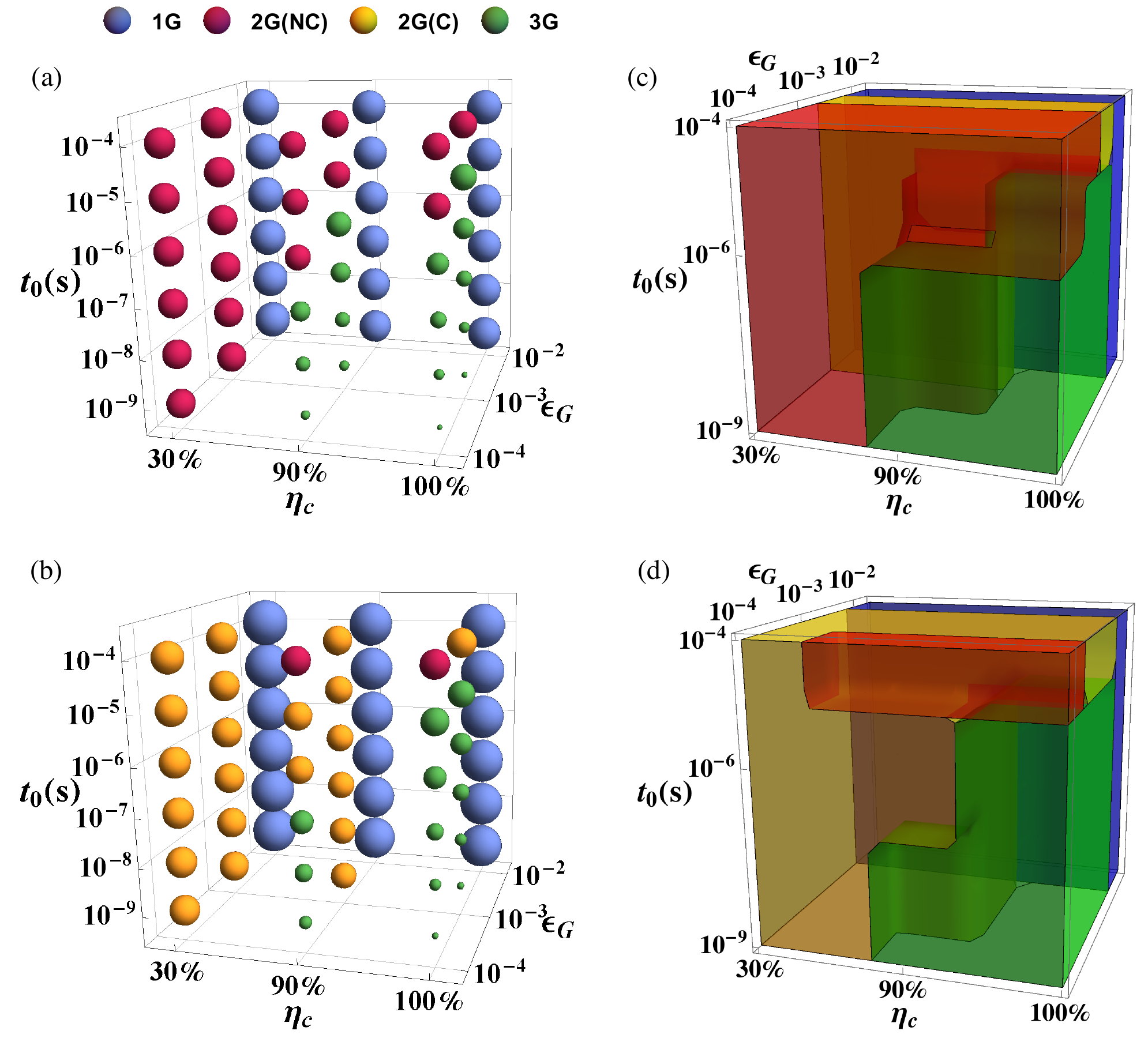}
\protect\caption{The bubble plot comparing various QR protocols in the three-dimensional parameter
space spanned by coupling efficiency $\eta_{\rm c}$, gate error probability  $\epsilon_{\rm G}$, and gate time $t_{0}$, for a)
$L_{\rm tot}=1000$\,km and b) $L_{\rm tot}=10,000$\,km. The bubble color indicates
the associated optimized QR protocol, and the bubble diameter is proportional
to the cost coefficient. The region plots (c) and (d) showing the distribution of different optimized QR protocol in the three dimensional parameter space for $L_{\rm tot}=1000$\,km and $L_{\rm tot}=10,000$\,km respectively. The region plot (c) contains a yellow region of second generation with encoding, which can be verified in a bubble plot with a finer discretization of $\epsilon_{\rm G}$. Figure from~\cite{Muralidharan2016}.
}
\label{fig:bubble}
\end{figure}

We may summarize the analysis of QRs based on the cost coefficient \cite{Muralidharan2016} using bubble and region plots in the three-dimensional parameter space, as shown in Fig. \ref{fig:bubble}, which compares representative protocols from three generations of quantum repeaters \cite{Briegel98,Jiang2008,Muralidharan17}. 
\footnote{The communication rate of the first generation of QRs can be boosted using temporal, spatial, and/or frequency multiplexing associated with the internal degrees of freedom for the quantum memory \cite{Sangouard2011,Afzelius2009}.}
The bubble color indicates the associated optimized QR protocol, and the bubble diameter is proportional to the cost coefficient. The parameter space can be divided into the following regions:
(I) For high gate error probability $(\epsilon_{\rm G}\gtrsim 1\%)$, the first generation dominates;
(II.A) For intermediate gate error probability, but poor coupling efficiency or slow local operation {[}$0.1\frac{L_{\rm att}}{L_{\rm tot}}\apprle\epsilon_{\rm G}\apprle1\%$ and ($\eta_{\rm c}\apprle90\%$ or $t_{0}\apprge1\mu$s){]}, the second generation \textit{with} encoding is more favorable;
(II.B) For low gate error probability, but low coupling efficiency or slow local operation {[}$\epsilon_{\rm G}\apprle0.1\frac{L_{\rm att}}{L_{\rm tot}}$ and ($\eta_{\rm c}\apprle90\%$ or $t_{0}\apprge1\mu $s){]}, the second generation \textit{without} encoding is more favorable; (III) For high coupling efficiency, fast local operation, and low gate error probability ($\eta_{\rm c}\apprge90\%$, $t_{0}\apprle1\mu $s, $\epsilon_{\rm G}\apprle1\%$), the third generation becomes the most favorable scheme in terms of the cost coefficient

\subsection{All-optical repeaters}
\label{sec:memoryless}

While the traditional repeater protocol necessitates physical memories---stationary quantum systems---to store quantum information during the long waits associated with long-distance entanglement generation, it is fairly nontrivial whether the protocol can be implemented all-optically just by replacing the memories with all-optical memories like ones in \cite{leung2006}. On the other hand, repeaters featuring QEC codes could preclude the necessity of such memories, as QEC codes can instead deterministically suppress the noise and loss affecting qubits. Indeed, error-corrected repeaters, which intersect with the second and third generations discussed above, are shown to be implementable all-optically; in this case,
the significant differences in analysis and implementation compared to matter-based repeaters warrant special attention, which we provide in this subsection.

To better understand all-optical or all-photonic repeaters, we first review
the operating principle of another quantum information protocol, \emph{measurement-based quantum computation (MBQC)} (sometimes referred to as one-way computation\footnote
{
``One-way'' has a special meaning in quantum communication, so we forego this terminology.
}), especially relevant for optical implementations. In a measurement-based quantum computer \cite{Raussendorf2001}, to be contrasted with a gate-based computer,
an entangled resource state, namely a cluster (or graph) state (Sec.~\ref{sec:graph_states}), is prepared initially and the computation proceeds by way of adaptive {\it single-qubit} measurements on this state. For physical platforms suffering from probabilistic entangling gates, among them discrete-variable dual-rail photonics (see Sec.~\ref{sec:photonic_encodings}),
this type of computer has the advantage that such probabilistic gates are involved only in the preparation of the initial resource states and are not necessary during the computation. This circumvents the exponential decay of the computational success with the number of entangling operations and dramatically reduces the resource costs \cite{Nielsen2004, Browne2005, Kok2007} compared to the gate-based scheme \cite{Knill2001}. Furthermore, the measurement-based approach allows for fixed-depth circuits where a physical qubit only undergoes a finite (and generally small) amount of gate operations before being consumed by a single-qubit measurement. This approach therefore accords well with flying qubits; it helps overcome the weakness of probabilistic entangling gates for certain photonic encodings, and drastically cuts down on the amount of loss each photon experiences. 

In measurement-based computation, universality---the ability to approximate any unitary on any number of data qubits arbitrarily well---is achieved through an appropriate choice of cluster state~\cite{Briegel2001}, as well as access to non-Clifford operations. Fault-tolerance---the exponential suppression of state preparation, gate and measurement errors---is obtained through an error-correcting code (Sec.~\ref{par:QEC}), which translates to a cluster state with a special shape and structure (the encoding); a prescription for implementing logical operations through adaptive single-qubit measurements; and a means of detecting and correcting the error, including an algorithm for extracting the outcomes of logical measurements (the decoding and recovery).

A common feature of recent architectures of all-optical repeaters is that they are realizable through measurement-based implementations of QEC codes. 
A measurement-based quantum repeater operates in much the same way as a measurement-based computer; however, there are a handful of salient distinctions, emblematic of the differences between computation and communication. First, gate-set universality is not necessary for communication, meaning Clifford operations suffice. Second, the dominant source of errors for the photonic states comprising optical repeaters---loss---is an even larger threat.
Third, in contrast to computation, which can be done locally, the goal of communication is inherently nonlocal---to entangle spatially distant objects. Since noise for physical qubits generally increases with time, it is important to take the classical communication time into account.

With these general notions out of the way, in the next subsections we overview the workings of several protocols for all-optical repeaters and describe promising schemes for the preparation of repeater graph states. We begin with a summary of the first all-photonic repeater proposal~\cite{Azuma2015}, as an instructive example.

\subsubsection{Original all-photonic repeaters}
\label{sec:apqr}

\begin{figure}
    \centering
    \includegraphics[width=0.5\linewidth]{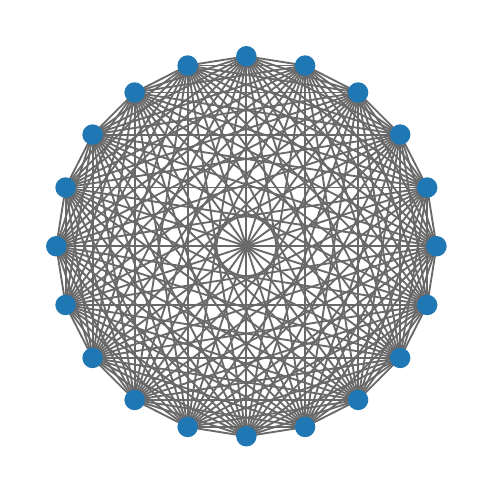}\includegraphics[width=0.5\linewidth]{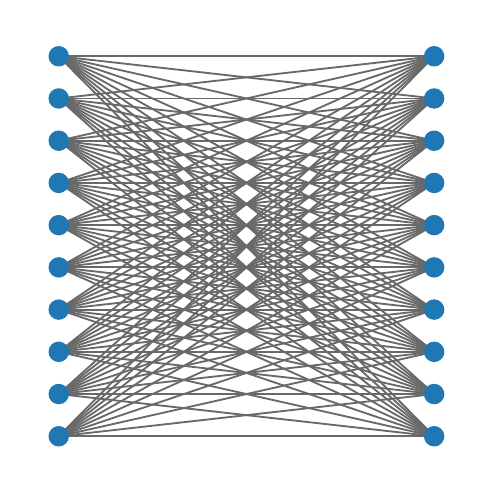}
    \caption{A clique (left) and biclique (right). In the former, each vertex is connected with every other. In the latter, each vertex from the left set is connected with a vertex on the right, but the sets are internally disconnected. These graphs can underlie repeater graph states. See Sec.~\ref{sec:graph_states} for more on graph states.}
    \label{fig:CBGS20}
\end{figure}

\begin{figure}
    \centering
    \includegraphics[width=\linewidth]{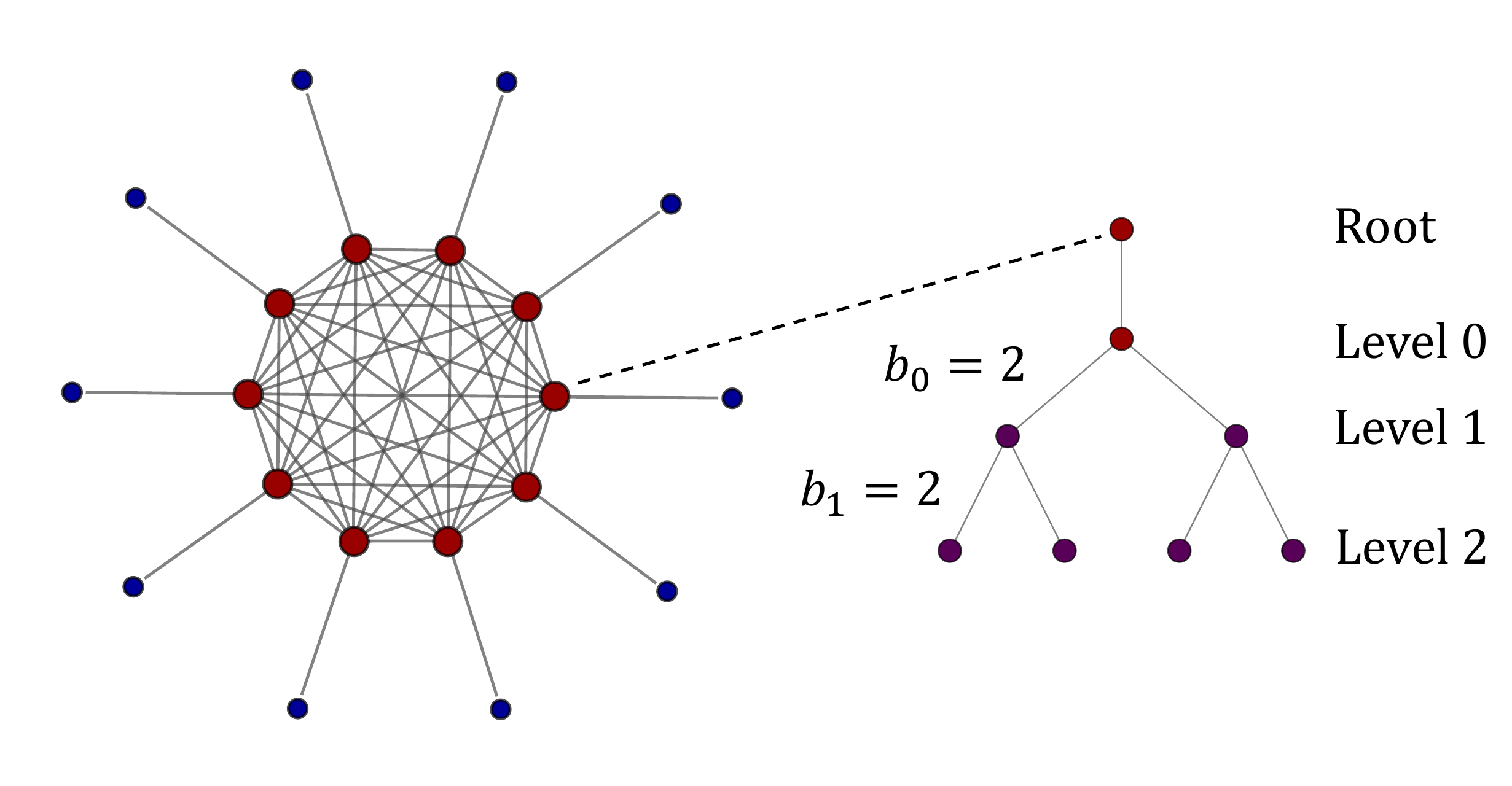}
    \caption{Encoded repeater graph state (RGS) proposed in~\cite{Azuma2015}. The RGS has two layers (left): the inner layer is composed of core qubits [big (red) vertices, closer to the center]; the outer layer is composed of outer qubits (or called leaves) [small (blue) vertices, further from the center].
    Each vertex in the clique (left) is a logical qubit, which can be encoded in, e.g., the Varnava tree code~\cite{Varnava2006} (right) to protect itself from loss (as well as general errors under the restriction of Pauli measurements). Displayed are the levels and branching parameters $\{b_0,b_1,\cdots,b_{d-1}\}$ of the tree ($d=2$ in this figure). Note the root and $0$th-level qubits
    [two upper (red) qubits] in the tree will be measured out in the $X$ basis, connecting the qubits in the first level with all of the neighbours of the root qubit. The inner logical qubits, conduits for the entanglement swapping, are connected to outer unencoded physical leaf qubits, which help effect the entanglement generation.}
    \label{fig:RGS}
\end{figure}

\begin{figure*}
    \centering
    \includegraphics[width=\textwidth]{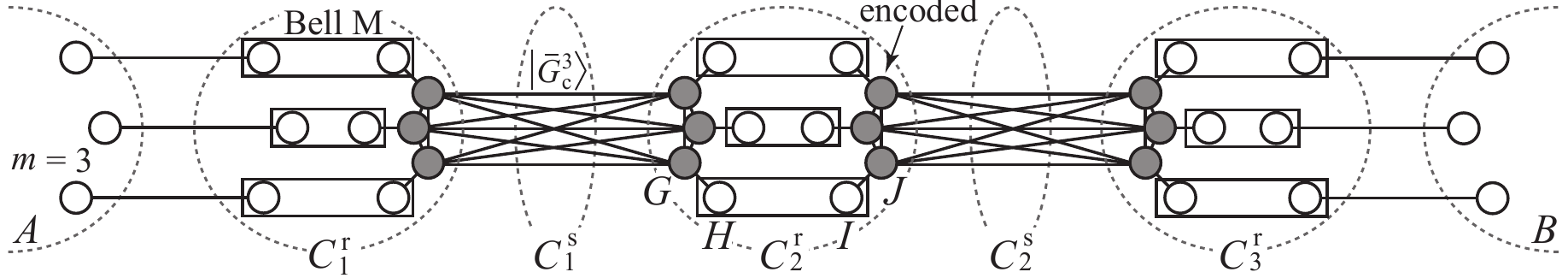}
    \caption{Summary of original all-photonic repeater scheme~\cite{Azuma2015}. Alice ($A$) and Bob ($B$) would like to establish one entangled pair; each prepares $m$ Bell pairs ($m=3$ in this figure) and sends them to a nearby receiver. Repeater graph states are created at $C^{\rm s}_1$ and $C^{\rm s}_2$ and their qubits are sent to adjacent receivers $C^{\rm r}_1$ and $C^{\rm r}_2$, respectively. The receivers perform $m$ simultaneous Bell state measurements on the outer qubits. In every receiver node, $X$-basis measurements is performed on a pair of inner qubits adjacent to outer qubits to which the Bell measurement is successfully applied, while $Z$-basis measurements are conducted on the other inner qubits. 
    Figure from~\cite{Azuma2015}.}
    \label{fig:apqr_scheme}
\end{figure*}

Our review of the all-photonic repeaters introduced in~\cite{Azuma2015} begins with a description of the repeater graph state (RGS). The ideal RGS that the authors propose has two layers. 

The inner or core layer is a \emph{complete graph} or \emph{clique} (Fig.~\ref{fig:CBGS20}), which is locally equivalent to a GHZ state of $n$ qubits from Eq.~\eqref{eq:GHZ}. The qubits in the inner layer are tailored to play the same role as quantum memories in a second-generation quantum repeater protocol. Recall the assumption of the second-generation QR protocol about quantum memories which allow us to apply {\it deterministic} Bell measurements only on quantum memories which have successfully shared entanglement with adjacent repeater nodes (see \ref{sec:2GQR}). To make photonic qubits play this role, the core qubits in the RGS of Fig.~\ref{fig:RGS} are prepared in a complete-graph state\footnote{However, it was shown in~\cite{Russo2018, Tzitrin2018, Tzitrin2020} that some of the connections in the clique comprising the RGS are unnecessary, so that 
some variant, such as the biclique in Fig.~\ref{fig:CBGS20}, is sufficient as core qubits.} as in Fig.~\ref{fig:CBGS20} (to overcome the probabilistic nature of the linear optical Bell measurements). 
In particular, if we apply $X$-basis measurements to two of them and $Z$-basis measurements to the other qubits, it works as the Bell measurement on the two qubits and decouples the others (although we only use single-qubit measurements). To achieve these $X$-basis or $Z$-basis measurement deterministically even under photon loss, the qubits in the inner layer are encoded into a larger graph state with sufficient redundancy. In~\cite{Azuma2015}, a tree-graph QEC code proposed by Varnava~{\it et al.} \cite{Varnava2006} is considered for this purpose, as demonstrated schematically on the right panel of Fig.~\ref{fig:RGS}. 
This code places a qubit to be encoded at the root of a tree graph state composed of physical qubits. Then, it allows one to execute an arbitrary logical single-qubit measurement on the encoded qubit deterministically, even under loss, via single-qubit measurements on the physical qubits. Increasing the size of the tree graph state with increasing losses will ensure the correction succeeds, as long as the loss probability per physical qubit is less than 50\%, a threshold consistent with the no-cloning theorem.

The other layer of the RGS consists of outer qubits or leaves appended to the vertices of the core graph;
these are analogous to photons entangled with quantum memories for the purpose of HEGP in the second-generation QR protocol. In fact, a pair of outer qubits, each of which belongs to different RGSs, will be subject to a linear-optical Bell measurement in order to entangle their neighboring core qubits.
Combining these layers of the RGS, the final state proposed by \cite{Azuma2015} is shown in Fig.~\ref{fig:RGS}.

With an understanding of the RGS, we can now overview the precise operations required for Alice and Bob to establish an entangled pair in a given clock cycle of the all-photonic repeater protocol. The scheme is illustrated in Fig.~\ref{fig:apqr_scheme}. We use the notation from before: $L$ is the total channel length; $N$ is the number of repeater stations (sources or major nodes), not including Alice and Bob; $m$ the number of parallel pulses. This means that there are $N+1$ measurement stations (receivers or minor nodes), and $M=2m$ is the size of the RGS, if it is symmetric.

Let us assume that an RGS is available at each source node (leaving the various preparation mechanisms for Sec.~\ref{sec:apqr_state_gen}). Then, each of the two nodes neighboring the source node receives half of photons in the RGS prepared and sent by the source. On arrival of the photons, every receiver first conducts simultaneous BSMs (Fig.~\ref{fig:BellM}~(a)) on $m$ pairs of leaf photons of RGSs from different source nodes; this connects their adjacent inner qubits. Although each such BSM can only succeed with a probability of at most $\nicefrac{1}{2}$ (and it is less than $\nicefrac{1}{2}$ in practice, because of the the losses experienced by the leaves), with $m$ large enough, at least one BSM per station would be guaranteed to have succeeded. Then, depending on the outcomes of the BSMs, every receiver node applies $X$-basis measurements on a pair of the inner qubits whose adjacent leaves have been subject to a successful BSM, and $Z$-basis measurements on the other inner qubits. Since inner qubits are encoded into the tree-graph code, these single-qubit measurements succeed almost deterministically (as long as the loss is below the threshold of 50\%). As a result, the $Z$-basis measurements transform the total state into a linear cluster state between Alice and Bob, which is then converted into a Bell pair between them by the $X$-basis measurements, according to the effects detailed in Sec.~\ref{sec:graph_states}.

Importantly, the choice of the measurement on an inner encoded qubit, and accordingly on the physical qubits composing the tree cluster, depends on the measurement outcomes from the outer qubits. This means it is necessary to convey classical information from the outer qubits to the inner qubits. However, this can be done locally at each receiver node---that is, just by using a local active feedforward technique---as the inner qubits are transmitted together with their adjacent outer qubits. Therefore, the amount of necessary signalling is designed to be minimal, reducing time-dependent loss and errors for the photons.

Although loss is the dominant source of noise, one cannot dismiss other sources of error. Aided by a majority vote protocol, the tree-graph code of Varnava~{\it et al.} is robust against general errors under the restriction of $X$-basis or $Z$-basis measurements on the encoded qubit, in contrast to other single-qubit measurements. In Sec.~\ref{subsubsec:other_optical_repeaters}, we overview an optical repeater protocol that instead makes use of parity codes~\cite{Ewert2016}. The existence of a better code specifically suited to an all-photonic repeater---in terms of error tolerance and overheads---is an important open question.

The all-optical protocol needs no quantum memories, including qubits held by Alice and Bob, for the applications in which entanglement for Alice and Bob, once generated, is consumed immediately to generate classical output strings, such as QKD \cite{Mayers2001,Lo1999,Shor2000,Koashi2009,Renner2008,Christopher2022,Bennett1992}, non-local measurements \cite{Vaidman2003,Clerk2008}, and cheating strategies in position-based quantum cryptography \cite{Lau2011,Kent2011,Buhrman2011}.
However, for applications that demand strictly a quantum output state to Alice and Bob, such as quantum teleportation and distributed quantum computation \cite{Gottesman1999a,Eisert2000,Collins2001}, the applications themselves require Alice and Bob to have quantum memories with memory time on the order of classical communication time between Alice and Bob, because of the necessity of classical signaling. See Sec.~\ref{sec:stages-of-QI} or Ref.~\cite{Azuma2015} for detail.

\subsubsection{Other optical repeaters}
\label{subsubsec:other_optical_repeaters}
\paragraph{Modified all-photonic repeaters.}
Although Ref.~\cite{Pant2016} aims to analyze the performance of the all-photonic repeaters of Ref.~\cite{Azuma2015}, the authors make several modifications that warrant discussion.

First, so-called \emph{boosted} Bell state measurements~ (BBSMs) are employed in~\cite{Pant2016}. The previously cited maximal linear-optical Bell measurement success rate of $\nicefrac{1}{2}$ can be increased with additional resources, such as ancillary photons in separable~\cite{Ewert2014} or entangled~\cite{Grice2011} states, weak nonlinearities~\cite{Barrett2005}, and pre-detection squeezing~\cite{Zaidi2015, Kilmer2019}. However, BBSMs are no panacea: they increase experimental complexity and overhead, and infinite resources are still needed for unit probability, in line with a no-go theorem \cite{Lutkenhaus1999}. The specific BBSMs\cite{Ewert2014} employed in~\cite{Pant2016} succeed $\nicefrac{3}{4}$ of the time. The analysis shows that they result in a net improvement to the overheads.

A more crucial design change is in the treatment of the inner qubits.
In the original proposal, photons forming the clique of the RGS---the encoded inner qubits---are sent to neighbouring receiver nodes, together with their adjacent leaves, while Pant~{\it et al.} assume they are stored locally at the source nodes in fiber spools. In the original proposal, signalling from the leaves to the inner qubits can be done via local active feedforward; however, all the physical qubits in the encoding must be sent, necessitating a large number of fiber connections. While resulting in fewer fiber connections, the approach of Pant~{\it et al.} comes at the expense of an increased loss, which stems from the necessity of signalling from the leaves to the inner qubits over the associated distance.

Finally, there is also a modification of the original scheme in Pant~{\it et al.} with regards to the multiplexing strategy in state generation, which is discussed briefly in Sec.~\ref{sec:apqr_state_gen}.

\paragraph{Repeaters based on encoded Bell measurements.}
In \cite{Ewert2016,lee2019fundamental}, all-optical repeater protocols are presented based on \emph{parity codes} \cite{Ralph2005}.
Specifically, the authors in \cite{Ewert2016} make use of Bell states with parity encoding. The graph states locally equivalent to the encoded Bell states look remarkably like the RGS from the original protocol: they are bicliques (complete bipartite graphs) with multiple leaves per node~\cite{Ewert2017}. However, the protocol of Ewert {\it et al.} itself is conceptually different from the all-photonic repeaters of \cite{Azuma2015} \cite{zwerger2016measurement}; it sends an encoded qubit from a sender to a receiver, directly, which makes it closer to the third-generation schemes of \cite{Munro2015,Zwerger2014,Varnava2007,muralidharan2014ultrafast,knill1996concatenated} based on quantum error correction than the protocol of \cite{Azuma2015}, which can be regarded as a time-reversed version of a second-generation quantum repeater protocol. In their
protocol \cite{Ewert2016}, Bell measurement efficiency and loss tolerance improves as the size of the parity code increases. Furthermore, their scheme does not require active feedforward techniques, lowering local operation times, reducing losses, and facilitating on-chip integration. The concatenated Bell measurement scheme in~\cite{lee2019fundamental} reaches the fundamental limits for Bell measurement efficiency and loss tolerance under the constraints of linear optics and the no-cloning theorem. Regarding loss-tolerance, this scheme also saturates the fundamental loss tolerance limits for logical Bell measurements based on adaptive linear-optical physical Bell measurements~\cite{Hilaire2023}. However, recent Bell measurement schemes~\cite{Hilaire2021, Bell2022}, based on an adaptive combination of physical two-photon Bell measurements and single-qubit measurements, exhibit an even stronger loss tolerance (saturating the no-cloning limit). So far, the performances of these new logical Bell measurement schemes remain to be evaluated in a quantum repeater scheme.

\paragraph{Bosonic repeaters.}
\label{sec:bosonic_repeaters}

Certain repeaters based on continuous-variable states have been proposed~\cite{Fukui2020, Rozpedek2021}. They leverage the inherent error-correction properties of bosonic encodings along with higher-level qubit codes to create what can be viewed as concatenated CV-DV error-correcting codes. Recall from Sec.~\ref{sec:photonic_encodings} that there are several advantages to the GKP encoding in particular. First, it can tolerate small displacement errors; since any continuous error can be decomposed into displacements, it can natively treat loss errors as well. In fact, it was discovered that GKP states far better against loss errors in certain settings than codes tailored to handle losses~\cite{Albert2018}. Furthermore, for GKP states, entangling gates and Bell measurements are deterministic contingent on the availability of Gaussian resources, with the only probabilistic component being state generation. Finally, additional (analog) information obtained from the GKP-level error correction can be used to improve the logical error rates at the qubit code level~\cite{fukui2017analog, noh2019b}.

The repeater architecture in~\cite{Rozpedek2021} leverages the above advantages of GKP encodings and uses two types of repeaters: those consisting purely of GKP states, which can correct small displacement errors, and those comprised of GKP states concatenated with small qubit-level codes. In a related work, the authors in~\cite{Fukui2020} compare using GKP encoding by itself, in a one- and two-way scheme, as well as with higher-level encodings.

\subsubsection{Repeater graph state generation}
\label{sec:apqr_state_gen}

Producing a large, high-quality optical graph state for measurement-based quantum information protocols is a tall order.
In all-optical approaches, the stochasticity of entangling operations in some encodings (e.g., dual-rail) and of state preparation in others (e.g., GKP states) can result in large overheads; in matter-based approaches, effects like decoherence and inhomogeneity between emitters can result in significant decay of entanglement with the size of the target state. Nevertheless, there has been steady theoretical and experimental progress towards high-probability, high-fidelity cluster state generation. Let us discuss some promising ways of preparing optical graph states here.

\paragraph{General framework.}

Optical graph state generation can be understood in a general framework that involves the ``stitching'' of smaller resource states into iteratively larger states. Measurement-based entangling operations, such as those used for dual-rail encodings, are more formally referred to as \emph{fusion gates} \cite{Browne2005}, introduced in \cite{pittman2001}. Fusion gates on two optical modes, each of which may have a single photon, come in two varieties: \emph{Type-I fusions}, which consume a single photon to create larger one-dimensional cluster states, and \emph{type-II fusions} (essentially rotated Bell measurements) which consume two photons to grow cluster states in higher dimensions. As with BSMs, fusion probabilities may also be boosted with additional resources, a fact that has been exploited for RGS generation in~\cite{Pant2016}; as before, this introduces tradeoffs with experimental complexity and overheads~\cite{MercedesThesis}. For completeness, we also mention fusion-based quantum computation (FBQC)~\cite{bartolucci2021fusion}, a proposed alternative framework to MBQC where the fusion operations serve both to create entanglement and perform logical operations.

The schema for generating optical graph states is as follows:

\begin{enumerate}
    \item \emph{Unit resource production.} First, an optical circuit produces the smallest unit states. These can be single-qubit states or small entangled states, such as Bell pairs, $n$-partite GHZ states for $n \geq 3$, or few-qubit linear cluster states.
    \item \emph{Growth into meta-units.} As an optional intermediary step, the unit resources can be combined into larger meta-units. The utility of this extra step is to leave open the possibility, for example, of generating dual-rail $n$-partite GHZ states directly from single photons, or instead from photonic Bell pairs (see, for example, ~\cite{MercedesThesis}).
    \item \emph{Stitching.} Units or meta-units are entangled iteratively until the desired graph state is created. For dual-rail encodings, this can be achieved with type-II fusions; for GKP states, this can be done with continuous-variable CZ gates.
    \end{enumerate}
    
A few notes are in order. First, the framework accommodates matter-based optical graph state generation; in this case, the entanglement in the growth or stitching stages can be achieved either directly at the optical level, or assisted by the interaction between emitters. Second, each step carries an associated probability and fidelity that depends on the choice of encoding, the scheme for generating and entangling the resources, and the particular hardware implementation. Other considerations that will affect the architectural design include how much of the state can be made spatially (i.e., with the state sources arranged space) or temporally (i.e., with entanglement between states generated at different time steps). This is related to the question of how much of the graph state (e.g., how many layers in a regular cluster state) must exist at one time.  

\paragraph{Dual-rail graph states.} We review two different approaches to produce a graph state of dual-rail encoded qubits: one all-optical but probabilistic, the other relying on matter qubits but deterministic.

\subparagraph{Probabilistic (optical) generation}

The original all-photonic repeater proposal \cite{Azuma2015} relies on the approach taken in~\cite{Varnava2007, Varnava2008} for generating a tree graph state specified by a branching parameter $\{b_0,b_1,\cdots,b_{d-1}\}$ (Fig.~\ref{fig:RGS}), where the root qubit of the tree graph state is connected to a 0th-level qubit, the 0th-level qubit is connected to $b_0$ 1st-level qubits by edges and every $i$th-level qubit is connected to $b_{i}$ $(i+1)$th-level qubits by edges $(i=0,1,\cdots,d-1)$. As the encoding, the root and the 0th-level qubits are measured offline in the $X$ basis.
The tree graph state can be transformed into an RGS.
The protocol of Varnava {\it et al.} proceeds as follows.

First, six single photons are prepared with single-photon sources. The photons are then sent to an optical circuit composed of beamsplitters, a type-I fusion gate and a type-II fusion gate, which produces a $3$-partite GHZ state with probability \nicefrac{1}{32}. Thanks to the design of this circuit, even if single-photon sources and detectors do not have unit efficiency, the generated $3$-partite GHZ state is affected only by individual (uncorrelated) loss \cite{Varnava2008}.
This GHZ state then becomes the unit resource to produce the RGS. In particular, two 3-partite GHZ states are converted to a 4-partite GHZ state by a type-II fusion gate, and this 4-partite GHZ state corresponds to a three-qubit tree, i.e., $\{2\}$-tree, with a redundant root qubit composed of two qubits. Then, from these elementary $\{2\}$-trees, one can efficiently generate an arbitary $\{b_0,b_1,\cdots,b_{d-1}\}$-tree from the bottom ($d$-th level) to the top ($0$-th level), with the help of type-II fusion gates.

Several generalizations or modifications are possible for this procedure. In~\cite{Pant2016}, the authors choose the more efficient generation scheme of~\cite{Li2015}, consider boosted fusion gates, improve the multiplexing strategy, and reorder the local measurements unconditioned on BSM outcomes. Furthermore, it is possible to create $n$-partite GHZ resource states with probability $\nicefrac{1}{2^{2n-1}}$, and this number can theoretically be increased with Bell-state inputs rather than single-photon inputs, as well as boosted BSMs~\cite{Zhang2008, Varnava2008, Joo2007, MercedesThesis}. For optical repeaters based on other error-correcting codes---which correspond to other graph states, these resource states can be stitched according to the different, tailored procedures.

\subparagraph{Deterministic (matter-based) generation}
Unlike fusion-based approaches, which are fundamentally probabilistic, the protocol of Buterakos~{\it et al.} \cite{Buterakos2017}, which uses emitter and ancilla qubits to generate the RGS, is---at least in principle---deterministic. The generation of linear cluster states from a single emitter was proposed by Sch\"on~{\it et al.} for atomic systems \cite{Schoen2005}
and by Lindner and Rudolph for quantum dots (QDs) \cite{Lindner2009}. More complex graph states, including a 2D square lattice cluster state, can be created by a linear chain of emitters with nearest-neighbor coupling \cite{Economou2010,Gimeno-Segovia2019}. Indeed, any graph state can be created with these ingredients \cite{Russo2019}. In \cite{Buterakos2017}, the key mechanism for generating the RGS is to entangle the emitter with an ancilla and pump it to produce one arm of the RGS, which emerges entangled to both the emitter and ancilla. The emitter is then measured and thus removed from the graph and the process is repeated until all the photonic arms are connected to the ancilla, which is assumed to have longer coherence time compared to the emitter. Measurement of the ancilla in the $Y$ basis disentangles it from the graph and connects all the inner photons to each other, completing the RGS.

An attractive feature of the protocol of Ref.~\cite{Buterakos2017} is that it is quite economical in terms of resources, which are quantified by the number of required matter qubits: To generate the unencoded version of the RGS, only one emitter and one ancilla are needed, irrespective of the size of the graph. In addition to the unencoded version, Buterakos~{\it et al.} provide a recipe for the deterministic creation of arbitrarily large \textit{encoded} RGSs in which the inner qubits are encoded using trees of depth 2 or 3. These protocols only require three matter qubits, including two emitters and one ancilla. Hilaire {\it et al.} \cite{Hilaire2020} give a more general recipe for generating RGSs with arbitrarily deep tree encodings of the core photons in which the requisite number of matter qubits scales linearly with the tree depth $d$ ($d-1$ emitters and 2 ancilla qubits). In this case, the number of required CZ gates is $2 m \left(2+\sum_{k=0}^{d-2} \prod_{j=0}^k b_j \right)$, where $b_j$ denotes the branching vector component of the tree at level $j$ and $2m$ is the number of arms in the RGS. These ideas for the deterministic generation of entangled photonic states were generalized in \cite{li2021entangled}, where a recipe for the generation of an arbitrary graph, using the minimal number of emitters, was provided. 

Buterakos {\it et al.} also introduced a recipe for producing tree graphs of arbitrary depth $d$ with $k$ arms at each vertex using $d-1$ emitters and one ancilla. The number of CZ gates required in this case is $\frac{b^d+(-1)^{d+1}}{k+1}-1$. This approach for creating tree-encoded photonic qubits is a powerful capability in its own right and can be applied to quantum repeaters of any generation. For example, Borregaard~{\it et al.} \cite{Borregaard2019} employ this tree generation procedure in their proposed scheme to implement third-generation repeaters using SiV defects in diamond as memory qubits.

The deterministic RGS protocol can be applied to any type of dual-rail encoding. Many of the proposals for graph state generation, especially with quantum dots, consider photon polarization encoding, but time-bin has also been proposed with these systems \cite{Lee2019}. In the case of time-bin, an alternative deterministic way of generating graph states is to use a single emitter and time-delayed feedback, as proposed by Pichler~{\it et al.} \cite{Pichler2017}, and adapted for RGS generation in \cite{Zhan2020}. In order to implement a maximally entangling gate, however, these approaches require the experimentally challenging capability of strong coupling between the emitter and the photonic waveguide where the photons propagate.

For the physical implementation of deterministic RGS generation schemes, modest-sized registers of well-controlled emitters and ancilla qubits are needed. The emitters need to be of high quality, especially in terms of brightness, so that the photon is emitted in the desired mode and successfully collected. This is critical for the protocol to be classified as deterministic. The register should also feature ancilla qubits with long coherence times, albeit not as long as what is required for quantum memories in first- and second-generation repeaters, along with the ability to perform high-fidelity gates between emitters and ancillae.

Self-assembled QDs are leading contenders for RGS generation. Indeed, the first experimental demonstration of an emitter-based cluster state generation protocol \cite{Schwartz2016} employed exciton-biexciton transitions in these systems. QDs are excellent photon emitters. They have a very efficient optical (excitonic) transition with a timescale of ~1\,ns (100\,ps) without (with) coupling to a cavity. The QD community has made rapid progress over the last several years to improve the brightness, indistinguishability, and purity of QD photon sources \cite{Senellart2017}. On the other hand, QDs have relatively low coherence times compared to point defects and atomic qubits and lack a long-lived quantum memory to act as the ancilla. Nevertheless, promising recent work \cite{Gangloff2019, Jackson2021} suggests that the dense nuclear spin environment (more than $10^4$ spinful nuclei) could potentially be cooled and controlled enough to play this role.

Other candidates for deterministic RGS generation are optically active point defects in wide bandgap materials, such as the nitrogen-vacancy or silicon-vacancy centers in diamond and the silicon-carbon divacancy or silicon vacancy in silicon carbide. These systems have longer coherence times than quantum dots and feature a small number of nuclear spins (natural abundance $\sim1\%$ in C and $\sim 4\%$ in Si), which can be isolated and controlled well and are thus already being explored as memory registers for quantum repeater nodes \cite{Taminiau2012,Nguyen2019,Bourassa2020}. On the other hand, defects are not as efficient and bright as QDs, and they tend to emit into unwanted modes a large fraction of the time. Atomic systems, such as trapped ions and atoms in optical lattices or cavities, have long coherence times and can be controlled with high fidelity. While their photon emission is not as fast, their other attractive properties could possibly compensate for the lower rates~\cite{Thomas2022}.
Interestingly, hybrid strategies combining deterministic generation based on quantum emitters and linear-optical fusion are particularly appealing when quantum emitters cannot interact with each others~\cite{Herrera2010, Hilaire2022}. In that setting, we can use quantum emitters to generate one dimensional clusters and GHZ states deterministically and fuse them probabilistically using linear-optical boosted fusion gates to generate graph states of arbitrarily complex topologies.

\paragraph{GKP-encoded graph states.}
While entangling gates for GKP encodings are deterministic and readily accessible experimentally, state preparation is a bigger challenge. There are several existing proposals to this end, with a recent focus on modified Gaussian Boson Sampling (GBS) devices, which use Gaussian optics combined with photon-number-resolving (PNR) detection~\cite{Sabapathy2018, Su2019a, Quesada2019, Tzitrin2020}. Once the GKP states are produced, they may be stitched together deterministically with passive and static optical resources, namely beamsplitters, phase shifters, and delay lines \cite{Tzitrin2021}.

\paragraph{Performance and overheads.}

The overheads of the various optical repeater protocols are highly sensitive to the chosen state generation scheme. In this section we review the resource requirements and performances of the repeaters discussed in this section.  

In the original all-photonic repeater protocol \cite{Azuma2015}, the total number of photons consumed to produce an entangled pair between Alice and Bob scales polynomially with the total distance. The average rate to produce an entangled pair with a single-repeater system is on the order of the repetition rate of the slowest device among single-photon sources, photon detectors, and active-feedforward techniques. The resource costs for the repeaters in~\cite{Ewert2016} and~\cite{Ewert2017} scale linearly or less-than-quadratically per the number of photons per encoded qubit.

Hilaire~{\it et al.} \cite{Hilaire2020} analyze the performance of repeaters based on the deterministic RGS generation of \cite{Buterakos2017} by calculating a bound on the secret key rate per matter qubit and comparing it to direct transmission and to ``memory-based'' (i.e., first- and second-generation) repeaters. To compare to the latter, the figure of merit is defined as the rate of a Bell state generation between the end nodes (Alice and Bob) divided by the number of matter qubits per node. In the case of memory-based repeaters, there is an upper bound on this quantity that originates from the need for classical heralding between nodes and which is given by $c/(4L)$.  This bound is used throughout Ref.~\cite{Hilaire2020} for memory-based repeaters; further reductions in the rate, originating from swap gates between the emitter and memory qubits, are ignored.

The deterministic RGS generation based on matter qubits in \cite{Buterakos2017} relies on entangling CZ gates between emitter and ancilla qubits, which enable us to create complicated photonic graph states. For realistic systems, the longest timescale in the deterministic RGS generation is the duration of these gates, $T_{\rm CZ}$, compared to which the photon generation and single-qubit gate times are negligible. It is therefore $T_{\rm CZ}$ that sets the bound for the secret key rate for repeaters based on deterministic RGS generation. In Ref.~\cite{Hilaire2020}, the authors fix the tree encoding depth to 2 for the inner RGS photons and optimize over the RGS size (number of arms), the branching vector $b_0,b_1$ of the tree encoding, and the number of nodes to maximize the key rate for a total distance of ~$10^3$ km. For these distances, it is found that for $T_{\rm CZ}\leq 60$\,ns the RGS approach always outperforms memory-based repeaters. In this case, the distance between adjacent nodes is approximately 3.5\,km. These are most likely conservative estimates, since memory-based repeaters also require entangling gates between matter qubits, which will further lower their rates. More research into quantifying the performance of deterministic RGS protocols is needed. For example, Hilaire~{\it et al.} kept the tree encoding depth fixed throughout their treatment ($d=2$). While deeper trees offer higher protection against photon loss, contributing to an increase of the rate, they also require a larger number of emitter-ancilla CZ gates, thus decreasing the rate. An analysis of optimal encoding depths is an open problem with deterministic RGS generation.

Sometimes, one is not limited by the number of photons, but rather by the number of optical modes available for communicating between neighboring repeater stations (i.e., by the optical channel bandwidth, as in classical communication). Then, it is important to choose good mode-efficient encoding schemes. In the low-loss regime, we may use continuous variable codes to encode multiple qubits per bosonic mode; for example, the GKP encoding can almost approach the quantum channel capacity of the one-way pure loss channel~\cite{noh2019}. In addition, other CV codes, like cat codes, can also boost the secure communication rate per mode when compared to DV encodings~\cite{LiL17}. Moreover, for CV-DV concatenated encoding, we may further reduce the resource overhead by optimizing the distribution of two different types of repeaters associated with CV and DV error correction, respectively~\cite{Rozpedek2021}.

\section{Milestones: Outperforming point-to-point optical communication}
\label{sec:milestones}
Point-to-point communication schemes allow for quantum communication over intracity distances even with the use of a standard optical fiber and they are ready for practical use (see, e.g., \cite{Lo2014,Xu2020}).
However, those schemes have a fundamental limitation on their achievable distances [which are about 400 km in practice, i.e., in the case of the use of a standard optical fiber \cite{boaron2018secure} (see Sec.~\ref{sec:IdealQR})].
This limitation is now explicitly given as the form of upper bounds \cite{Takeoka2014,Pirandola2015} on the two-way private capacity of a lossy bosonic channel, which are proportional to the transmittance $\eta$ of the channel for small $\eta$.
The two-way private capacity represents how many private bits can be obtained per use of a given channel, in an asymptotically faithful manner, with the free use of LOCC. In the case of the lossy bosonic channel of Eq.~(\ref{eq:lossch}), this quantity is given by the PLOB bound \cite{Pirandola2015}, $-\log_2 (1-\eta)$ (see Sec.~\ref{sec:internet}).

On the other hand, as one can see from Sec.~\ref{sec:qrepeaters}, a quantum repeater scheme has no fundamental limitation on their achievable distances. Indeed, it enables us to perform quantum communication efficiently even over intercontinental distances, but its realization is still challenging. Therefore, there is a technological gap between quantum repeater schemes for intercontinental distances and point-to-point communication schemes for intracity distances. 

To bridge the gap, intermediate quantum communication schemes, especially for the application to QKD, for intercity distances have been proposed \cite{abruzzo2014measurement,Panayi2014,AzumaIntercity,Lucamarini2018,luong2016overcoming,rozpkedek2019near, xie2022breaking,zeng2022mode}.
In particular, the schemes use only a single node $C$ which is located at the center between a sender, Alice, and a receiver, Bob,
and is connected to them with optical fibers. The goal of the schemes is basically to double the achievable distances of point-to-point QKD schemes, by making the secret key rate proportional to $\sqrt{\eta}$, outperforming the two-way private/quantum capacities proportional to $\eta$ (for small $\eta$), where $\eta$ is the transmittance of a pure-loss channel between Alice and Bob (see also Ref.~\cite{curty2021quantumleap} which contextualizes this approach from the viewpoint of security for QKD).
This expected secret key rate has the same scaling of the private capacity of single-repeater communication schemes with the use of pure-loss channels \cite{pirandola19,Azuma2016,azuma2017aggregating,rigovacca2018versatile} (see Sec.~\ref{sec:internet} for detail).
The schemes are divided into three categories: one is based on two-photon interference with dual-rail encoded qubits (Secs.~\ref{se:AMDIQKD} and \ref{sec:postpairingMDIQKD}) at the central node $C$, another is based on single-photon interference with single-rail encoded qubits (Sec.~\ref{se:TFQKD}), while the third one is a time-reversed version of these (Sec.~\ref{se:trAMDIQKD}) to work without optical Bell measurements.
In this section, we review these schemes, whose realizations are regarded as good and natural milestones towards quantum repeaters.

\subsection{Adaptive measurement-device-independent QKD}\label{se:AMDIQKD}

To double the communication distance by utilizing a central node $C$ between communicators, an adaptive measurement-device-independent (MDI) QKD scheme has been proposed with matter quantum memories \cite{abruzzo2014measurement,Panayi2014} or with all-optical quantum non-demolition (QND) measurements \cite{AzumaIntercity}, based on a dual-rail encoding.
Although these schemes have originally been proposed to perform QKD,
its use as an entanglement generation protocol (or, a coherent version) can be summarized as follows (Figs.~\ref{fig:AdaptiveMDImemory} and \ref{fig:AdaptiveMDIall}):
(i) Each of Alice and Bob sends $m$ optical polarization qubits (by using $2m$ bosonic modes), each of which is maximally entangled with a local qubit, to the central node $C$.
(ii) On receiving the pulses, the node $C$ essentially performs QND measurements to the pulses to confirm the arrival of single photons over lossy channels. (iii) Then, qubits of single photons that have successfully arrived from Alice are paired with ones from Bob at the node $C$. (iv) The node $C$ then performs a linear-optical Bell measurement of Fig.~\ref{fig:BellM}~(a) relying on two-photon interference on each of these pairs. (v) Node $C$ then announces the pairings and the measurement outcomes of the Bell measurements. (vi) Finally, Alice and Bob keep their local qubits which are supposed to be entangled with each other from the announcement of step (v).
The essence of this protocol is to perform the Bell measurement {\it only} on pairs of pulses which still have single photons even after the travel over the lossy optical channels.

If the protocol is used for QKD like the original proposals \cite{abruzzo2014measurement,Panayi2014,AzumaIntercity}, Alice and Bob perform at random $Z$-basis or $X$-basis measurement on each of their local qubits just after step (i), and their measurement outcomes are regarded as their choice of random bits in QKD. Then, the step (i) is replaced by the random preparation of BB84 signals $\{\ket{0},\ket{1},\ket{+},\ket{-}\}$ \cite{Bennett1984}. This would also imply that Alice and Bob could use phase-randomized weak coherent states emitted by lasers, instead of single-photon sources, by using the decoy-state method \cite{hwang2003quantum,wang2005beating,lo2005decoy}. The security simply follows from that for the original MDI QKD \cite{Lo2012,curty2014finite}, because it relies only on the trust for Alice and Bob.

The communication efficiency of the above protocol scales with $\sqrt{\eta}$, rather than $\eta$,
where $\eta$ is the transmittance of a direct lossy bosonic channel between Alice and Bob.
This can be understood as follows. First notice that the success probability of the QND measurement in step (ii) is proportional to $\sqrt{\eta}$, because
the polarization qubit emitted by Alice (or Bob) just travels over a lossy bosonic channel connecting between the central node $C$ and Alice (between the central node $C$ and Bob), rather than between Alice and Bob. This means that if the number $m$ of multiplexing, defined in step (i), is on the order of $(\sqrt{\eta})^{-1}$, the probability with which
the QND measurement in step (ii) finds the arrival of non-zero single photons from Alice and from Bob is pretty high. Then, the node $C$ can have nonzero pairs in step (iii), to which the Bell measurements are applied in step (iv). Thus, as long as the inherent success probabilities of the QND measurement and the Bell measurement are constant (or, precisely, independent of the transmittance $\sqrt{\eta}$ of the channels), Alice and Bob would have an entangled pair with a finite probability, through steps (v) and (vi). Therefore, $m \sim (\sqrt{\eta})^{-1}$ is enough to present an entangled pair to Alice and Bob, implying that the communication efficiency, that is, the secret key rate per pulse\footnote{Notice that an optical pulse here is regarded as being composed of two bosonic modes, i.e., a mode for horizontally polarized photons and a mode for vertically polarized photons. Hence, for this optical pulse, the PLOB (upper) bound on achievable secret key rates of point-to-point QKD between Alice and Bob per pulse (composed of the two bosonic modes) is $-2\log_2 (1-\eta)$, which is approximated to $2\eta/\ln 2 \approx 2.89 \eta$ for very small $\eta$.}, of the protocol scales with $\sqrt{\eta}$.

\subsubsection{Memory-assisted implementation}\label{se:MAI}

\begin{figure}[t]
    \centering
    \includegraphics[width=\linewidth]{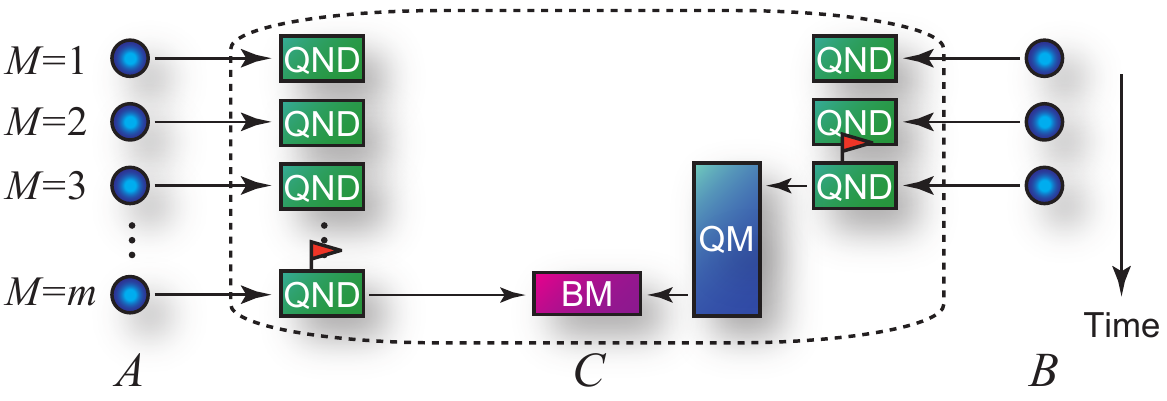}
    \caption{The concept of memory-assisted MDI QKD. In this protocol, once the node $C$ confirms the arrival of an optical polarization qubit either from Alice's side or from Bob's side with QND measurement [which is described by a (red) flag on a box labeled ``QND'' in the figure], it keeps it in a quantum memory (QM) until an optical polarization qubit arrives at the node $C$ from the other side, followed by its release to be subjected to Bell measurement (BM).}
    \label{fig:AdaptiveMDImemory}
\end{figure}

\begin{figure}[b]
    \centering
    \includegraphics[width=60mm]{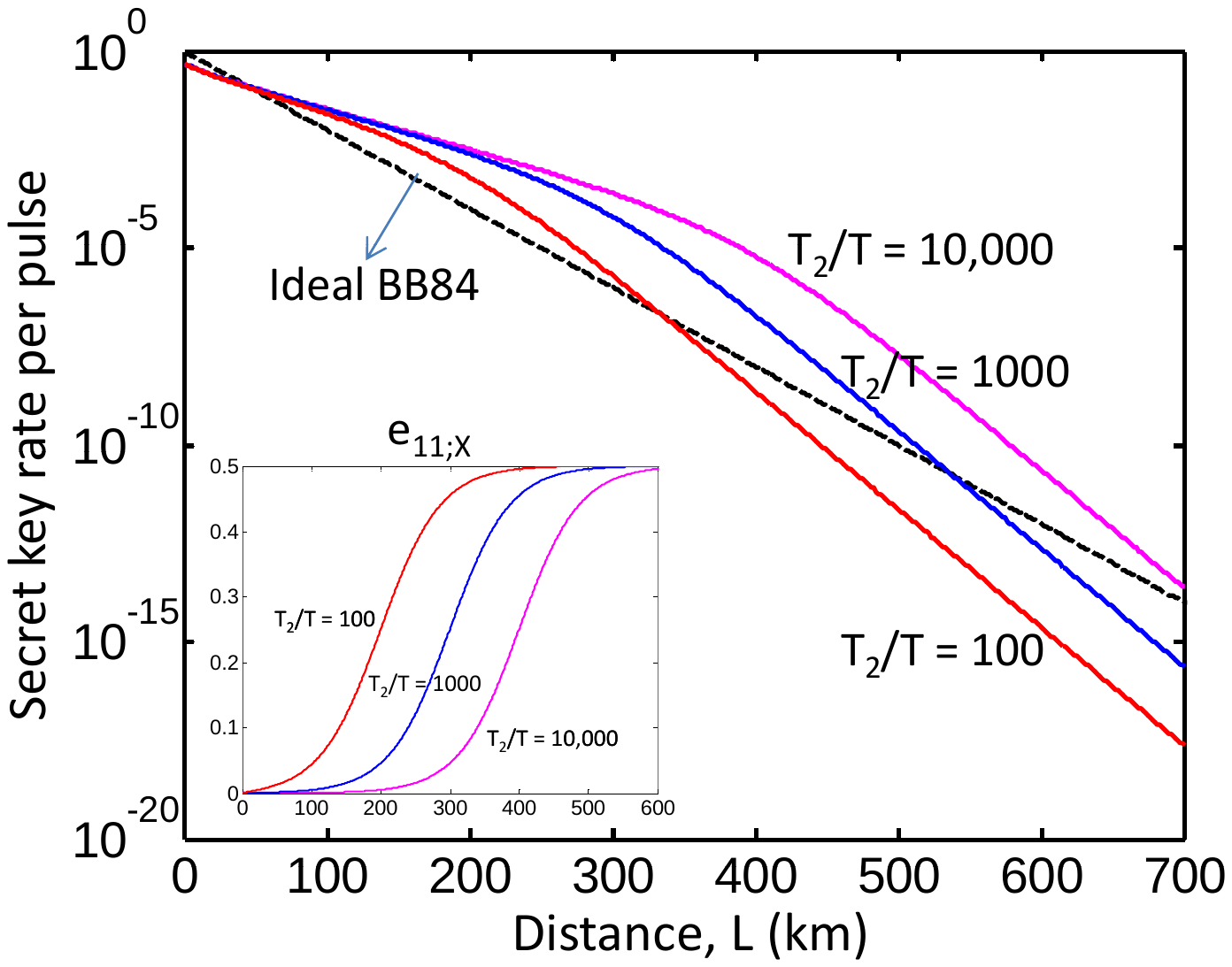}
    \caption{Secret key rate (per pulse) of an adaptive MDI QKD protocol based on matter quantum memories with heralding storage and on Alice's and Bob's use of ideal single-photon sources. The secret key rate of the ideal BB84, which scales linearly with $\eta=e^{-L/L_{\rm att}}$ ($L_{\rm att}=22$\,km), is also shown as a reference. $T_2$ is the dephasing time for the matter quantum memories, $1/T$ is the pulse generation rate of Alice and Bob, and $e_{11;{\rm X}}$ is the phase error rate for Alice's and Bob's raw key. $T_2/T$ corresponds to how many attempts, each of which needs time $T$, are possible for the matter quantum memory to successfully store a single photon within its coherence time $T_2$, that is, the allowed number $m$ of time multiplexing in the protocol. The secret key rate scales linearly with $\sqrt{\eta}$ as long as $T_2/T \ge (\sqrt{\eta})^{-1}$, but it then converges to $\eta$ as $\eta$ decreases. This is because the increase of phase error $e_{11;{\rm X}}$ for the case of $T_2/T \le (\sqrt{\eta})^{-1}$ nullifies the benefit of time multiplexing from the use of matter quantum memories, as shown in the panel.
Figure adapted from~\cite{Panayi2014}.}
    \label{fig:AdaptiveMDImemory-graph}
\end{figure}

The memory-assisted MDI QKD protocol \cite{abruzzo2014measurement,Panayi2014} corresponds to an implementation of the above protocol (in Sec.~\ref{se:AMDIQKD}) by utilizing the functionality of matter quantum memories (Fig.~\ref{fig:AdaptiveMDImemory}). In particular, the protocol assumes that the central node $C$ uses matter quantum memories to achieve steps (ii)-(iv), and $m$ optical polarization qubits in step (i) are sent by Alice and Bob in a time-multiplexing manner.
If we can use a matter quantum memory that heralds the successful storing of a received optical polarization qubit, this heralding signal is regarded as the signal of the success of the QND measurement in step (ii). To achieve step (iii), the node $C$ just uses one memory for Alice and one memory for Bob. Each of these memories receives optical pulses from Alice or from Bob until it successfully stores a single photon. Once this storage succeeds, each memory keeps the qubit information until the other memory heralds the successful storage. If both memories herald the successful storage of a single photon, they load the stored photons to perform the linear-optics-based Bell measurement of Fig.~\ref{fig:BellM}~(a) on them as step (iv). The secret key rate of this protocol is exemplified in Fig.~\ref{fig:AdaptiveMDImemory-graph}, which shows $\sqrt{\eta}$-scaling when the required memory time in the protocol is shorter than the coherence time of quantum memories.

Although we have assumed that the matter quantum memories have a function of heralding the storage, this method works even with a matter quantum memory which can just compose a Bell state with an optical polarization qubit. In particular, in this case, as step (ii),
the node $C$ just needs to perform the linear-optical Bell measurement of Fig.~\ref{fig:BellM}~(a) on this polarization qubit emitted by a quantum memory and a received pulse from Alice (or Bob). Since this Bell measurement provides the signal of the success only when it receives two (or more) photons, the signal of the success of this Bell measurement implies that the qubit information held by the pulse from Alice (or Bob) is successfully teleported into the other half of the Bell state, i.e., into the matter quantum memory. That is, this is essentially the success of the QND measurement required in step (ii). Hence, a matter quantum memory which can compose a Bell state with an optical polarization qubit allows the node $C$ to implement the QND measurement in an indirect manner, which is also enough to implement the memory-assisted MDI QKD protocol.

This memory-assisted implementation uses time multiplexing by utilizing matter quantum memories. The dominant noise of matter quantum memories is dephasing and/or amplitude damping (which is sometimes treated as a depolarizing channel to simplify theoretical treatment), any of which increases exponentially with time.
Therefore, the noise would significantly limit the allowed number $m$ of time multiplexing in the memory-assisted MDI QKD protocols.

In fact, the secret key rate of a memory-assisted MDI QKD protocol using matter quantum memories with dephasing is limited by the allowed number $m$ of multiplexing, that is, by $T_2/T$ in Fig.~\ref{fig:AdaptiveMDImemory-graph} which corresponds to how many attempts, each of which needs time $T$, are possible for the matter quantum memory to successfully store a single photon within its coherence time $T_2$.
In the graph, as $\eta$ decreases, the secret key rate scales linearly with $\sqrt{\eta}$ as long as $T_2/T \ge (\sqrt{\eta})^{-1}$, but it then converges to $\eta$.
This implies that the required coherence time $T_2$ is on the order of $(\sqrt{\eta})^{-1} T= e^{L/(2 L_{\rm att})} T$ with $\eta=e^{-L/L_{\rm att}}$ ($L_{\rm att} =22$\,km), and thus, it scales exponentially with $L/2$.
However, as long as the period $T$ of Alice's and Bob's pulse generation can be taken to be small, the required coherence time could be smaller \cite{Panayi2014} than even the minimum coherence time $L/c$ required by multiplexed first generation quantum repeaters \cite{Razavi2008}.

\subsubsection{All-optical implementation}

\begin{figure}[t]
    \centering
    \includegraphics[width=\linewidth]{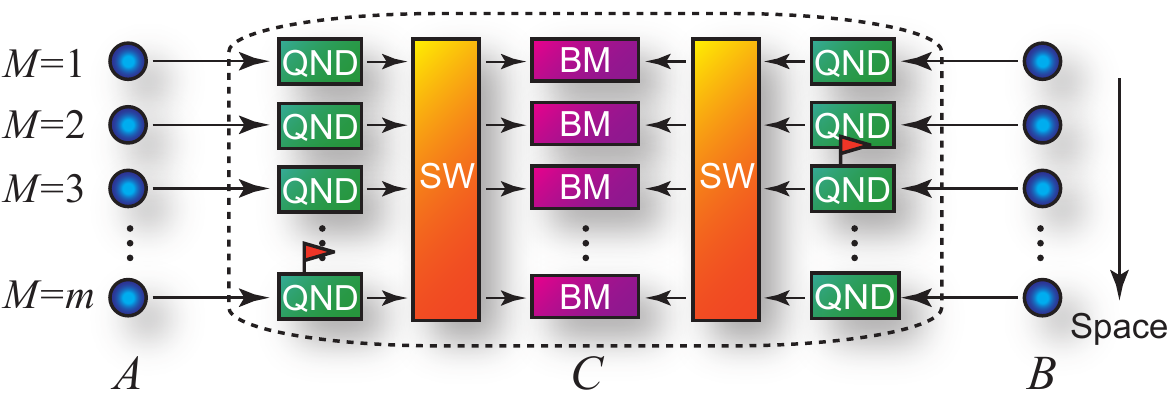}
    \caption{The concept of all-photonic adaptive MDI QKD. In this protocol, the node $C$ first performs QND measurements to confirm the successful arrival of single photons [which is described by a (red) flag on a box labeled ``QND'' in the figure], followed by optical switches (SW) to send the surviving photons to Bell measurement (BM) modules. Figure adapted from~\cite{AzumaIntercity}. }
    \label{fig:AdaptiveMDIall}
\end{figure}

\begin{figure}[b]
    \centering
    \includegraphics[width=65mm]{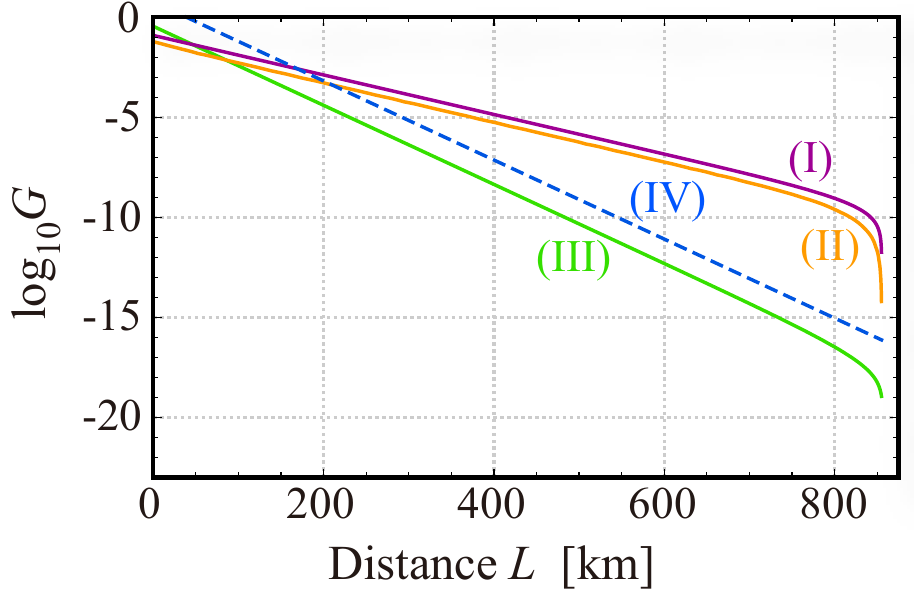}
    \caption{Secret key rate (per pulse) $G$ of an all-photonic adaptive MDI QKD protocol. $\eta$ is rephrased by the distance $L$ between Alice and Bob, with $\eta=e^{-L/L_{\rm att}}$ ($L_{\rm att}=22$\,km), and $c=2.0\times 10^8$\,m/s.
    Lines (I)–(IV) represent the performance of the protocol with active optical switches, that of the protocol with a passive Hadamard linear optical circuit, that of the original MDI QKD protocol \cite{Lo2012}, and the TGW bound \cite{Takeoka2014}, respectively.
    This graph is described under the following assumptions \cite{AzumaIntercity}: a single active feedforward can be completed within time $\tau_{\rm a}$, during which photons run in optical fibers, being subject to the corresponding photon loss; heralded single-photon sources emit pulses with duration $\tau_{\rm s}$, with efficiency $\eta_{\rm s}$, and they are multiplexed \cite{migdall2002tailoring,ma2011experimental,collins2013integrated,christ2012limits,bonneau2015effect} to produce high-fidelity telecom single photons with the repetition rate of the slowest optical device at the expense of the use of (at least) one active feedforward; single-photon detectors have quantum efficiency $\eta_{\rm d}$ and dark count rate $\nu_{\rm d}$; Bell pairs for the all-photonic QND measurements can be generated in constant time $\tau_{\rm a}$ with single-photon sources rather than a Bell-pair photon source, by paralleling a probabilistic procedure \cite{Browne2005} with the active feedforward technique. In particular, they are assumed to be
    $\eta_{\rm s}=0.90$ \cite{migdall2002tailoring,giustina2013bell,christensen2013detection}, $\tau_{\rm s}=100$\,ps \cite{shibata2014quantum}, $\eta_{\rm d}=0.93$ \cite{Marsili2013}, $\nu_{\rm d} =1$\,s$^{-1}$ \cite{Marsili2013,shibata2010single}, and $\tau_{\rm a}=67$\,ns \cite{ma2011experimental}. Figure from~\cite{AzumaIntercity}.}
    \label{fig:AdaptiveMDIall-figure}
\end{figure}

The all-photonic adaptive MDI QKD protocol could be understood as an all-optical implementation of the above protocol in Sec.~\ref{se:AMDIQKD} \cite{AzumaIntercity} (Fig.~\ref{fig:AdaptiveMDIall}). In the protocol,
the QND measurement in step (ii) is assumed to be performed by using a quantum teleportation, similar to the memory-assisted MDI QKD protocol, but it is implemented by using only optical devices\footnote{An idea similar to this, called a qubit amplifier, is also used in the context of the device-independent QKD in order to close the detection loophole problem \cite{GPS10,CM11}.}. 
In particular, to achieve the QND measurement in step (ii), the node $C$ first prepares optical polarization qubits in a Bell state locally, and applies the linear-optical Bell measurement of Fig.~\ref{fig:BellM}~(a) on the half of this Bell pair and the optical pulse sent by Alice or Bob.
The success of this Bell measurement teleports the qubit information of the surviving single photon into the other half of the Bell pair, corresponding to the success of the QND measurement.
Since this protocol does not assume to use matter quantum memories, $m$ optical polarization qubits in step (i) of this protocol are assumed to be sent by Alice and Bob simultaneously in a spatial-multiplexing manner. Thus, the above all-optical QND measurements in step (ii) are performed at the same time on all the pulses sent by Alice and Bob, and then the pairing in step (iii) is made by using an optical switch. The performance of this protocol is exemplified in Fig.~\ref{fig:AdaptiveMDIall-figure}, which shows $\sqrt{\eta}$-scaling of the secret key rate.

This all-optical implementation uses spatial multiplexing by utilizing optical switches. The dominant noise of optical switches is the photon loss. However, in contrast to memory-assisted implementation, this loss increases only logarithmically with the number $m$ of spatial multiplexing \cite{AzumaIntercity}.
Note that the all-optical protocol can achieve the $\sqrt{\eta}$-scaling even if it uses only an $m\times 1$ optical switch and a Bell measurement module at the middle node $C$. Thus, if we implement an $m\times 1$ optical switch by concatenating $2\times 1$ optical switches with transmittance $\eta_{\rm sw}$ in a knockout tournament manner with depth $\lceil \log_2 m \rceil$, the transmittance of the $m\times 1$ optical switch decreases as $\eta_{\rm sw}^{\lceil \log_2 m \rceil}$, and it thus scales only logarithmically with the number $m$.
This is a merit to use the spatial multiplexing, rather than time multiplexing.
Such combination of an $m\times 1$ optical switch and a Bell measurement module is also implementable without using such a large-scale optical switch, that is, by
using only single-mode on/off switches, a passive Hadamard linear optical circuit and single-photon detectors \cite{AzumaIntercity}. The performance in this case is also described in Fig.~\ref{fig:AdaptiveMDIall-figure}.

\subsubsection{Challenges}

The question of whether a two-mode squeezed state, which can be produced with practical systems, can directly be used as the Bell state to implement the teleportation-based QND measurement in step (ii) has been answered to be negative, so far.
For instance, if we use atomic-ensemble quantum memories for the memory-assisted MDI QKD protocol,
the memory can naturally compose a two-mode squeezed state with an optical pulse \cite{Duan2001,Sangouard2011}. However, this entanglement cannot directly be used as a resource to implement the teleportation-based QND measurement in step (ii) \cite{piparo2014measurement}, because the multi-photon component of the two-mode squeezed state makes the success probability of the QND measurement depend on the transmittance $\sqrt{\eta}$ of the channels. This result is made stronger by assuming that the node $C$ is allowed to use photon number-resolving detectors \cite{Trenyi2019}, rather than threshold detectors assumed in \cite{piparo2014measurement}. In particular, the paper shows that the polarization entanglement produced by a spontaneous parametric down-conversion (SPDC) process cannot directly be used to implement the QND measurement in step (ii) of the all-photonic adaptive MDI QKD protocol, by deriving necessary conditions on photon-number statistics of the entanglement photon sources.

As a result, a single matter qubit, such as a single ion, a quantum dot or a nitrogen-vacancy center in a diamond, inside a cavity is proposed as a candidate for the memory to realize the memory-assisted MDI QKD protocol \cite{piparo2017measurement,piparo2017memory}, while a source emitting an entangled photon pair with a low multi-photon component, such as one assumed in the original paper \cite{AzumaIntercity} (see the caption of Fig.~\ref{fig:AdaptiveMDIall-figure}) or an entanglement photon source \cite{Eisaman2011}, is needed to implement the all-photonic adaptive MDI QKD protocol. In the case where multi-photon emission is highly suppressed, threshold detectors without having the function of photon-number resolving are sufficient for implementing the teleportation-based QND measurement.

As for the all-photonic adaptive MDI QKD protocol, since it needs only QND measurements on the photon number, it could adopt different types of QND measurements, such as one in \cite{imoto1985quantum} based on an optical Kerr effect and one in \cite{brune1990quantum} based on a dispersive atom-field coupling (see textbooks \cite{scully1999quantum,walls2007quantum}). It is thus an important open question whether the all-photonic adaptive MDI QKD keeps their merit on communication efficiency even if we replace the teleportation-based QND measurement with an alternative one.
As for the memory-assisted MDI QKD, a proof-of-principle experiment of the key element has been performed with a single solid-state spin memory integrated in a nanophotonic diamond resonator \cite{bhaskar2020experimental}, based on an encoding on the phase difference between sequential two pulses (like one used in a differential phase shift QKD \cite{PhysRevLett.89.037902}) (see also Sec.~\ref{se:pocofQR}).

\subsection{Twin-field QKD}\label{se:TFQKD}

To double the communication distance by utilizing a central node $C$ between communicators, another idea is also focused on especially in the field of QKD, thanks to the proposal of a twin-field (TF) QKD protocol based on a single-rail encoding \cite{Lucamarini2018}. The scaling improvement of the TF QKD protocol is essentially brought by the following point: like entanglement generation processes in quantum repeater protocols \cite{Duan2001,childress2006fault,azuma2012quantum}, the protocol makes the node $C$ use a simple linear-optical Bell measurement of Fig.~\ref{fig:BellM}~(b) based on single-photon interference, rather than two-photon interference used in the original MDI QKD \cite{Lo2012}, and Alice and Bob encode their qubit information into a {\it single} optical mode (i.e., a single-rail encoding), rather than two modes (i.e., a dual-rail encoding, such as polarizations and time bins). This aims to utilize the feature that this Bell measurement---to project a given state into a Bell state $(\ket{0}\ket{1}\pm \ket{1}\ket{0})/\sqrt{2}$ with the vacuum state $\ket{0}$ and the single-photon state $\ket{1}$ as shown in Fig.~\ref{fig:BellM}~(b)---succeeds if a single photon reaches node $C$ either from Alice or from Bob. For instance, in the case of the DLCZ protocol \cite{Duan2001}, states of each local memory of Alice and Bob are entangled with the number states (i.e., the Fock states) of a single optical mode, while in the case of hybrid quantum repeater protocols \cite{childress2006fault,azuma2012quantum}, the computational basis states of each of Alice and Bob's local qubits are entangled with two coherent states of a single optical mode (corresponding to a cat-state encoding).
As a result, the efficiency of this type of entanglement generation schemes \cite{Duan2001,childress2006fault,azuma2012quantum} scales with $\sqrt{\eta}$, rather than $\eta$, without requiring any challenging devices at the node $C$, thanks to the use of single-photon interference.
This scaling improvement in the entanglement generation might be reasonable because it relies on the following technical challenges: 
\begin{itemize}
    \item[(a)] those entanglement generation schemes need intense phase stabilization regarding the channels between Alice and the node $C$ and between Bob and the node $C$, in contrast to ones based on two-photon interference at the node $C$;
    \item[(b)] those schemes require Alice and Bob to use matter quantum memories which could be used to prepare nontrivial optical states, such as number states \cite{Duan2001} and cat states \cite{childress2006fault,azuma2012quantum}.
\end{itemize}

A bold claim was given in the original proposal of the TF QKD protocol \cite{Lucamarini2018}: it had argued that if we borrow the idea of the decoy-state method \cite{hwang2003quantum,lo2005decoy,wang2005beating}, coherent states  are enough to achieve QKD with $\sqrt{\eta}$ scaling, without the necessity of any device which has a potential to prepare nontrivial optical states (in contrast to the entanglement generation schemes with requirement (b) above).
The idea was stemmed \cite{Lucamarini2018} from making a decoy-state phase-encoding BB84 protocol be in the form of an MDI QKD setup, namely, attaching the decoy-state method to a phase-encoding MDI QKD protocol \cite{tamaki2012phase}. However, despite the extremely appealing claim, a rigorous security proof against the most general type of eavesdropping strategies was missing in the original proposal \cite{Lucamarini2018}: only security over restricted eavesdropping was proven.
This triggered a lot of interest to develop variants of the TF QKD protocol, as well as their security proofs over arbitrary eavesdropping attacks in asymptotic scenarios \cite{tamaki2018information,ma2018phase,lin2018simple,curty2019simple,cui2019twin,wang2018twin} and in finite-size scenarios \cite{jiang2019unconditional,maeda2019repeaterless,lorenzo2019tight,yu2019sending,Xu2020}.
Here we focus on a variant \cite{curty2019simple} of the TF QKD protocol, as it is explicitly related with entanglement generation protocols in quantum repeaters, to see why coherent states are enough to achieve QKD.

Before introducing the variant protocol, let us introduce its coherent version, which is essentially equivalent to an entanglement generation protocol \cite{azuma2012quantum}.
The coherent version is described as follows.
(i) Each of Alice and Bob prepares an optical pulse entangled with a local qubit, whose state is described as $(\ket{0}\ket{\alpha}+\ket{1}\ket{-\alpha})/\sqrt{2}$, where $\ket{0}$ and $\ket{1}$ are orthogonal states of the local qubit and $\ket{\pm \alpha}$ are coherent states of the optical pulse with an amplitude $\alpha>0$.
(ii) Each of them sends the prepared optical pulse to the node $C$ over a lossy bosonic channel (\ref{eq:lossch}) with the transmittance $\sqrt{\eta}$.
(iii) On receiving the pulse $a$ in coherent state $\ket{\pm \sqrt[4]{\eta} \alpha}_a$ from Alice and the pulse $b$ in coherent state $\ket{\pm \sqrt[4]{\eta} \alpha}_b$ from Bob, the node $C$ performs a linear-optical Bell measurement of Fig.~\ref{fig:BellM}~(b) relying on single-photon interference on them. 
(iv) The node $C$ then announces the measurement outcome of the Bell measurement. (v) Finally, Alice and Bob keep their local qubits if they know that one of two detectors for the Bell measurement announces arrival of photons, through the announcement in step (iv).

Notice that the 50:50 beamsplitter of the Bell measurement in step (iii) (Fig.~\ref{fig:BellM}~(b)) converts received states $\ket{\pm \sqrt[4]{\eta} \alpha}_a\ket{\pm \sqrt[4]{\eta} \alpha}_b$ into coherent states $\ket{\pm \sqrt{2} \sqrt[4]{\eta} \alpha}_c \ket{0}_d$ and $\ket{\pm \sqrt[4]{\eta} \alpha}_a\ket{\mp \sqrt[4]{\eta} \alpha}_b$ into coherent states $\ket{0}_c \ket{\pm \sqrt{2} \sqrt[4]{\eta} \alpha}_d$, respectively, where $c$ and $d$ are the outputs having received constructive interference and destructive interference, respectively. Since the detection of photons in the number basis erases the phase information $\pm$ of the coherent states $\ket{\pm \sqrt{2}\sqrt[4]{\eta} \alpha}$, the successful detection of photons defined in step (v) works as nondestructive parity measurement, i.e., projection measurement $\ket{00}\bra{00}+\ket{11}\bra{11}$ or $\ket{01}\bra{01}+\ket{10}\bra{10}$ on Alice and Bob's local qubits \cite{azuma2012quantum}, which entangles their local qubits in the protocol.

To see the scaling, suppose that the Bell measurement is performed by using ideal threshold detectors, for simplicity. Then,
the success probability of the Bell measurement is $r=1-e^{-2 \sqrt{\eta} \alpha^2}$, while the Bell pair obtained at step (iv) includes only the phase error with probability $e_Z =(1-e^{-2 \alpha^2(2-\sqrt{\eta})})/2$ \cite{azuma2012quantum}. This performance as entanglement generation is shown to be optimal in various scenarios \cite{azuma2009optimal,azuma2010tight,azuma2012optimal,azuma2022optimal}.
If we maximize an asymptotic key rate formula $G=r(1-h(e_Z))$ with the binary entropy function $h$ over $\alpha$, we can easily confirm that $G$ scales with $\sqrt{\eta}$. However, this merely means that the key rate $G$ could scale $\sqrt{\eta}$ when Alice and Bob use matter quantum memories to realize their local qubits, as considered in Ref.~\cite{azuma2012quantum}.

\begin{figure}[b]
    \centering
    \includegraphics[width=\linewidth]{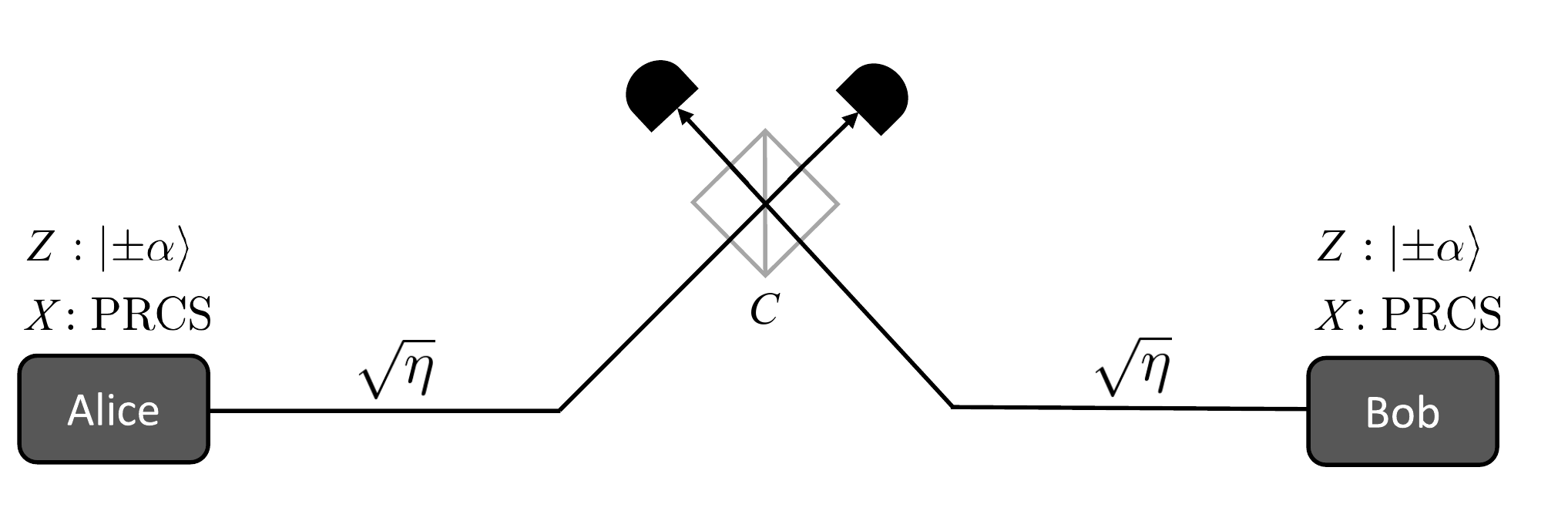}
    \caption{Schematic of TF-type QKD protocol \cite{curty2019simple}. Each of Alice and Bob chooses $Z$ basis or $X$ basis, randomly. If $Z$ basis is selected, Alice and Bob prepare coherent state $\ket{\alpha}$ or $\ket{-\alpha}$ at random, and send it to the central node $C$. If $X$ basis is selected, Alice and Bob prepare a phase-randomized coherent state (PRCS) whose intensity is chosen randomly from a predefined set (so as to be able to use the decoy-state method \cite{hwang2003quantum,lo2005decoy,wang2005beating}), and send it to the central node $C$. On receiving pulses from Alice and Bob, the central node $C$ performs the Bell measurement based on single-photon interference (Fig.~\ref{fig:BellM}~(b)).  The secret key is distilled only from instances where both of Alice and Bob choose $Z$ basis and the Bell measurement at the node $C$ succeeds. 
    Figure adapted from~\cite{lorenzo2019tight}.}
    \label{fig:TFtype}
\end{figure}

\begin{figure}[t]
    \centering
    \includegraphics[width=\linewidth]{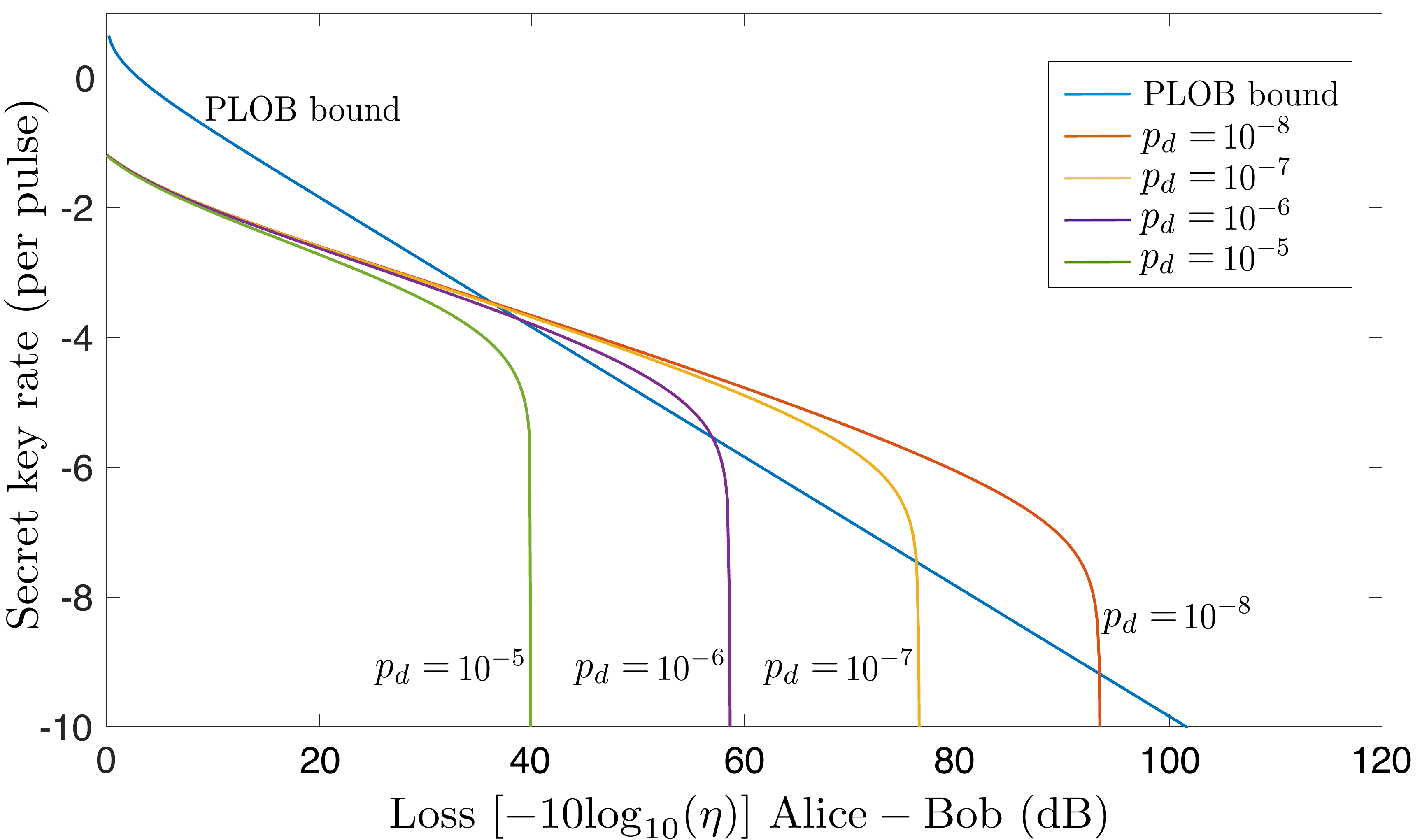}
    \caption{Secret key rates (per pulse) of a TF-type QKD protocol for different dark count rates, $p_{\rm d}$, in logarithmic scale as a function of the overall loss between Alice and Bob. The PLOB bound is the private capacity of a lossy bosonic channel \cite{Pirandola2015}.
    This figure is described by assuming a misalignment of 2\% in each channel Alice-$C$ and Bob-$C$, and the inefficiency function for the error
correction process $f = 1.16$.
    Figure adapted from~\cite{curty2019simple}.}
    \label{fig:TFtype-figure}
\end{figure}

To make the protocol composed of steps (i)-(iv) a prepare-and-measure scheme, Alice and Bob are supposed to perform $Z$-basis or $X$-basis measurement randomly on each of their local qubits just after step (i) and before step (ii). Here, the $Z$-basis measurement prepares the optical pulse in coherent state $\ket{\alpha}$ or $\ket{-\alpha}$ at random, while the $X$-basis measurement prepares it in cat state $\ket{C_+}:=(\ket{\alpha}+\ket{-\alpha})/(2\sqrt{p_+})$ with probability $p_+$ or $\ket{C_-}:=(\ket{\alpha}-\ket{-\alpha})/(2\sqrt{p_-})$ with probability $p_-$, where $p_\pm= (1\pm \langle-\alpha|\alpha\rangle)/2$.
The preparation of coherent states $\ket{\pm \alpha}$ regarding the $Z$-basis measurement can be done easily.
In contrast, the preparation of cat states $\ket{C_\pm}$ regarding the $X$-basis measurement is problematic, because it requires a challenging device in practice.
However, this preparation is not necessary,
if Alice and Bob will distill a key only from the outcomes of the $Z$-basis measurements. In particular, in the case of this QKD, the $X$-basis measurements are used only to estimate the actual phase error rate $e_Z$ for privacy amplification,
and estimation of its upper bound by Alice and Bob through a protocol is enough to prove the security \cite{Mayers2001,Lo1999,Shor2000,Koashi2009,Renner2008,Christopher2022,Bennett1992}. In fact, it turns out that the estimation of an upper bound on the phase error $e_Z$ can be done just by sending phase-randomized coherent states in the case of the choice of $X$ basis and by invoking a decoy-state method, without preparing the cat states $\ket{C_\pm}$ \cite{curty2019simple}. As a result, the protocol is described as in Fig.~\ref{fig:TFtype}, and the conjecture in the original proposal that
the coherent states (and their phase-randomized ones) are enough to achieve QKD with $\sqrt{\eta}$-scaling is concluded to be true, as shown by the performance in Fig.~\ref{fig:TFtype-figure}.

The TF QKD protocol and its secure variants omit technical challenge (b) as unnecessary for QKD, but they still include technical challenge (a). Nonetheless, various experiments \cite{minder2019experimental,zhong2019proof,wang2019beating,pittaluga2021600,chen2020sending,clivati2022coherent,chen2021twin,wang2022twin,zhong2021proof,zhong2022simple}
to overcome this have already been performed, towards the full implementation of the TF-type QKD protocols in practical scenarios.
These trials are important even for quantum repeaters, because they
represent a good milestone towards the realization of a quantum repeater protocol based on single-photon interference, which involves
the same technical challenge (a) (like Refs.~\cite{Duan2001,childress2006fault,azuma2012quantum}).

In TF QKD, to achieve the phase stability required for single-photon detection based entanglement swapping, there are two general strategies.
The first strategy is to use only one laser and
employ auto-compensation with a Sagnac loop where
optical signals go through the same path either clockwise or counter-clockwise \cite{zhong2019proof,zhong2021proof,zhong2022simple}. The second strategy allows two independent lasers to be used,
but may require a combination of techniques including, for
example, frequency locking,
using a reference pulse for compensation and ensuring that
the optical path lengths
of the two optical fibers do not differ too drastically \cite{minder2019experimental,wang2019beating,pittaluga2021600,chen2020sending,clivati2022coherent,chen2021twin,wang2022twin,li2022twin}.

\subsection{The single sequential quantum repeater}
\label{se:trAMDIQKD}
A third alternative is to invert the previous schemes and place a quantum device with a quantum memory in the central node and two detectors in the end nodes. This scheme was proposed by \cite{luong2016overcoming}. 

In this scheme, the central node sends a photon entangled with a memory qubit to one of the end nodes until the end node confirms successful detection of the photon. Then, the central node repeats the same process with the other end node and thus emits a photon entangled with a memory qubit until success. Once the second end node confirms the successful detection of a photon, the central node performs a Bell measurement and heralds the measurement outcome to the two end nodes. 

The advantage of this scheme is the simplicity of the setup, requiring a single node holding two memory qubits and no optical Bell measurement. On the other hand, this setup is not measurement-device independent and it requires qualitatively long coherence times when compared with memory-based adaptive MDI QKD. In particular, the coherence time should be large when compared with the sum of the travel time of a photon from the center node to an end node plus the corresponding heralding signal, multiplied by the average number of times required for a successful event. 

The feasibility of this setup for outperforming the point-to-point limits was analyzed for different hardware parameters in \cite{luong2016overcoming}, \cite{rozpkedek2018parameter} and \cite{Rozpcedek2019}. An experimental demonstration of the setup was recently reported in \cite{langenfeld2021quantum} with Rubidium atoms in an optical cavity. While below the fundamental limit for direct transmission, the scaling of the key rate in the experiment was shown to be proportional to the square root of the transmittance of an optical fiber connecting two end parties.

\subsection{Post-pairing measurement-device-independent QKD}\label{sec:postpairingMDIQKD}

\begin{figure}[t]
    \includegraphics[width=\linewidth]{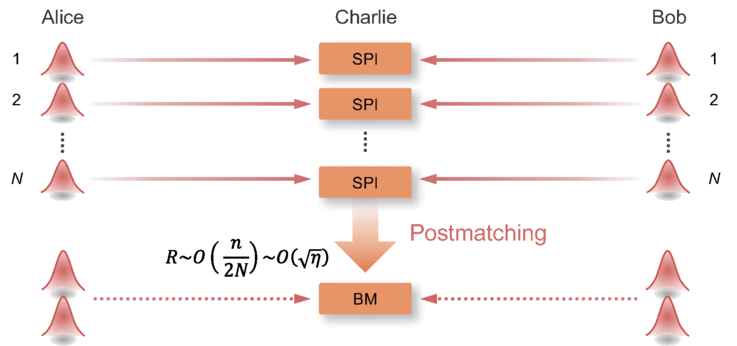}
    \caption{Post-encoding measurement-device-independent QKD. In this protocol, Alice and Bob send $N$ pulses to the middle node $C$, Charlie, to perform the linear-optical Bell measurement of Fig.~\ref{fig:BellM}~(b) based on single-photon interference (SPI), and a two-photon Bell state is obtained by postmatching two successful SPI events. Here $n$ represents the number of successes of the Bell measurement based on SPI and $\sqrt{\eta}$ represents the transmittance of pure-loss channels between Alice and Charlie and between Charlie and Bob. Figure adapted from \cite{xie2022breaking}.}
    \label{fig:PEMDIQKD}
\end{figure}

Recently, Xie {\it et al.} and Zeng {\it et al.} have proposed a variant of MDI QKD protocol \cite{xie2022breaking,zeng2022mode} which may be conceptually intermediate between adaptive MDI QKD and TF QKD and whose secret key rate can scale with $\sqrt{\eta}$, rather than $\eta$, where $\eta$ is the transmittance of a pure-loss channel between Alice and Bob. In this protocol (Fig.~\ref{fig:PEMDIQKD}), the middle node $C$ still uses linear-optical Bell measurement of Fig.~\ref{fig:BellM}~(b) based on single-photon interference like TF QKD, while Alice and Bob send $N$ optical pulses in coherent states to the middle node sequentially, that is, in a time-multiplexing manner, like adaptive MDI QKD. The main aim here is to make a protocol rely on the application of a Bell measurement to project into Bell states $(\ket{01}_{a_ia_j}\ket{10}_{b_ib_j}\pm \ket{10}_{a_ia_j}\ket{01}_{b_ib_j})/\sqrt{2}$, based on {\it two-photon interference} between $i$th and $j$th time bins ($i,j=1,2,\cdots,N$ and $i \neq j$) at the middle node $C$, where $a_i$ and $b_i$ are $i$th time bins sent by Alice and Bob, respectively. This is implemented by postselecting time slots $i$ and $j$ to which the Bell measurements based on single-photon interference at the node $C$ are successfully applied, under the assumption that the phase correlation between such possibly long time separated $i$th and $j$th time bins is kept in the implementation. This keeping of the phase correlation is a technological challenging part if the number $N$ of multiplexing is large. Nonetheless, since this protocol can be regarded as relying on two-photon interference at the middle node $C$, rather than single-photon interference, like adaptive MDI QKD, an intense phase stabilization regarding the channels between Alice and the node $C$ and between Bob and the node $C$ could be unnecessary in contrast to the TF QKD.
In the protocol, Alice and Bob send Charlie optical pulses in coherent states whose phases are chosen randomly from $[0,2\pi)$ and whose intensities are chosen randomly from a predefined set. This is designed so that time bins $a_ia_j$ and $b_ib_j$, postselected by the middle node $C$, can be regarded as a BB84 signal and a decoy state, that is, a signal used in the normal MDI QKD with time-bin encoding \cite{Ma2012mdi}. This postselection includes the matching between Alice's and Bob's random choices of phases in some cases (although, in contrast, it was shown to be unnecessary in the case of TF QKD \cite{cui2019twin,curty2019simple,lin2018simple,lorenzo2019tight,maeda2019repeaterless}).

For a large number $N$ of the multiplexing, $n=\mathcal{O}(N\sqrt{\eta})$ Bell measurements based on {\it single-photon} interference would succeed, where $\sqrt{\eta}$ represents the transmittance of pure-loss channels between Alice and the middle node $C$ and between the middle node $C$ and Bob. Hence, there would be $\mathcal{O}(n/2)=\mathcal{O}(N\sqrt{\eta}/2)$ instances to which the target Bell measurements based on {\it two-photon} interference are successfully applied. Since the success of the target Bell measurement could produce an entangled state between Alice's virtual qubit and Bob's virtual qubit, the secret key rate of the protocol could scale with $\sqrt{\eta}$.

According to the proposals, called mode-pairing QKD \cite{zeng2022mode} and asynchronous
MDI-QKD \cite{xie2022breaking}, experimental demonstrations have been performed in Refs.~\cite{zhu2022experimental} and \cite{Zhou2023}, respectively.

\section{Experimental progress towards repeaters}
\label{sec:exps}
Long-distance quantum communication is enabled by low-loss media for photon transfer. Free-space communication~\cite{Ursin2007} and satellite-based communication~\cite{Liao2017, Yin2017} have unique experimental challenges; in this section, we chiefly describe the practical advances towards optical-fiber-based quantum communication schemes featuring quantum repeaters. We organize our discussion roughly according to the requirements of each generation of repeaters from Sec.~\ref{sec:gens} and of memoryless repeaters from Sec.~\ref{sec:memoryless}.

Almost all quantum repeater architectures require the implementation of efficient interfaces between quantum memories and photons. In first-generation repeaters, a quantum memory must be capable of storing quantum information for a long time (Sec.~\ref{subsec_memory}) and emitting photons that are entangled with the memory degrees of freedom (Sec.~\ref{subsec_emission}). These photons are then coupled into optical fibers that connect distant repeater nodes. The intermediate entanglement between distant quantum memories (Sec.~\ref{subsec_spin_spin}) is finally used to create end-to-end entanglement links between Alice and Bob with a rate ideally much higher than direct transmission over fibers.

In first-generation repeaters, unavoidable memory errors are dealt with through entanglement distillation (Sec.~\ref{subsec_purification}). In the second generation of quantum repeaters, memory errors are corrected through quantum error correction. Therefore, quantum registers of many quantum memories are required at each repeater node to encode logical memory qubits (Sec.~\ref{subsec_error_correction}). In the third generation of repeaters, loss errors are also dealt with through QEC. Since any QEC code can only tolerate a probability of erasure (a common model for loss) of 50\% (see Sec.~\ref{sec:no-cloning} and \ref{subsec_error_correction}), advanced engineering is required to obtain high transmissivities as well as collection, coupling, and detection efficiencies for the photons (Sec.~\ref{sec:loss_mitigation}).
In addition to the experimental progress aligning with the three generations, we review the headway that has been made towards memoryless repeaters (Sec.~\ref{subsec_qr_progress}), whose all-photonic implementations require the efficient generation of highly-entangled states of many photons. Finally, we overview the experimental demonstrations of trusted QKD networks and small quantum networks (Sec.~\ref{sub_sec_exp_quantum_network}) that exist as important milestones on the way to a quantum internet.

\renewcommand{\arraystretch}{1.0}
\begin{table*}
\footnotesize
    \begin{tabular}{c | c || c | c | c | c | c | c  | c | c |  c | c | c | c}
   \toprule
   \multicolumn{2}{c||}{Quantum emitter} &
   \multicolumn{3}{c|}{Quantum memory} &
   \multicolumn{2}{c|}{Quantum register}  &
   \multicolumn{6}{c|}{Emitting properties} & Refs. \\ [0.5ex]
    \multicolumn{2}{c||}{ } &
     $T_2$ & $T_2^*$ & F (gate)
     & $N$ & $T_2$
     &  $T_1$ & $\eta_{\rm eff}$ & $\iota$ & $\eta_{\rm DW}$ & $F$ (atom-phot.) & $\lambda_{\rm QE}$ (nm) & \\
    \hline
    Atomic ensemble & $^{87}$Rb & 16\,s  &   &   &  &  &    & 87\%  &   &   &  $\geq 93.3\%$  & 780  &\footnote{\cite{Hosseini2011, Dudin2013, Xu2013b, Park2019}}  \\
    \hline
    Single atoms / & $^{87}$Rb & 2.6\,ms & $400\, \mu$s  & $\geq 97.5 \%$  &    &   & $300$\, ns & 60\%  &   &   & 89\%   & 780 & \footnote{\cite{Ebert2015, Levine2019, Langenfeld2020, Daiss2021, Van2020}}\\
    trapped ions & $^{171}$Yb$^+$  & $>1$\,h &   &   &    &   &    &   &   &   &    &  369 & \footnote{\cite{Wang2021}}\\
      & $^{128}$Ba$^+$  & 4\,ms  &   &   & $\geq 4$ ($^{171}$Yb$^+$)  &  $>1$\,h   &   &   &   &   & $\geq 86\%$   &  493 & \footnote{\cite{Inlek2017}}\\
     \hline
    Quantum dot &  & 3\,$\mu$s & 39\,ns  & 95\%  & 1  & $1\,\mu$s  &  0.6-0.8\,ns & 57\%  & 99.5\%  & 90\%  &  $\geq 80\%$ &   900-1565 & \footnote{\cite{Bechtold2015,Ethier2017,DeGreve2011,Gangloff2019,Jackson2021, Tomm2021, Somaschi2016,Matthiesen2013,Olbrich2017}}\\
    \hline
    Defects& NV  & 0.6\,s  & $5$-$36\,\micro$s  & $>99\%$   & 9  & 75\,s  & 13\,ns & 37\% & $98.6\%$  & 4\%  & $96\%$ &  637 & \footnote{\cite{Bar-Gill2013, Bradley2019, Arroyo2014, Bauch2018, Pompili2021,Aharonovich2011,Hensen2015, Ruf2021}} \\
     (diamond) & SiV  & 10\,ms & 115\,ns  &   &  1 & 100\,ms  & 1.6\,ns &  85\% & 72\%  & 75\%   & 94\%   &  737 & \footnote{\cite{Sukachev2017,Nguyen2019,Becker2018,bhaskar2020experimental,Neu2011,Neu2011b, Sipahigil2014, Ruf2021, Pingault2017}} \\
    & GeV  &  &   &   &   &   & 5.5\,ns & 0.72\% &  & 60\% &   & 602 & \footnote{\cite{Iwasaki2015,Palyanov2015, Ruf2021, Wan2020}}\\
    & SnV  &  & 540\,ns  &    &   &   & 4.5\,ns &  &  & 57\% &   & 620 &  \footnote{\cite{Trusheim2020, Gorlitz2020,Ruf2021}}\\
    \hline
    Defects & $\text{V}_{\text{Si}}$  & 0.8-20\,ms  &    &   & &   & 37\,ns    &   & 69\%  & $6-9\%$  &    & 862-917 & \footnote{\cite{Lukin2020} and references therein} \\
      (in SiC) & $\text{V}_{\text{Si}}\text{V}_{\text{C}}$  & 64\,ms & 375\,$\mu$s  & 99.98\%  &   $\geq 1$   &   & 91\,ns  &   &   &  $7\%$  &    & 1078-1132 & \\
       & V$^{4+}$  &  &   &   &     &   & 45\,ns  &   &   &  50\%  &    &1278-1388 & \\
       &  NV  &  & 1\,$\mu$s  &   &     &   & 13ns  &   &   &    &    & 1180-1468 & \\
    \hline
    Defects& G  &   &    &   &  &   &   34\,ns  &   &   & 15\%  &    & 1269 &  \footnote{\cite{Durand2021}}\\
      (in Si) & T  &  &   &   &     &   &  802\,ns &   &   & 23\%   &    & 1326 & \footnote{\cite{Bergeron2020, Kurkjian2021}}\\
    \hline
    Rare-earth & $\text{Eu}^{3+}\ignore{:\text{Y}_2\text{SiO}_5}$  &  8.1\,ms   &   &   &  $\geq 1$  & 6\,h      &  0.8-1.2\,ms &   &   &   &    & 579  & \footnote{\cite{Zhong2015b, Zhong2019b}} \\
     ions & $\text{Er}^{3+} \ignore{:\text{Y}_2\text{SiO}_5}$  &     &   &   &  $\geq 1$ &   1.2\,s  & 1.5-8.7\,ms  &   &   &   &   & 1532  & ~\footnote{\cite{Ranvcic2018, Zhong2019b}}\\
    & $\text{Pr}^{3+} \ignore{:\text{Y}_2\text{SiO}_5}$  &  $880\,\mu$s   &   &   & $\geq 1$  &      & $140\,\mu$s  &   &   &   &   & 606  & \footnote{\cite{Lago-Rivera2021, Zhong2019b}} \\
     & $\text{Nd}^{3+}\ignore{:\text{YVO}_4/\text{YSO}}$  &   &   &   &  $\geq 1$  &  &   &   &   &   &   $80\%$*   &  883 &\footnote{\cite{Liu2021, Zhong2015}} \\
     \bottomrule
  \end{tabular}
  
\caption{Properties of selected memory qubits for quantum repeater applications. Results for most systems were generally obtained in separate experiments. We distinguish the properties of the qubit emitters with those of the potential $N$-qubit registers they are coupled to. Also included are the properties of the systems as single-photon emitters, including the emission time $T_1$, the end-to-end collection efficiency $\eta_{\rm eff}$, the photon indistinguishability $\iota$, the Debye-Waller factor $\eta_{\rm DW}$, the fidelity $F$ of the atom-photon entanglement, and the emission wavelength $\lambda_{\rm QE}$. * Heralded entanglement generation fidelity between two quantum memories.}
  \label{table_exp}
\end{table*}

\subsection{Long-lived quantum memories}
\label{subsec_memory}

The success of most quantum repeater schemes critically relies on the performance of their quantum memories. The \emph{coherence time}, $T_2$, of the memory (sometimes called the \emph{memory time}) is the relevant figure-of-merit: it characterizes the time during which quantum information can be stored in the memory before being degraded by the environment. For example, when generating entanglement between two quantum memories at nodes separated by a distance $L_0$ in a heralded manner (Sec.~\ref{sec:HEGP}), high entanglement fidelities can only be achieved if $L_0 \ll c T_2$, with $c$ the speed of light in fiber.
A quantum memory needs also to  have characteristics beyond the coherence time, namely fast, efficient and high-fidelity initialization, gate application, and photon retrieval and read-out. For brevity, we restrict our discussion to the coherence time, and refer interested readers to Refs.~\cite{Lvovsky2009, Simon2010, Heshami2016} for the other important features of quantum memories.

Several candidate quantum memories are under development, among them atomic ensembles ($T_2 = 0.2$-$16\,\second$~\cite{Dudin2013, Yang2016}) including Bose-Einstein condensates~\cite{Riedl2012}, and single natural or artificial atomic systems such as cold atoms, trapped ions ($T_2=4\, \milli\second$ for $^{128}$Ba$^+$~\cite{Inlek2017}), colour centres in diamond ($T_2=1\,\second$~\cite{Bar-Gill2013, Abobeih2018}), and quantum dots ($T_2 = 3\,\micro\second$~\cite{Greilich2007}).
All of these platforms are also quantum emitters, making them suitable candidates for atom-photon interfaces; other systems may have superior coherence times but cannot emit photons. To benefit from these extremely long-lived memories, hybrid strategies can be chosen in which the quantum memory is indirectly interfaced with photons through its coupling to an efficient quantum emitter. This occurs naturally in NV centers, for example, where the electron spin is coupled via hyperfine interaction with nearby $^{13}$C nuclear spins ($T_2 = 75 \,\second$)~\cite{Bradley2019}.
The same strategy is also taken with trapped ions, where ionic species with good emission properties, such as $^{128}$Ba$^+$, are interfaced at the same quantum node with $^{171}$Yb$^+$, the latter of which have much longer coherence times ($T_2 > 1\,\hour$~\cite{Wang2017b, Wang2021}). Using these ions, \cite{Hucul2015} showed two-ion entanglement which persists over more than $1\, \second$. Recent results also show that a typically short-lived quantum dot spin can be efficiently coupled to a single magnon excitation of its nuclear environment, which consists of $10^4 - 10^5$ nuclear spins that behave as a long-lived memory ($T_2^* \approx \micro\second$~\cite{Gangloff2019, Jackson2021}, compared to the $T_2^* = 39\, \nano\second$ for the electron spin~\cite{Ethier2017}).
Rare-earth $\text{Eu}^{3+}$ ions in Y$_2$SiO$_5$ crystals have the longest coherence time experimentally observed with $T_2 = 6$\,h~\cite{Zhong2015b}. This platform has an optical memory which can store a time-bin encoded photonic qubit for 1\,h~\cite{Ma2021}.

Another important criterion for these platforms is the temperature at which they operate. It implies potentially the use of different cooling strategies that can be technologically demanding, from dilution refrigerator or liquid helium temperature cryostats to laser cooling.
Interestingly, several studies propose to soften this requirement through the use of ``room temperature'' quantum repeaters  based on either hybrid optomechanical systems with NV centers \cite{Ji2022}
or warm atomic vapors~\cite{Katz2018, Shaham2021, Borregaard2016, Pang2020, Li2021, Dideriksen2021}.

In Table~\ref{table_exp}, we summarize the experimental performance of long-lived quantum memories together with their emission properties.
In addition to the coherence time, several other figures of merit are also important for quantum repeater applications.
These include the quantum emitter control gate fidelity ($F$) and dephasing and relaxation times ($T_2^*, T_1$), and the availability of an additional quantum register and its properties.
The photonic properties of the quantum emitters are also important, namely the photon collection efficiency ($\eta_{\rm eff}$), the Debye-Waller factor in the case of solid-state defect qubits (i.e., probability of emitting photon into the zero-phonon line) $\eta_{\rm DW}$,
the indistinguishability $\iota$ and the quality of the spin-photon entanglement ($F$ (atom-phot.)). The photon wavelength also plays a crucial role in quantum communication since the best transmission rates are achieved for telecom wavelengths.
We include well-established quantum emitters alongside more recent but promising systems, such as rare-earth ions and new defects in diamond and silicon.

\subsection{Emission of photons entangled with the quantum memory}
\label{subsec_emission}

\begin{figure}
    \includegraphics[width=\linewidth]{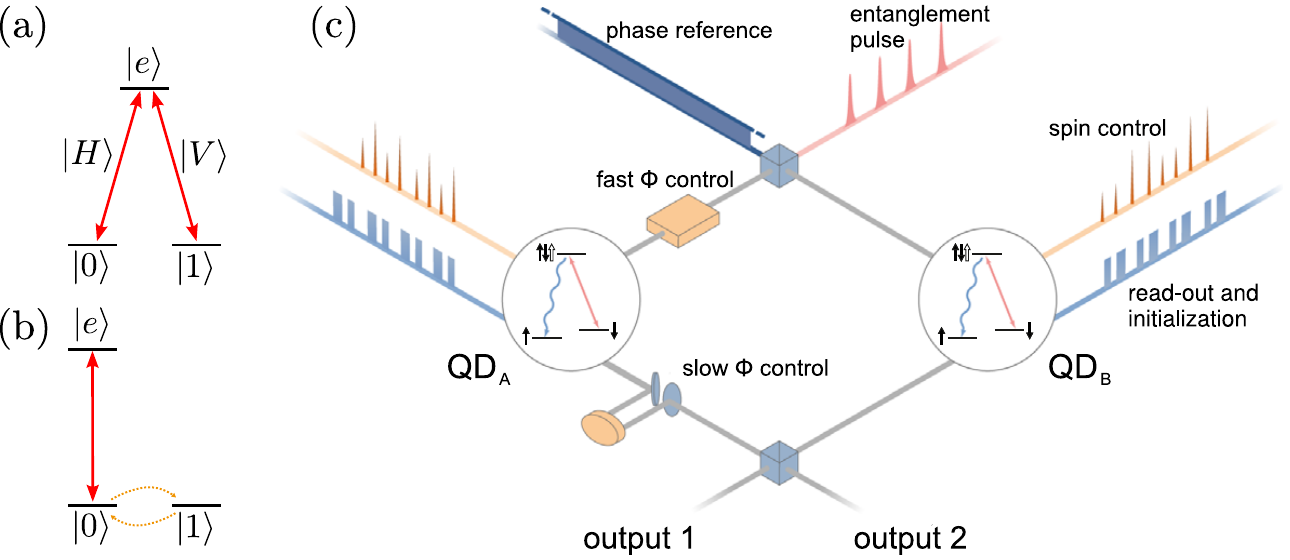}
    \caption{Level structure and heralded entanglement generation. In (a), a Lambda level structure with states $\ket{e}$ (excited), $\ket{0}$ (connected to $\ket{e}$ by horizontally polarized light, $\ket{H}$) and $\ket{1}$ (connected to $\ket{e}$ by vertically polarized light, $\ket{V}$). In (b), a level structure for time-bin entanglement. $\ket{e}$ only connected to $\ket{0}$; control of the qubit states is required. In (c), a setup for spin-spin heralded entanglement generation. Figure from Ref.~\cite{Stockill2017}.}
    \label{fig:spin_entangled_photon}
\end{figure}

Quantum memories should have---or should be coupled to quantum emitters which have---optical transitions that allow the emission of photons entangled with the memory qubits.
The emitted photonic qubits are to be encoded in one of the degrees of freedom discussed in Sec.~\ref{sec:photonic_encodings}.
The emission of spin-entangled qubits encoded into photonic frequency, polarization, emission time bin, and spatial modes has already been experimentally demonstrated with the help of trapped ions, NV centers, and quantum dots.
Many schemes exist for the production of photons entangled with the memory's degrees of freedom, varying in details depending on the photonic encoding and the energy level structure of the emitter.
For concreteness, we will review two of the most common examples of such schemes.

Polarization-entangled photons can be produced in a system with a $\Lambda$-shaped level structure (that is, a \emph{$\Lambda$ system}), where the qubit ground states $\ket{0}$ and $\ket{1}$ are both optically coupled to a single excited state $\ket{e}$ by orthogonally-polarized transitions (say, horizontally-polarized and vertically-polarized photons, respectively).
This type of level structure is present in most quantum emitters, including some species of trapped ions~\cite{Blinov2004} and atoms~\cite{Volz2006}, in atomic ensembles~\cite{Chen2007b}, NV centers~\cite{Togan2010}, and in quantum dots when a transverse magnetic field is applied~\cite{Gao2012, DeGreve2012, Schaibley2013}.
A quantum memory prepared in the excited state will spontaneously emit a single photon with either horizontal or vertical polarization ($\ket{H}$ or $\ket{V}$), as shown in Fig.~\ref{fig:spin_entangled_photon}~(a). After this emission, the  total memory--photon system is in the entangled state $\ket{0, H} + \ket{1, V}$. For this scheme to successfully produce such a maximally entangled state, the coupling strength of the two optical transitions ought to be the same. If the transitions differ in energy ($E_H \neq E_V$), as in quantum dots, the final state might instead be $\ket{0, (H, E_H)} + \ket{1, (V, E_V)}$,
where $\ket{(A, E_A)}$ for $A=H,V$ denotes the redundant encoding of the photonic qubit on its polarization and frequency degrees of freedom.
The demonstration of bipartite entanglement is therefore challenging in this case, since it requires that this redundancy be erased, but such a quantum erasure of the photon frequency has been for example demonstrated in~\cite{Yu2015}.

Despite its relative simplicity, the previous scheme may not be available for all quantum memories, as it requires a Lambda-level structure. There is an alternative approach~\cite{Lee2019, Tchebotareva2019, Hensen2015, Vasconcelos2020}, which requires only one strong optical transition, that results in a photon whose emission time bin is entangled with the memory qubit. The minimal level structure required for this scheme is illustrated in Fig.~\ref{fig:spin_entangled_photon}~(b); it corresponds to a three-level system ($\ket{0}$, $\ket{1}$, $\ket{e}$), where only one state of the qubit states, e.g., $\ket{0}$, is optically coupled to the excited state $\ket{e}$.
The memory is initialized in a superposition state $\ket{0} + \ket{1}$, and then the optical transition $0 \leftrightarrow e$ is excited by a $\pi$-pulse such that the system ends up in $\ket{e} + \ket{1}$. If in the excited state, the memory emits a photon in the early time bin $\ket{t_1}$; otherwise, it emits no photons, resulting in the state $\ket{\rm vac}$: $\ket{0, t_1} + \ket{1, {\rm vac}}$.
The memory qubit is then flipped in its qubit subspace (yielding $\ket{1, t_1} + \ket{0, {\rm vac}}$) and the $0 \leftrightarrow e$ transition is excited again, leading to the emission of a photon in the time bin $t_0$ if the excited state was populated: $\ket{1, t_1} + \ket{0, t_0}$. We see that a single photon is always emitted, and that its emission time bin is indeed entangled with the quantum memory. This strategy requires the preparation of the memory in a superposition state and more control pulses; however, it has the advantage of operating with only a single optical transition, making it particularly convenient in the case of, e.g., NV centers~\cite{Bernien2013}, when one specific optical transition has better properties than the others. This approach has also been demonstrated in quantum dots, when a certain transition is made more favorable through cavity (Purcell) enhancement~\cite{Lee2018}.

\subsection{Distant entanglement generation}
\label{subsec_spin_spin}

It is possible to generate heralded entanglement between distant qubits mediated by the detection of photons. The implementation of these schemes is usually based on interference of photons within a linear-optical setup. To optimally interfere and hence create maximal entanglement, the photons emitted by two distant quantum memories should be perfectly indistinguishable~\cite{Senellart2017, Aharonovich2016}.

The scheme of Cabrillo~{\it et al.}~\cite{Cabrillo1999} based on single-photon interference (see also \cite{bose1999proposal} for a similar proposal) for distant entanglement generation has been demonstrated with trapped ions~\cite{Slodivcka2013}, quantum dots~\cite{Delteil2015, Stockill2017}, NV centers in diamond \cite{humphreys2018deterministic} and atomic ensembles~\cite{Chou2007}. The
experiment in~\cite{Stockill2017}, based on two quantum dot spins separated by a few meters, resulted in a postselected entanglement generation rate of $7.3\,\kilo\hertz$~\cite{Stockill2017}.

Let us illustrate how the scheme works experimentally. Two quantum dots, $A$ and $B$, situated at two separated nodes are prepared in a Voigt configuration (in-plane magnetic field) to exhibit a $\Lambda$-level structure with similar optical transition energies. The two quantum dots are prepared initially in the state $\ket{0_A, 0_B} = \ket{\downarrow_A, \downarrow_B}$ and are excited by the same weak phase-stabilized laser so that a photon may be produced by each quantum dot through Raman scattering with a probability $p \ll 1$ (see Fig.~\ref{fig:spin_entangled_photon}~(c)). The photonic modes are then mixed on a 50:50 beamsplitter at a central node to erase the which-path information---that is, to make it impossible to tell which quantum dot emitted the photon (essentially to perform the Bell measurement of Fig.~\ref{fig:BellM}~(b)). The state before the photon detection is:
\begin{equation}
  \begin{aligned}
  \ket{\Psi} & = (1 - p) \ket{\downarrow_A, \downarrow_B} \ket{0_1, 0_2} \\
  & + \sqrt{p(1-p) / 2} \left( e^{i \Phi_a} \ket{\uparrow_A, \downarrow_B} + e^{i \Phi_B} \ket{\downarrow_A, \uparrow_B} \right) \ket{1_1, 0_2} \\
  & + \sqrt{p(1-p) / 2} \left( e^{i \Phi_a} \ket{\uparrow_A, \downarrow_B} - e^{i \Phi_B} \ket{\downarrow_A, \uparrow_B} \right) \ket{0_1, 1_2} \\
  & + p/\sqrt{2} e^{i (\Phi_A + \Phi_B)} \ket{\uparrow_A, \uparrow_B} \left(\ket{0_1, 2_2} - \ket{2_1, 0_2} \right),
  \end{aligned}
\end{equation}
where $\ket{i_1, j_2}$ (with $i,j$ integers) corresponds to the number of photons in the first and second output modes of the beamsplitter, and $\Phi_A$ and $\Phi_B$ are the optical phases along the different optical paths corresponding to qubit $A$ and $B$. If a single photon is detected, the quantum dot system is projected with probability $\approx p$ into the maximally entangled state $e^{i \Phi_a} \ket{\uparrow_A, \downarrow_B} \pm e^{i \Phi_B} \ket{\downarrow_A, \uparrow_B}$, with the sign depending on the output mode of the beamsplitter in which the photon was detected. In practice, $p$ cannot be as high as desired because the quantum dot spins undergo two spin flip processes with probability $p^2$, resulting in the emission of two photons. In that case, if only one of the two photons is detected---either due to imperfect collection and detection efficiencies or transmission losses---the heralding single-photon process leads to a state with fidelity that decreases with higher $p$.
Refs.~\cite{Stockill2017, Yu2020, Pompili2021, Lago-Rivera2021} used this methods to demonstrate heralded entanglement generation. Ref.~\cite{Stockill2017} demonstrated the highest rate for distant spin-spin entanglement with postselection and
Ref.~\cite{Yu2020} demonstrated the longest fiber distance between two remotely entangled quantum memories using atomic ensembles.  However, while the two memories were separated by 50 kilometers of fiber, this was achieved using a spooled fiber of that length, the actual distance between the systems was a meter.

Other methods for generating distant heralded entanglement exist, namely the \textit{Barrett and Kok} scheme~\cite{Barrett2005} based on two-photon detection. This scheme has been demonstrated with NV centers~\cite{Bernien2013} and trapped ions~\cite{Moehring2007}.
The longest-distance entanglement between separated systems reached using this approach, $1.3\,\kilo\meter$, was achieved also with NV centers, in a loophole-free Bell test experiment~\cite{Hensen2015}. In Ref.~\cite{Yu2020}, the authors have also demonstrated a field-deployed heralded entanglement generation between two atomic ensembles separated by 11 kilometer ($22\, \kilo\meter$ of fibers) using two-photon interference. The latter was achieved by increasing the collection and detection efficiencies of the photons as well as converting the optical photons to the telecommunication frequency, which enjoys the highest transmissivity in optical fibers (see Sec.~\ref{sec:loss_mitigation} for more details).

Cabrillo {\it et al.}'s scheme is required to operate in the low photon emission probability regime to obtain high fidelity heralded entanglement. In comparison, Barrett and Kok's scheme can operate in the high fidelity regime even with high emission probability. Therefore, it should be better suited for efficient quantum emitters and short distance between the nodes. However, for longer distances, the fiber losses becomes dominant and having a single-photon heralding like the Cabrillo {\it et al.} protocol leads to a better scaling with distance compared to the two-photon heralding of the Barrett and Kok scheme [similar to the relation between the TF QKD and the original MDI QKD (see Sec.~\ref{se:TFQKD})].

\subsection{Entanglement distillation}
\label{subsec_purification}

During the generation of entanglement between remote nodes, operation errors or the decoherence of quantum memories can lead to a reduced fidelity of Bell states shared between distant nodes. For first-generation quantum repeaters, the fidelity of Bell pairs can be increased through entanglement distillation (Sec.~\ref{sec:1GQR}). Starting with two imperfect copies of a Bell pair, it is possible to produce a single Bell pair with improved fidelity with a success probability of at best 50\%.
Entanglement distillation has been demonstrated with photonic Bell pairs~\cite{Pan2001,yamamoto2003experimental,yamamoto2001concentration,pan2003experimental}, atoms~\cite{Reichle2006}, and NV centers~\cite{Kalb2017}.

Photonic realizations differ in success rate because it is impossible to perform a deterministic CNOT gate with linear optics. Instead, the entanglement distillation protocols are performed using solely linear optics with a success rate limited to 25\% at best~\cite{Pan2001,yamamoto2001concentration}. Ref.~\cite{Reichle2006} demonstrated the first experimental entanglement distillation with quantum memories. They distilled two Bell pairs of $^9{\rm Be}^+$ ions, confined in the same Paul trap, with an overall success probability above 35 \%. Yet, because the pairs of entangled atoms were not spatially separated, this scheme is not particularly useful to enable long-distance quantum communication applications. Using two NV centers with two $^{13}$C nuclear spins, Kalb {\it et al.} demonstrated entanglement distillation of a $65\pm 3\%$-fidelity Bell state in NV centers that were spatially separated by 2 meters. The highest reported heralded entanglement rate was $182 \hertz$~\cite{Stephenson2020} with trapped ions separated by 2 meters using a two-photon interference scheme. In this work, the authors expect a distilled Bell pair fidelity of 99 \% is within experimental reach.

\subsection{Multi-qubit quantum registers and error correction}
\label{subsec_error_correction}

Multiple memory qubits will be required per repeater node, either for increasing the communication rate via multiplexing~\cite{Collins2007}, or for enabling error correction in repeaters beyond the first generation. A quantum register extends the architecture from Sec.~\ref{subsec_memory} to a quantum emitter with good optical properties coupled to a large number of long-lived quantum memory qubits. This arrangement naturally occurs in colour centres in diamond, where the defect is coupled by hyperfine interaction to tens of $^{13}$C nuclear spins~\cite{Bradley2019}, forming the register of qubits.
There have been several advances in this line of research, e.g., in experiments where the nuclear spins are individually controlled using the electron spin~\cite{Dutt2006, Childress2006, Balasubramanian2009, Fuchs2011, Taminiau2012, Bradley2019}.
Similarly, in the trapped ion setting, a quantum register of many qubits has been realized using one quantum emitter coupled to many memory qubits in the same optical trap.
For example, dual species quantum nodes based on pairs of different ionic species such as $^{128}$Ba$^{+}$-$^{171}$Yb$^{+}$~\cite{Inlek2017} or $^{25}$Mg$^{+}$-$^{9}$Be$^{+}$~\cite{Tan2015} are being investigated.
In a quantum dot, however, the spin is only coupled to one (potentially two) different magnon species~\cite{Jackson2021}, imposing limits on  the size of the register. An alternative strategy for obtaining more qubits at each repeater node could be to vertically stack quantum dots~\cite{Stinaff2006}.

For repeaters from the second and third generation, a quantum register at each node can be seen as a quantum processor used to logically encode the quantum information transferred between nodes and to correct errors. A QEC code has recently been implemented in trapped ions~\cite{Egan2020}. Here, 9 physical $^{171}{\rm Yb}^+$ qubits (with 4 additional qubits for stabilizer measurements) are associated with one logical qubit of the Bacon-Shor code in a fault-tolerant design. A recent experiment using superconducting qubits~\cite{google2023suppressing} demonstrated experimentally a logical error rate reduction through increasing the size of the QEC code being used.
There is also an effort to pursue error-corrected repeater nodes with solid-state spins \cite{waldherr2014quantum,Cramer2016}. In particular, with defects in diamond \cite{abobeih2021}, it has recently been shown
the experimental fault-tolerant operation of a logical qubit using the 5-qubit code together with a flag protocol \cite{chao2018quantum,chamberland2018flag} requiring a total of seven qubits. Yet, this proof-of-principle demonstration remains still above the break-even point for which logical qubit operations have higher fidelities than physical qubit operations.

Importantly, the logical qubits in error-corrected repeaters must be interfaced optically.
For several platforms investigated for the realization of multi-qubit processors, such as superconducting circuits, a major challenge for quantum communication applications revolves around the emission of optical photons, which requires quantum transduction from microwave to optical energies~\cite{Lauk2020, Mirhosseini2020, Ang2022}.

The realization of logical photonic qubits is also being pursued; they are required in the third generation of repeaters and in all-photonic quantum repeaters in order to correct for loss errors.
Error detection has been demonstrated on a photonic platform~\cite{Bell2014}, and recently a proof-of-concept photonic 9-qubit Shor code has been experimentally implemented together with an all-photonic quantum repeater proposal~\cite{Zhang2022}.

\subsection{Loss mitigation, quantum frequency conversion, and photonic source efficiency}
\label{sec:loss_mitigation}

\begin{figure}
    \includegraphics[width=\linewidth]{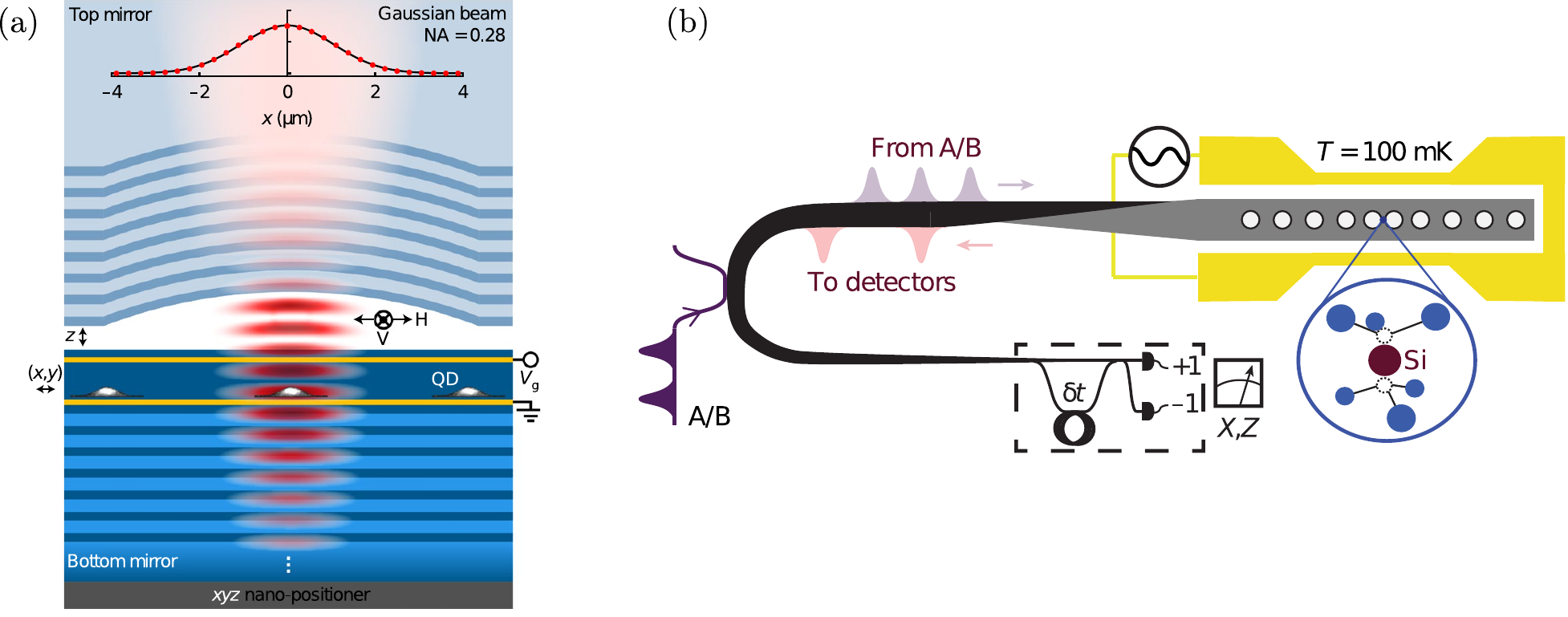}
    \caption{
    { State-of-the-art cavity-QED devices. In (a), a quantum dot coupled deterministically to an open Fabry-Pérot cavity. In (b), a silicon vacancy center in diamond in a photonic crystal cavity evanescently coupled to a fiber. Figure (a) from Ref.~\cite{Tomm2021} and figure (b) from Ref.~\cite{bhaskar2020experimental}.}}
    \label{fig:source}
\end{figure}

A stringent requirement on correcting photonic errors is imposed by the no-cloning theorem (Sec.~\ref{sec:no-cloning}), which implies that it is impossible to correct physical qubit losses of more than 50\% with QEC. In light of this, reducing the photon losses throughout a quantum network is critical for the implementation of those repeaters where the loss is handled via QEC. Loss occurs at every optical component, with the main sources being propagation and coupling losses due to the intrinsic properties of fibers and photonic chips. Loss also occurs at the detectors and during the collection of photons produced by quantum emitters.

Losses in fibers are chiefly caused by infrared absorption and Rayleigh scattering, as well as imperfections introduced in manufacturing. Minimal loss is obtained at the telecom wavelength (1550\,nm), where the loss coefficient is 0.2\,dB per km, with few prospects of improvement. Even though there exist ultra-low-loss fibers with losses of 0.16\,dB per km~\cite{boaron2018secure}, they are not widely available, and would require complete modification of the existing infrastructure. It is therefore crucial to use quantum emitters that emit at the telecom wavelength, such as some engineered quantum dots~\cite{Benyoucef2013} or rare-earth ions~\cite{zhong2019proof} and color centers in silicon~\cite{Redjem2020, Bergeron2020}.

An alternative strategy consists of using a quantum frequency converter. The objective is to change the frequency of the photonic qubits while preserving the quantum information they encode (and the single-photon statistics if required for the scheme)~\cite{Tanzilli2005, Mcguinness2010, Ikuta2011}. Frequency converters are generally based on a non-linear $\chi^{(2)}$ crystal (or possibly $\chi^{(3)}$) pumped by a laser pulse with frequency $\omega_l$ chosen such that the frequency $\omega_i$ of an input photon is modified into $\omega_f = \omega_i - \omega_l$. This strategy has been used to convert the frequency to a telecom wavelength of photons emitted by NV centers~\cite{Tchebotareva2019},
quantum dots~\cite{Zaske2012, DeGreve2012},
single atoms~\cite{Van2020, Van2022},
ions~\cite{Bock2018, Krutyanskiy2019, Krutyanskiy2022},
rare-earth-doped crystals~\cite{Maring2017},
and
atomic ensembles~\cite{Dudin2010, Ikuta2018, Yu2020}.

The efficient collection of light produced by quantum emitters is another important technological challenge. Since spontaneous emission is non-directional, photon collection efficiencies tend to be quite low. To obtain a high efficiency source of single photons, the electromagnetic environment of the quantum emitter ought be engineered to force its emission into one specific mode that can then be coupled into a fiber. This can be achieved using waveguides, which inhibit the emission outside of the waveguide mode~\cite{Arcari2014}, or with micro-cavities, which enhance the coupling between the quantum emitter and the electromagnetic mode confined in the cavity. In these two cases, the emission of a single photon is much more probable inside a particular mode (of the cavity or the waveguide) than in all the others. This photonic mode can then be efficiently coupled to the transmission fiber. Cavity enhancement also has the important advantage of increasing the probability of emission of indistinguishable coherent photons~\cite{Riedel2017} as compared to incoherent phonon-assisted emission. Two examples of state-of-the-art cavity-QED devices are reviewed in Fig.~\ref{fig:source}. The single-photon collection efficiency has drastically improved over the years for all quantum emitters, through technological and material improvement of cavity-QED devices~\cite{Barros2009, Somaschi2016, bhaskar2020experimental, Wang2019, Tomm2021, Uppu2020, Maiwald2012}; in quantum dots, trapped ions and defects in diamond, it has now risen above the 50\%  threshold.

While not making use of quantum emitters, it is also worth mentioning that spontaneous parametric downconversion sources have seen their effective collection efficiency increase to 67\% through large-scale multiplexing and active switching~\cite{Kaneda2019}.
While it is not possible to use these sources to realize an efficient light-matter interface in quantum repeater protocol based on matter qubits, they nevertheless show great potential for all-photonic approaches, as detailed in the following section.

The single-photon detection efficiency~\cite{Hadfield2009} has also been significantly increased through the development of superconducting nanowire single-photon detectors (SNSPDs). Devices with detection efficiencies as high as 95\% are now commercially available and superconducting nanowire detectors with efficiencies as high as 99 \% have been demonstrated at telecom frequencies~\cite{Hu2020, Chang2021}. Transition edge sensors also enjoy high detection efficiencies, with the bonus that they can resolve photon numbers~\cite{Lita2008}, which can be useful for some heralded entanglement schemes.

\subsection{Progress towards memoryless quantum repeaters}
\label{subsec_qr_progress}

\begin{figure}
    \includegraphics[width=\linewidth]{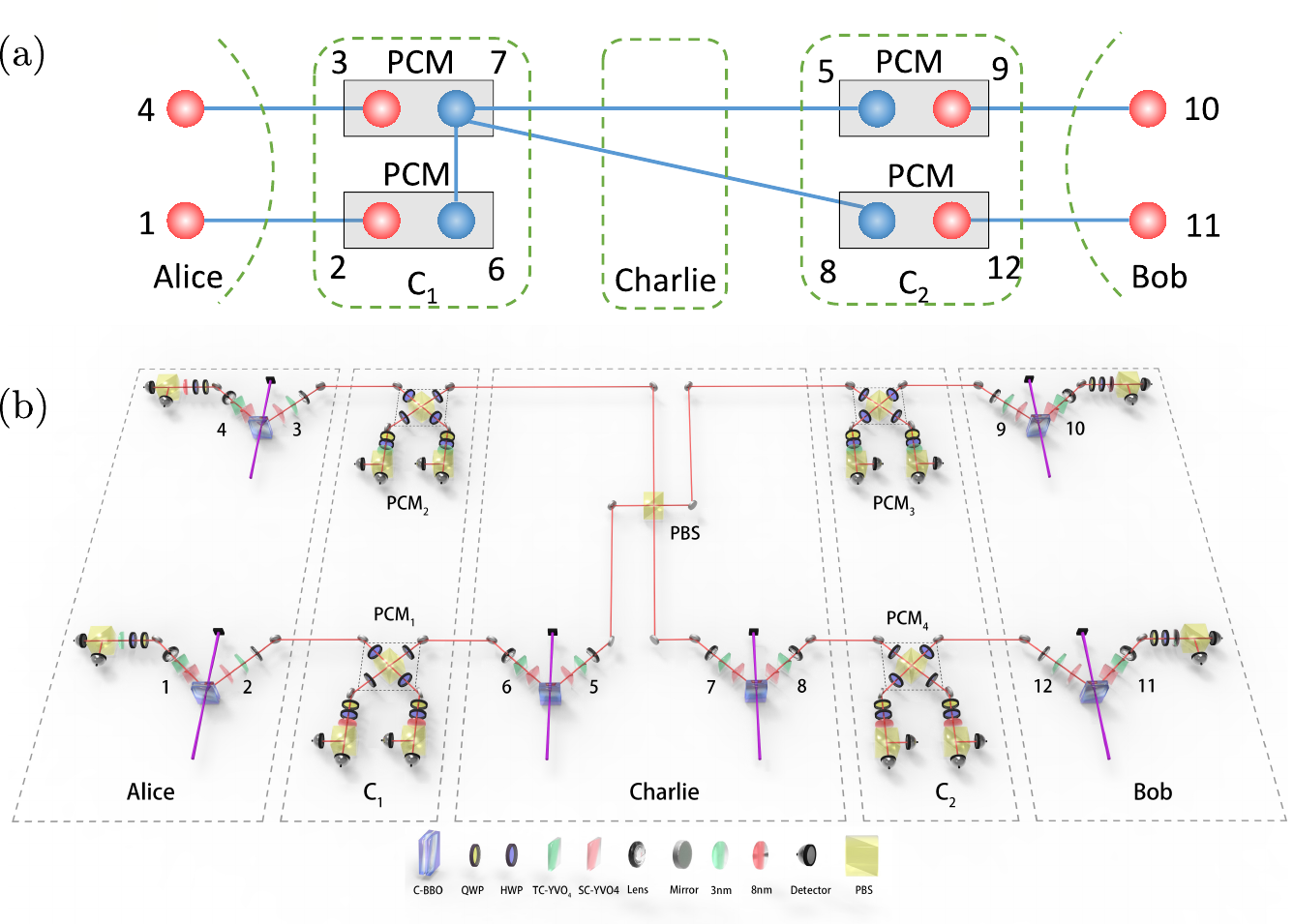}
    \caption{
    {A proof-of-principle experiment for an all-photonic quantum repeater. PCM stands for ``passive choice measurement,'' which automatically performs an entangling Bell measurement (in case of a coincidence detection) or a disentangling local $X$ measurement (in case of a single-photon detection). Figures from~\cite{Li2019}.}}
    \label{fig:mid_point_all_photon_QKD}
\end{figure}

In all-photonic quantum repeaters, error-correction and loss-tolerance are achieved through photonic codes, so that these protocols do not require quantum memories. The technological requirements of such repeaters are therefore considerably different from the other approaches. The primary challenge revolves around the creation of large, highly entangled photonic states, namely graph states.

Several different approaches have been suggested for photonic graph state generation. Until very recently, the largest entangled states of photons have been produced experimentally using spontaneous parametric downconversion sources and fusion gates~\cite{Browne2005}.
The probabilistic nature of the fusion gates is the main limitation to the number of photons in the graph state that can produced with this approach, the current maximum being 12~\cite{Zhong2018}.

Proof-of-principle experiments of all-photonic quantum repeaters have already been realized~\cite{Hasegawa2019, Li2019}. In both cases, the original protocol in \cite{Azuma2015} was replaced by a variant in order to facilitate its experimental realization. In this new all-photonic communication scheme, introduced in Ref.~\cite{Hasegawa2019}, Alice and Bob prepare $n$ photonic Bell pairs each, sending one half of every one of them through a lossy fiber to a central node (Charlie). Prior to the arrival of the photons, Charlie prepares a $2n$-qubit GHZ state (equivalent to the complete graph state from Sec.~\ref{sec:apqr}) and performs photonic Bell state measurements between the incoming photons and the corresponding photons in the GHZ state. The first key concept behind this scheme is a time-reversed adaptive Bell measurement, which Li~\textit{et al.} refer to as a passive choice measurement. If the photon $a_i$ ($i=1,2,\cdots, n$) emerging from Alice arrives at Charlie's node, and the joint measurement with photon $c_i$ from Charlie's GHZ state is successful, then Charlie achieves a Bell state projection. However, if the photon $a_i$ does not make it to Charlie's node, or if the measurement is unsuccessful, the Bell state analyzer passively adapts to an $X$-basis measurement on $c_i$, which disconnects photon $c_i$ from the GHZ state. This leads to the second important idea in~\cite{Hasegawa2019}: the outer qubits from the original repeater graph state in~\cite{Azuma2015} can be removed, leaving a bare GHZ state in its place.

In their work, Li~\textit{et al.} demonstrated the above scheme with a four-qubit GHZ state and $n=2$ multiplexed communication channels. We illustrate the experiment in Fig.~\ref{fig:mid_point_all_photon_QKD}. Alice, Bob, and Charlie each prepare two Bell pairs using spontaneous parametric down-conversion sources.
Alice and Bob send one qubit from each Bell pair---each corresponding to a communication channel---to Charlie's node. Charlie mixes his two Bell pairs to produce a four-qubit GHZ state and the protocol proceeds as explained previously with $n=2$. Although the experiment did not surpass the PLOB bound~\cite{Pirandola2015}, Li~\textit{et al.} demonstrated an enhancement in communication rates between Alice and Bob compared to the case where Charlie uses a Bell pair for each communication channel (that is, does not multiplex the channels). These results attested to the interest and experimental feasibility of all-photonic solutions for quantum communication.

In principle, the above modifications simplify the original all-photonic repeater, making it attainable with current technology. However, the protocol only works if a single QR node is used, consequently leading to a $\eta^{1/2}$ scaling, at best, and limiting the communication distances to, at most, about $800~\kilo\meter$ in practice\footnote{For instance, with a twin-field-type QKD protocol which utilizes a single node between communicators, Wang {\it et al.} have successfully generated a secret key with $4.572\times 10^{-1}$ secret bits per second over 786.67\,km of fiber and with $1.399\times 10^{-2}$ secret bits per second over $833.80$\,km of fiber \cite{wang2022twin}, experimentally.} (in the sense explained in the footnote in Sec.~\ref{sec:IdealQR}). Going beyond this limit would require cascading multiple QR nodes and using photonic states with much more photons, such as the RGS in original protocol (Sec.~\ref{sec:apqr}). Furthermore, the protocol is particularly sensitive to local losses at Charlie's node, as demonstrated in Ref.~\cite{Hasegawa2019}. Delaying the preparation of the GHZ state only goes part of the way to mitigate this issue, with a more complete scheme requiring loss-tolerant error correction. Recently, Ref.~\cite{Zhang2022} demonstrated a 9-qubit Shor code, with a new all-photonic quantum repeater approach which could be cascaded. They have also shown its tolerance to single-photon losses. Among the remaining steps to be fully operable, this Shor code should be generated in a heralded fashion rather than being postselected.

To move to higher photon numbers, the all-optical strategy requires probabilistic fusion gates combined with high-speed feedforward (Sec.~\ref{sec:apqr_state_gen}) to grow bigger and bigger graphs based on small graph resources. Having efficient feedforward techniques is thus crucial~\cite{Zanin2021}. This is achievable only with ultrafast optical switches and electronics.

The technological challenges of bosonic repeaters (Sec.~\ref{sec:bosonic_repeaters}) are somewhat different than the discrete-variable repeater that we have focused on. For the particular case of encoding qubits into momentum-squeezed or GKP states, one can deterministically combine modes into large graph states with Gaussian operations (linear optics and squeezing). However, the production of photonic GKP states is challenging, and is not yet to be implemented on photonic platforms. On the other hand, Gaussian states of light are now a well-mastered technology~\cite{Asavanant2019}.

An alternative strategy for producing photonic graph states is to use light-matter interfaces in generation procedures such as~\cite{Buterakos2017} or~\cite{Pichler2017, Zhan2020, Zhan2023}, based on the initial work of Refs.~\cite{Schoen2005, Lindner2009, Economou2010}. This strategy is more demanding experimentally but has the advantage of being (in principle) deterministic. Indeed, with unity collection efficiency of the photons and perfect control of the quantum emitters, the generation procedure does become completely deterministic: the entanglement between photons is produced through the control of the quantum emitter rather than through probabilistic fusion gates. A proof-of-concept experiment has been realized by Schwartz~\textit{et al.}~\cite{Schwartz2016} where a linear cluster state is produced by manipulating and optically pumping the spin of a quantum dot. The authors produced a three-qubit linear cluster state and showed that entanglement persists for up to five photons. More recently, this group showed that entanglement persists over 10 photons, with indistinguishability above 90 \%, using also the deterministic generation from a hole spin quantum dot emitter~\cite{Cogan2020b}. Quantum dot-based sources of entangled photons have also been inserted inside microcavities to generate linear-cluster states at much higher rates~\cite{Coste2022}. A similar generation scheme using a single atom trapped in a cavity was used to demonstrate a $12$-photon linear cluster state and a $14$-photon GHZ state~\cite{Thomas2022}, which to date constitutes the record largest entangled photonic state demonstrated experimentally.
In these experiments, the emitters produce  polarization-entangled photons, but strategies involving time-bin entanglement have also been explored~\cite{Lee2018, Vasconcelos2020, Appel2022, Vezvaee2022}.

To go beyond linear cluster state generation, one can either use multiple solid-state qubits or the strong non-linear interaction induced by atoms for light to effect entangling gates. For the generation procedures of Refs.~\cite{Pichler2017, Zhan2020}, one needs to implement spin-photon CZ gates, where a phase shift is induced onto a photon depending on the spin state. Cavity-QED devices increase the spin-photon interaction such that such spin-photon gates are within reach with many cavity-QED platforms~\cite{Reiserer2014, Arnold2014, Sun2016, Androvitsaneas2019, Wells2019, bhaskar2020experimental, Javadi2018}.

\subsection{Experimental realization of quantum networks}
\label{sub_sec_exp_quantum_network}

In this section, we review experiments that go beyond two-node quantum communication to inch closer to the quantum internet. We first start by presenting the experimental realizations of trusted large-scale repeater networks for QKD applications based on trusted relays. We then discuss experimental progress towards the realization of quantum repeaters to actualize long-distance untrusted nodes. Finally, we discuss the experimental realization of untrusted quantum networks.

\subsubsection{Trusted large-scale repeater networks}\label{sec:trusted}

\begin{figure*}
    \includegraphics[width=15cm]{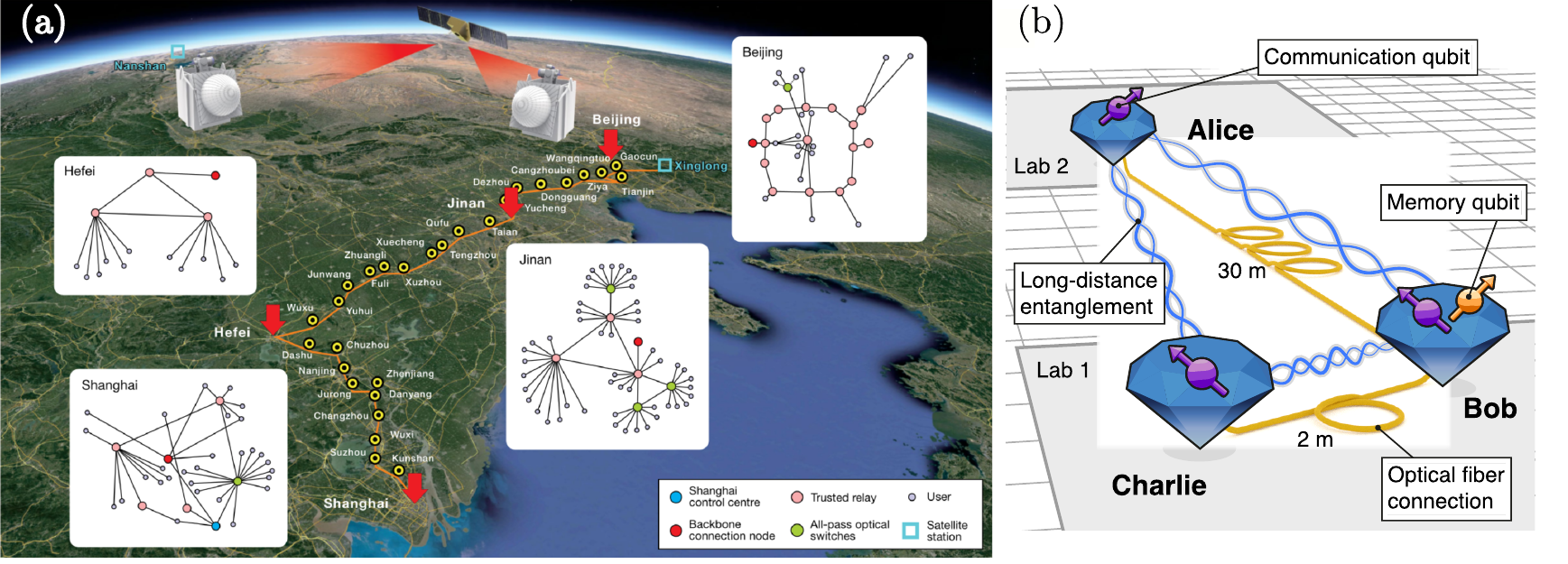}
    \caption{
    Quantum networks. (a) Shanghai-Beijing QKD network. (b) Experimental quantum network composed of  NV centers acting as quantum memories. Figure (a) from Ref.~\cite{chen2021integrated} and figure (b) from Ref.~\cite{Pompili2021}.}
    \label{fig:quantum_network_exp}
\end{figure*}

Several intercity QKD networks have already been realized, such as the SECOQC network~\cite{peev2009secoqc} in Austria, the Tokyo QKD network~\cite{Sasaki2011} in Japan, the SwissQuantum network~\cite{Stucki2011} in Switzerland, the Illinois Express Quantum network~\cite{Chung2021} in the USA, and the Shanghai-Beijing QKD network~\cite{chen2021integrated} in China. In all of these networks, cryptographic keys are distributed between nodes separated by long distances using relay nodes. Assuming that the relay nodes are trusted, a secure key can be established at rates much higher than what is possible through direct fiber transmission~\cite{Pirandola2015}, thereby enabling efficient QKD over very long distances.

In Fig.~\ref{fig:quantum_network_exp}~(a), we illustrate the Shanghai-Beijing QKD network, the largest QKD network to date. This network links four metropolitan areas---Shanghai, Hefei, Jinan, and Beijing---using a backbone of 32 trusted relays in a linear topology. If any one of the 32 relay nodes is compromised, the generated key may be insecure. The trusted relays allow for efficient long-distance quantum communication between these metropolitan areas. Each of these cities is comprised of small QKD networks with different topologies, where end users with reduced capabilities (only requiring a QKD source) can connect to the network. This network incorporates both fiber- and satellite-based communication: the nodes at a Nanshan and Xinglong are separated by $2600\, \kilo\meter$, communicating through free-space via a satellite node that also acts as a trusted relay. A similar strategy has also been used to distribute a secret key over intercontinental distances---between Graz in Austria, and Nanshan and Xinglong in China, covering a total distance of $7600\,\kilo\meter$~\cite{Liao2018}. Thanks to this combination of fiber- and satellite-based quantum communication, the Shanghai-Beijing QKD network covers a total distance of $4600\,\kilo\meter$ and provides a typical secret rate between each node of 50 kilobits per second (kbps) and a minimum inter-node secret key rate of 28\,kbps in the entire network. Such large key rates achieved at such distances are completely out-of-reach for direct transmission over a fiber. Given its covered distance, complex topology, and the different quantum channels used, this QKD network can be considered as a genuine prototype of the quantum internet for QKD applications, albeit at the cost of having to trust the network provider.

\subsubsection{Proof-of-concept of a quantum repeater}\label{se:pocofQR}

Improving on a trusted repeater network requires \emph{device-independent} QKD, which can be realized through the distribution of Bell pairs to the end nodes.
Recent experimental demonstrations of fibered device-independent QKD based on quantum memories (respectively single $^{87}$Rb atoms~\cite{Zhang2022b} and $^{88}$Sr$^+$ ions~\cite{Nadlinger2022}) constitute significant improvements as they close the detection loophole in the violation of Bell's inequality.
In an experiment using an untrusted satellite node to share private keys with the help of the Ekert 91 protocol~\cite{Ekert1991}, Yin~\textit{et al.}~\cite{Yin2017} set the record distance of $1200\,\kilo\meter$ for distribution of entangled photons. Realizing a long-distance device-independent multi-node network would also crucially require the practical implementation of efficient quantum repeaters in real networks. However, this major milestone is the subject of active research and remains to be demonstrated.
Note, however, device-independent QKD still suffers\footnote{This is because once a key has been generated, it is classical,
and as such is subject to copying. Therefore,
if a QKD system is reused in future
QKD sessions, then the key generated in a previous session might be
stored in some memories and be leaked.
Moreover, not only the QKD devices, but
also the conventional computers used in the classical post-processing
(e.g. error correction and privacy amplification) may leak key
information via covert channels.
} from attacks such as
memory attacks~\cite{Barrett2013} and covert channels~\cite{Curty2019foiling}.

Bhaskar {\it et al.}~\cite{bhaskar2020experimental} have demonstrated that the use of a single repeater node in an experiment increases the communication rate of \emph{measurement-device-independent} (MDI)-QKD compared to repeaterless communication. Similar to Li~\textit{et al.} \cite{Li2019}, they used a repeater scheme with a single repeater node; however, they were able to demonstrate an improvement over the PLOB bound in terms of a key rate in bits per channel use versus an effective channel 
transmission---a four-fold secret key rate increase over the original MDI QKD \cite{Lo2012}. The repeater node consists of a single silicon-vacancy center embedded in a diamond photonic crystal cavity. The cavity mode of this device is efficiently evanescently coupled to a fiber to minimize the photonic losses. A significant improvement toward the photon collection efficiency was also demonstrated, reaching 85\%. The silicon-vacancy system is positioned in a dilution refrigerator to achieve a coherence time $T_2 = 0.2\, \milli\second$. In their experiment, the quantum memory at Charlie's node does not emit photons, but receives weak coherent time-bin-encoded pulses from Alice and Bob. Using electromagnetically-induced transparency of their cavity-QED device, these weak pulses are reflected or not depending on the electronic spin state. The reflected photonic pulses are then detected by superconducting single-photon detectors. If a photonic pulse coming from Bob is detected shortly after a pulse from Alice, a key bit can be distributed between Alice and Bob when Charlie communicates the two-photon and spin measurement results. With these experiments it is possible to achieve a $\sqrt{\eta}$ scaling because the two coherent pulses do not need to arrive simultaneously at the repeater node, thanks to the quantum memory (see Sec.~\ref{sec:milestones} for detail). The role of the memory is to store the information of the first pulse during the time it remains coherent while waiting for the second pulse to be detected. While operating with only one quantum memory per node for the moment, these results foresee a promising route toward long-distance quantum communication. Indeed, a silicon-vacancy color center can, in principle, make use of their $^{13}{\rm C}$ neighbors to effect a quantum register of long-lived memories~\cite{Nguyen2019b}.  This may increase the protocol's performance by enabling longer storage time as well as the concatenation of multiple repeater nodes, in principle paving the way to obtain a polynomial scaling of the rate with the communication distance.

In a more recent experiment, \cite{langenfeld2021quantum} demonstrated a memory-enhanced quantum repeater node based on two ${}^{87}$Rb atoms in an atomic cavity. This node can in principle be cascaded can be at the core of a quantum repeater scheme that overcome the previous $\sqrt\eta$ limits of repeater nodes with a single memory such as \cite{bhaskar2020experimental}.  
Moreover, the single-qubit error rate was below $11\%$ ensuring that a secure key can indeed be transferred using this repeater node. 
Such a memory-enhanced repeater has also been demonstrated by performing entanglement swapping with two ${}^{87}$Rb atomic ensemble memories \cite{pu2021}. 

\subsubsection{Untrusted quantum networks}

Since a quantum internet for applications beyond QKD may look like a multi-node network where quantum information is stored and processed by quantum memories, a complementary route toward the development of long-distance multi-node networks is to create multi-qubit quantum networks at a small distance and to progressively increase their size when the quantum repeater technology becomes more mature.
Pompili~{\it et al.}~\cite{Pompili2021} is the first realization of such a small quantum network, where each node includes a quantum memory to process quantum information locally. This network is based on three nodes, with an inter-node distance of a maximum of seven meters (see Fig.~\ref{fig:quantum_network_exp} (b)). Each node includes one or two quantum memories based on a nitrogen-vacancy (NV) center electron spin, and potentially another proximal $^{13}$C nuclear spin.

Pompili~{\it et al.} used their network to perform non-trivial multi-node operations such as the generation of a 3-qubit GHZ state with a memory qubit at each node, and the generation of a Bell pair between quantum memories situated at nodes that were not directly connected. After the generation of heralded entanglement between an NV electron spin at Alice's node and the electron spin at Charlie's node, the information encoded in Charlie's electron spin qubit was swapped to a $^{13}$C nuclear spin so that the electron spin could be used again to generate entanglement with Bob's NV center. This entanglement generation step could be realized with the strategies introduced in Sec.~\ref{subsec_spin_spin}. Then, the entanglement was swapped by performing a Bell measurement between the electron and the nuclear spins at Charlie's node. This was the first demonstration of entanglement swapping between distant nodes that were not originally connected. The work required the cooperation of a multitude of experimental components. In this work, they have used the single-photon detection scheme proposed by Cabrillo {\it et al.} to herald entanglement generation between distant spins with 80\% fidelity and at rates of 7 and 9\,Hz, using phase-stabilized links between the three nodes. 
The quantum information initially stored in Charlie's electron spin qubit needed to be swapped into one of its proximal nuclear spins,
thereby requiring a nuclear spin register and a high level of control. In addition, since the entanglement ought to be stored for the time the three nodes were connected, dynamical decoupling sequences were used to further isolate the spins from their environment. Finally, an electron-nuclear spin Bell state was used to swap the entanglement at the central node and produce a Bell state between Alice's and Bob's spin qubits at a rate of 25\,mHz. This protocol had an overall fidelity of 55\%, which could potentially be improved by using better photonic interfaces, spin control, and readout techniques, as well as reducing the infidelities and increasing the rate of the distant spin-spin entanglement generation. 
Such a network has also been used to teleport quantum information between two nodes that are not immediate neighbors~\cite{Hermans2022}.

The interest of these results is also to provide a testbed for real life applications and to prepare the other technological aspects of the implementation of a quantum network, such as the communication protocols. There is also a considerable development of quantum network simulator software~\cite{netsquid2020, Matsuo2019, Wallnofer2022} to assist in this goal, for example, to envision a city-scale network~\cite{Yehia2022}.

\section{Quantum internet}
\label{sec:internet}
The goal of this section is to look beyond linear networks, i.e., chains of quantum repeaters, and discuss how they blend into the vision of a future quantum internet. 
We first present a set of communication tasks that can be implemented over a quantum network and we link these sample communication tasks with experimental requirements and associate the tasks with a taxonomy of stages of the quantum internet which summarizes the discussion in \cite{Wehner2018}.
Second, we introduce the elements of a quantum networks and place repeaters in the larger context of a quantum network architecture. 
Finally, we investigate how to evaluate the usefulness of quantum networks for these different tasks. For this, we introduce a simplified model of a network in terms of a graph. The evaluation is phrased in the form of network capacities, quantities that can be achieved in an idealized situation. We observe that in spite of the apparent additional difficulty of dealing with a network, in this abstract setting many of the tools from point-to-point links carry to the network setting (see ~\cite{azuma2020tools} for a review on tools for predicting quantum network performance).

\subsection{Applications of the quantum internet}
\subsubsection{A set of representative communication tasks}
\label{sec:repcomtasks}
Before we discuss how to quantify the usefulness of a quantum network, it is relevant to discuss the potential applications of quantum networks
and more generally of the quantum internet. 
In the following we discuss a representative set of the applications that we know today divided by area.
However, similar to the early days of the Internet, we should expect many new applications to be found as the number of users increases. 

First of all, a quantum internet can be used for transmitting information. The nodes in the network might want to transmit classical information or quantum information. The latter is obviously not possible without a quantum network, but also for the former the quantum internet can offer an advantage with respect to a classical network. In particular, both entangled channel inputs  \cite{hastings2009superadditivity} and joint quantum measurements \cite{sasaki1998quantum,guha2011structured} can enhance the transmission rate of classical communication. A quantum internet can also be used to transmit classical information between two parties that is secret to any third party \cite{devetak2005private}. In turn, this enables secret key distribution, a task that is possible with classical means only if the parties are willing to make assumptions on the communication channel, e.g., wireless physical layer security relies on a model of the conditional probability distribution associated with the wireless channel  \cite{bloch2008wireless}, or on the capabilities of a potential eavesdropper, e.g., the security of the RSA cryptosystem \cite{rivest1978method} relies on the difficulty of the factoring problem.

Second, a quantum network can be used to implement several cryptographic tasks beyond private communication, with qualitative advantages with respect to classical networks. The best known one is QKD. Some other tasks are byzantine agreement \cite{ben2005fast}, certified deletion \cite{broadbent2020quantum}, conference key agreement \cite{augusiak2009multipartite,murta2020quantum,Chen2007}, distribution of money \cite{wiesner1983conjugate}, leader election \cite{tani2005exact}, and secret sharing \cite{hillery1999quantum,cleve1999share}.
Then there are some important cryptographic tasks which cannot be implemented neither with classical nor with quantum resources, such as information-theoretically secure quantum bit commitment and two-party secure computation
\cite{mayers1997unconditionally,lo1997insecurity,lo1997quantum,lo1998quantum}. But, if one is willing to make an assumption on the amount of storage \cite{damgaard2008cryptography} or on the quality \cite{konig2012unconditional} of the storage of a potential attacker, then implementing these tasks with quantum resources is advantageous. In this category fall quantum protocols for bit commitment \cite{kent2011unconditionally,konig2012unconditional}, oblivious transfer \cite{schaffner2010simple,wehner2010implementation} and secure identification \cite{damgaard2007secure,dupuis2014entanglement}. Strikingly, quantum offers the possibility of implementing most of these cryptographic tasks without making any assumptions on the behavior of the devices held by the legitimate parties \cite{mayers1998quantum}. In consequence, these so-called device-independent implementations close by construction one of the most important sources of side channel attacks.

Third, as noted in the introduction (Sec.~\ref{sec:into}), the study of
quantum communication complexity tells us that by sending
quantum information (qubits), we can dramatically lower the amount of
communication required compared to sending classical information (bits).
Quantum fingerprinting  \cite{Buhrman2001} is an example of the quantum advantage in
communication.

A fourth important application of quantum networks is computation. In its more direct sense, an alternative paradigm to the monolithic construction of a quantum computer is the so called modular or distributed quantum computer \cite{nickerson2014freely}. In this paradigm high quality small quantum computers are linked via entanglement to build a larger quantum computer. A quantum network can also be used to perform quantum computation on a remote quantum computer without revealing information about the computation or the underlying data \cite{childs2001secure,aharonov2017interactive,broadbent2009universal}, to perform  multipartite computation \cite{cleve1997substituting}, or to obtain a computational advantage in distributed computation tasks \cite{le2019quantum}.

Finally, the entanglement distributed by a quantum network can boost the performance of sensing applications \cite{degen2017quantum}. Notable examples in this domain are in clock synchronization \cite{komar2014quantum} and in interferometry where entanglement can be used to extend the baseline of telescopes \cite{gottesman2012longer,khabiboulline2019optical}.

\subsubsection{Stages of the quantum internet} \label{sec:stages-of-QI}

The path to building the quantum internet will be long and difficult. The current standard viewpoint is that the quantum internet will probably develop in stages. There are different ways to divide it into stages. The classification proposed in \cite{Wehner2018} is based on the network functionality available to the end nodes. 

Interestingly, quantum networks where nodes have very limited functionality are already useful for applications and new tasks can be implemented as the functionality of the end nodes increases. This means, that even at the early stages of development, we expect quantum networks to be useful. We will briefly recap the discussion in \cite{Wehner2018}, linking the communication tasks introduced in \ref{sec:repcomtasks} to development stages.

In the first stage {\it trusted repeater networks} are built. 
In this stage, the nodes can prepare and transmit quantum states to adjacent nodes in the network. This functionality allows to implement prepare-and-measure quantum key distribution protocols between adjacent nodes. In this way, it is possible, for instance, to construct a network of individual quantum key distribution links, but it is not a fully quantum network in the sense that quantum information cannot be transmitted to non-adjacent nodes. This very limited functionality is nonetheless useful: in such a network, if two end nodes trust the behavior of the nodes in a path connecting them, then they can exchange keys that are secure under this assumption \cite{salvail2010security}. Existing quantum networks such as the Tokyo QKD network \cite{Sasaki2011}, the SECOQC network \cite{peev2009secoqc} and the Shanghai-Beijing network \cite{chen2021integrated} are in this stage (see Sec.~\ref{sec:trusted}).

In the second stage, end-to-end {\it prepare-and-measure networks} are built. In this stage, the nodes can prepare single qubits and transmit them to any other node in the network without any trust assumption and on the receiving side, nodes can measure incoming qubits.
A potential price to pay is the post-selection of the transmitted signals. 
Nonetheless, prepare-and-measure networks can still be useful for various additional applications including secure identification in two-party cryptography with noisy quantum memories and key distribution. This includes protocols where entanglement is used to guarantee security but the nodes do not share an entangled state at any moment. Instead, it is sufficient that the nodes can confirm whether entanglement could have been shared if the end nodes had run a coherent version of a prepare and measure protocol. For instance, communicators in a time-reversed entanglement distribution protocol \cite{biham1996quantum}, measurement-device-independent quantum key distribution (MDI QKD) \cite{Lo2012}, and twin-field quantum key distribution (TF QKD) \cite{Lucamarini2018} fall into this category, which remove assumptions about the measurement devices and highly limit the feasibility of side channel attacks  (see, e.g., \cite{curty2021quantumleap}).

In the third stage, {\it entanglement distribution networks} are achieved where two users can obtain end-to-end quantum entanglement in either a deterministic or a heralded fashion.  In this stage, the end nodes require no quantum memories. This added functionality enables, for example, device-independent QKD, when the loss is sufficiently low.

In the following we discuss the final three stages. These stages differ in the quality of the quantum computational capabilities of the nodes. 

In the fourth stage, quantum memory networks are built. In this stage, the end users can store quantum information in their memories
and teleport quantum information to each other. The minimum storage time is determined by the transit time between the
two end nodes. Note that in this stage, the operations are done directly on the physical qubits.  There is no fault tolerance. 
This functionality enables some blind quantum computation schemes, provided that there exists a remote quantum computer \cite{aharonov2017interactive,broadbent2009universal}. It also enables protocols for extending the baseline of telescopes \cite{gottesman2012longer,khabiboulline2019optical}, protocols for cryptographic tasks such as anonymous quantum communication \cite{christandl2005quantum}, secret sharing \cite{hillery1999quantum, cleve1999share}, simple leader election \cite{ambainis2004multiparty}, and some protocols for clock synchronization \cite{komar2014quantum}.

In the fifth stage, few-qubit fault-tolerant networks are built. Here, the end nodes can perform local quantum
operations fault-tolerantly on a few logical qubits. This ability allows more complex protocols to be executed.
More concretely, an end node can perform fault-tolerant execution of a universal gate set
on $q$ logical qubits such as the number $q \geq 1$ is small
enough that the local quantum processors can still be simulated efficiently by a conventional computer.
Since conventional computing power tends to increase exponentially with time,
what value of $q$ remains simulatable is a function of time and technology. 
This functionality enables the implementation of a distributed quantum computer by linking the end nodes.

In the sixth and final stage, quantum computing networks are built and large-scale fault-tolerant quantum computation
can be performed. The end node can perform large-scale quantum computation that cannot be simulated
efficiently by any conventional computer. This will be the ultimate quantum internet. 
With this functionality it is possible to implement protocols for leader election \cite{tani2005exact}, fast byzantine agreement \cite{ben2005fast}, quantum money \cite{gavinsky2012quantum} and weak coin flipping with arbitrarily small bias \cite{mochon2007quantum,chailloux2009optimal}.

We end the recap of the stages by noting that the placement of the tasks in a stage in \cite{Wehner2018} corresponds to the current theoretical state of the art. Future protocol proposals might allow to reduce the requirements to implement a given task. For a more thorough description of existing protocols and their relation to the development stages we point the reader to the quantum protocol zoo \cite{quantumprotocolzoo}. 

\subsection{Quantum networks}
\subsubsection{Elements of a quantum network}
The Internet connects user devices that we call end-systems or hosts. These devices are linked by communication channels to other nodes in the network. However, the hosts are not directly linked. Instead, they are connected via intermediate devices that are called routers. Routers in the internet receive packets of information on incoming links and depending on the content of the packet forward it through one outgoing link. Devices situated in a communication link that passively amplify the signal and do not take routing decisions are called relays. 

Similarly, a quantum network \cite{van2014quantum} connects end-systems linked by quantum channels. Intermediate nodes in quantum networks, in addition to taking routing decisions, participate in the generation of long-distance entanglement. The responsibilities associated with entanglement generation depend on the technology (see Sec.~\ref{sec:gens}). They might include generating entanglement with adjacent nodes, implementing a purification protocol, swapping entanglement or processing encoded quantum information. Moreover, quantum networks will also require classical nodes and links for their operation.

In this review, we have used the term quantum repeaters to denote all intermediate nodes in a quantum network. However, it is possible to make a finer classification. In analogy with classical networks, Munro {\it et al.} \cite{munro2022designing} differentiate between quantum relays and quantum repeaters depending on whether they process quantum information passively or actively. Another distinction can be made depending on whether the intermediate nodes participate in network management and decide how to swap entanglement or not. The former are called quantum routers and the latter automated quantum nodes \cite{Dahlberg2019,kozlowski2020architectural}.

\subsubsection{Network architecture}
The Internet provides an information-transmission service to the end-systems. To implement this service, most communication networks rely on a layered approach. Each layer of the so-called network stack uses the service from the layer below without requiring any knowledge about how it is implemented or what hardware components it relies upon and provides a more complex service to the layer above. 

A priori, the main service of the quantum internet will be the delivery of remote bipartite entanglement, which can then be used as a resource for applications \cite{van2014quantum}. Other proposals posit that the delivery of graph states will be the fundamental primitive of the quantum internet \cite{pirker2019quantum}.
Independently of the main service, for the quantum internet we can expect a similar layered architecture \cite{van2008system,van2013designing,Dahlberg2019,cacciapuoti2019quantum,pirker2019quantum,kozlowski2020architectural,kozlowski2020designing} to the Internet, see \cite{illiano2022quantum} for a survey on protocol stack proposals. Recently, Pompili {\it et al.} \cite{pompili2022experimental} demonstrated experimentally entanglement delivery using a network stack.

The quantum internet architecture will not be independent of the Internet since it is clear that the quantum internet will rely on classical communication for its functionality. However, the quantum internet could also support the functionality of the classical internet creating a complex interplay \cite{cacciapuoti2022quantum}. 

\subsection{The fundamental limits of communications over network}
In the following we discuss the usefulness of quantum networks from an information theoretic point of view.
First, we introduce a model of a network
in terms of a graph and the relevant notation. Then, we define the quantities that characterize the fundamental limits for communicating over quantum networks, i.e., the quantum network capacities. In the network setting, there is a richer set of quantities when compared with direct transmission depending, for instance, on how the communication rates are defined  or whether several sets of users concurrently want to perform a communications task.

Second, we show how to bound the network capacities both from above and from below. These bounds take a particularly simple form in some relevant cases: e.g., for general linear networks \cite{pirandola19} or for bounding the performance of DLCZ-like protocols (like the one in Sec.~\ref{sec:IdealQR}) in the presence of noisy memories \cite{Azuma2016}.

We end by discussing the computability of these bounds, and show that given bounds on the individual channel capacities, the bounds on the network capacities can be derived efficiently.

\begin{figure}[t]
\centering
\includegraphics[width=\linewidth]{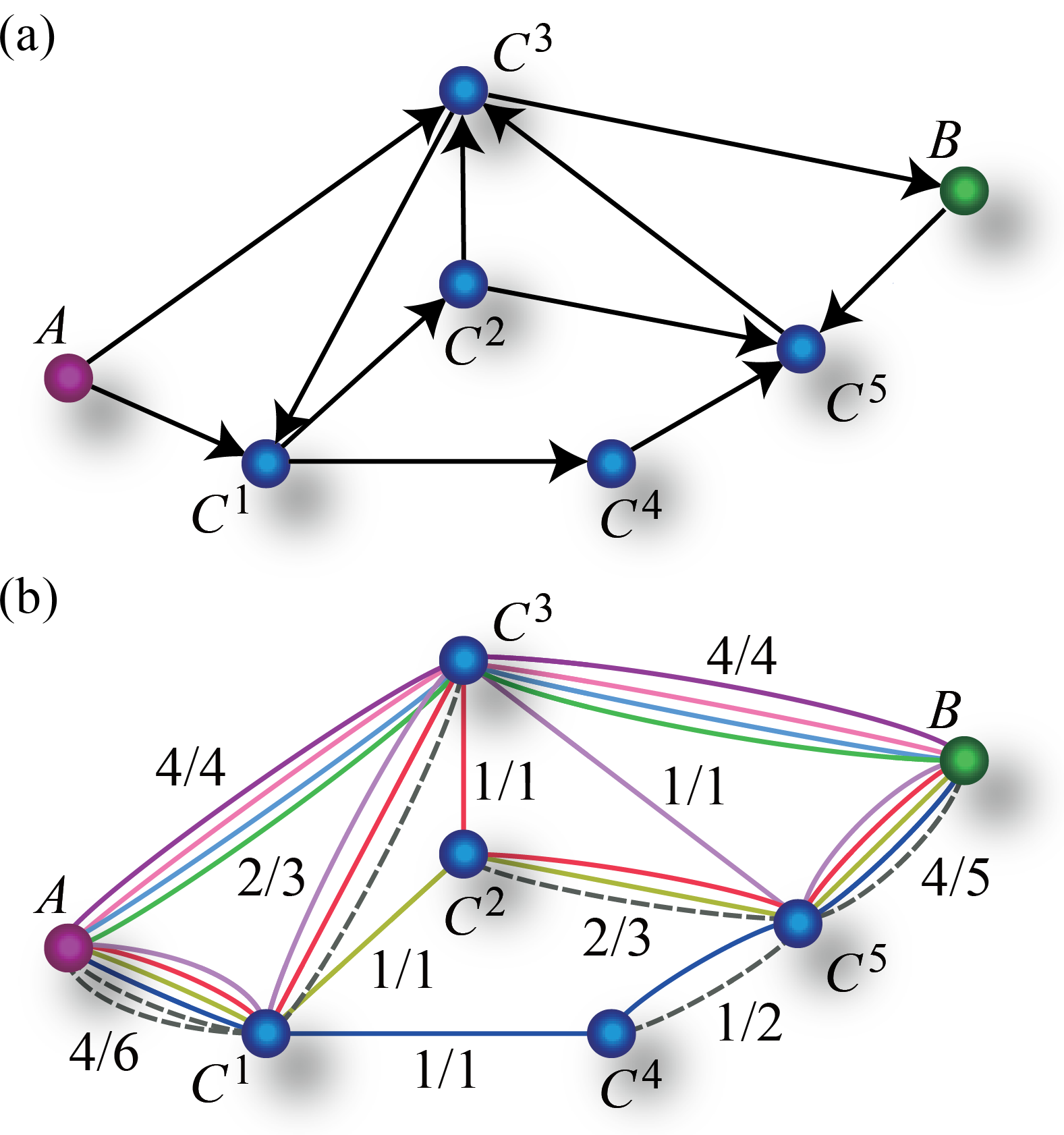}
\caption{ Quantum network and a Bell-pair network.
a) A quantum network as a graph. A quantum network can be abstracted by a directed graph,  $G=(V,E)$ with $V$ and $E$ the sets of vertices and edges. 
However, if two-way classical communication is considered a free resource, edge directions play no role as quantum teleportation can be used to revert the direction of the channel.
We associate with each vertex $v\in V$ a node in the quantum network and with each edge $e\in E$ a quantum channel $\mathcal N_e$. In this example, Alice's node $A$ and Bob's node $B$ are part of network with seven nodes also including intermediary nodes $C^1, C^2, C^3, C^4$ and $C^5$. 
b) A network of maximally entangled states. One approach to entanglement distribution between distant parties in a quantum network is the aggregated repeater protocol \cite{azuma2017aggregating}. In this protocol, adjacent nodes prepare maximally entangled states that then can be transformed into end-to-end entanglement between two distant parties by swapping the entanglement. In the figure, the graph from panel a) has been used to generate entanglement between adjacent nodes. Each edge is annotated with a fraction $x/y$, where the denominator $y$ denotes the number of entangled pairs, while the numerator $x$ denotes the number of entangled states used to establish entanglement between the end parties $A$ and $B$. In this example, a total of eight Bell pairs could be distributed between $A$ and $B$. Figure adapted from \cite{azuma2017aggregating}.
} 
\label{fig:aggrep}
\end{figure}

\begin{figure}[t]
\centering
\includegraphics[width=\linewidth]{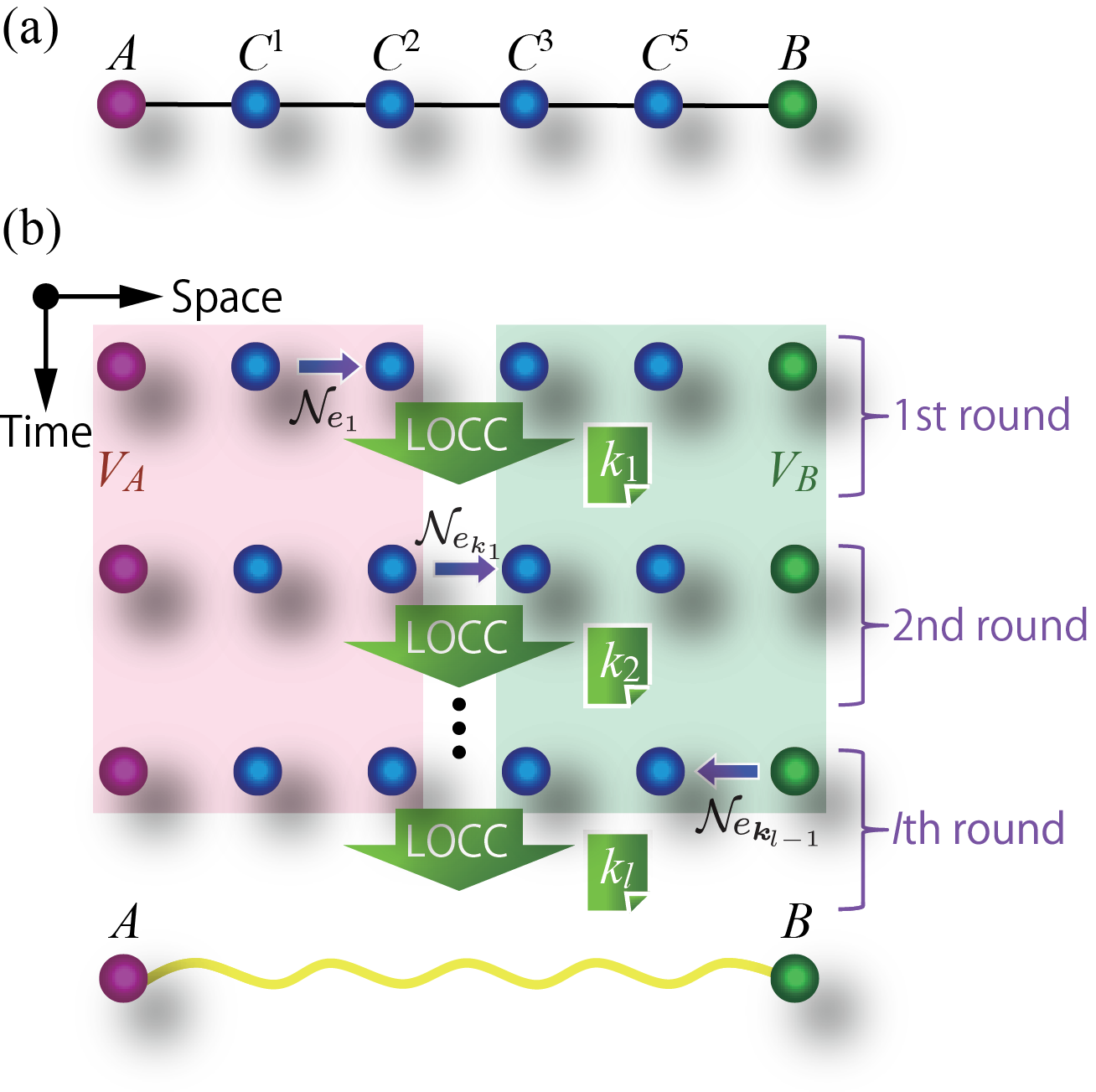}
\caption{Linear network and general protocol.
a) A repeater chain or linear quantum network is associated with a linear graph, i.e., a graph that can be described by a sequence of edges connecting distinct nodes. The linear network in the panel may be a subnetwork of the network in panel a) from Fig. \ref{fig:aggrep}. 
b) The general adaptive protocol \cite{Azuma2016} illustrated over the linear network from panel a). The goal of the protocol is to distribute Bell pairs between $A$ and $B$. The protocol begins with the network joint state represented by a separable state and proceeds iteratively until meeting a termination condition. On each round a node transmits a local subsystem through a quantum channel. Then all nodes perform an LOCC operation. The LOCC operation, the choice of a channel, and transmitted subsystem can depend on the history of the measurement outcomes of the protocol. 
The nodes of the linear network can be divided into two disjoint virtual nodes, $V_A$ [nodes on the left (pink) box] including $A$ and $V_B$ [nodes on the right (green) box] including $B$. The intuition behind the capacity upper bounds in \eqref{eq:qnetcapup} and \eqref{eq:qnetcapupN} is that distributing entanglement between these two virtual nodes is an easier task than distributing entanglement between $A$ and $B$ over the network. Figure adapted from \cite{Azuma2016}.} 
\label{fig:qnetwork}
\end{figure}

\subsubsection{An abstract depiction of networks}

Similar to classical networks, quantum networks will consist of many different components: end nodes, communication channels, routers, switches, multiplexers, etc. However for analysis purposes it is more convenient to restrict networks to two different components: nodes and communication channels.

We can represent this abstract network by $\mathcal G=(G, g)$, where $G=(V,E)$ is a directed graph (see Fig. \ref{fig:aggrep}~(a)) and $g$ a map from edges in the graph to quantum channels, i.e., completely positive and trace-preserving maps. 

We denote by $V$ the set of nodes in the graph and by $E$ the set of edges. Letting $e\in E$ be a directed edge from node $u$ to node $v$, we say that the tail and head of the directed edge $e$ are $u$ and $v$, respectively. We will denote the edge by $uv$ whenever it is useful to specify the tail and head of a node.

We associate with each node $v\in V$ a quantum information processing device. The capabilities of the quantum information processing devices sitting at network nodes can range from a source that can prepare a predefined set of quantum states to a fully-fledged universal quantum computer. For the rest of the section, we assume that nodes can perform noiselessly arbitrary local operations (LO). Since classical communication is qualitatively cheaper than quantum communication, it is common to assume free classical communication between nodes connected by a quantum channel, and sometimes between any two nodes in the network. With this additional assumption, the nodes in the network can implement local operations and classical communication (LOCC) without cost.

Finally, we associate with each edge $uv$ a quantum channel that receives a quantum system as the input from node $u$ and outputs a quantum system to node $v$, via the map: $g(uv)=\mathcal N_{u\rightarrow v}$. To simplify notation, when possible we denote the channel at edge $e$ by $\mathcal N_e$.

This abstract depiction of a network as a graph allows us to leverage tools from graph and network theory. One concept that will be useful for the following is a cut. Given a bipartition of the vertex set $V$, i.e., two sets $V'\subset V$ and $V''=V\setminus V'$, the associated cut-set or cut $\Delta(V')$ is the set of edges connecting $V'$ with $V''$. In particular, the cut associated with $V'$ is given by 
\begin{equation}
\Delta(V'):=\Delta^+(V')\cup \Delta^-(V') 
\end{equation}
with the outcut $\Delta^+(V')$ associated with $V'$,
\begin{equation}
\Delta^+(V'):=\{uv\in E: u\in V', v\in V\setminus V' \},
\end{equation}
and with the incut $\Delta^-(V')$ associated with $V'$,
\begin{equation}
\Delta^-(V'):=\{uv\in E: u\in V\setminus V', v\in V'  \}.
\end{equation}

Given two different vertices $A,B \in V$, we denote by $V_{A;B}$ the set of all bipartitions of $V$ separating $A$ and $B$, i.e., the set of all the subsets of $V$ that include node $A$ but do not node $B$.

\subsubsection{Quantum network capacities}
\label{sec:qncap}
While the applications of the quantum internet are very different, most of them can be implemented if the relevant nodes in the network share an appropriate entangled state. For instance, in order to transmit a $d$-dimensional quantum state it is sufficient to distribute a $d$-dimensional bipartite maximally entangled state,
\begin{equation}\label{eq:ebit}
    \Ket{\Phi_d} \equiv \frac{1}{\sqrt{d}}\sum_{i=1}^{d} \Ket{ii},
\end{equation}
called an edit (called an ebit when $d=2$), which then can be consumed to teleport the desired state (see Sec.~\ref{sec:teleportation}).

Similarly, to transmit secretly a message from a set of $d$ possible messages, it suffices to distribute a $d$-dimensional bipartite private state \cite{horodecki2005secure} or pdit (called pbit when $d=2$). The family of private states consists of the states that can be used to generate a $d$-dimensional secret key, i.e., a uniform probability distribution over $d$ values shared between two honest parties Alice and Bob and secret to any other user. The class of private states includes the class of maximally entangled states but is strictly larger. In fact, there exist states that cannot be distilled into a maximally entangled state but, nonetheless, can be used to distill a pdit \cite{horodecki2005secure}.

Formally, a pdit is a state shared between Alice who holds the systems $a_1a_2$ and Bob who holds $b_1b_2$ in the following form:
\begin{equation}\label{eq:pdit}
    \gamma_{d}\equiv U_{\text{twist}}\left(\ket{\Phi_d}\bra{\Phi_d}_{a_1b_1}\otimes \sigma_{a_2b_2}\right)U_{\text{twist}}^{\dagger},
\end{equation}
where $\sigma_{a_2b_2}$ is an arbitrary bipartite state and $U_{\text{twist}}=\sum_{i=1}^{d}\ket{ij}\bra{ij}_{a_1b_1}\otimes U^{(ij)}_{a_2b_2}$ is a so-called twisting controlled unitary: the systems $a_1b_1$ control the application of  $U^{(ij)}_{a_2b_2}$, arbitrary unitary operators on the systems $a_2b_2$.

GHZ states and multipartite private states \cite{augusiak2009multipartite} play a similar role as a resource for multiuser tasks such as secret sharing and conference key agreement. Hence, in order to study the usefulness of a quantum network for a given application, it suffices to study the rate at which the network can produce a desired resource state. In fact, for many tasks of interest both problems are equivalent.

For the sake of simplicity, we restrict the following discussion to bipartite target states, which we denote by $\theta^{(d)}_{AB}$. Typically the target state is a maximally entangled state or a private state: $\theta^{(d)}_{AB} = \ket{\Phi_d}\bra{\Phi_d}_{AB}$ or $\theta^{(d)}_{AB} = \gamma_{d}$.

As mentioned earlier, we assume that the nodes can apply noiselessly any LOCC operation. Let us now discuss a general protocol for distributing entanglement in a quantum network between nodes $A$ and $B$ (see Fig. \ref{fig:qnetwork}). Before the protocol, there is no entanglement between different nodes in the network. Therefore, the joint state is represented by a separable state as in Eq.~(\ref{eq:mulisep}). Iteratively, first a node transmits a local subsystem through a quantum channel and then all nodes perform an LOCC operation. The LOCC operation, the choice of a channel, and transmitted subsystem can depend on the history of the protocol, e.g., on measurement outcomes obtained through LOCC in previous rounds.

We denote the reduced state between $A$ and $B$ at the end of the protocol by $\sigma_{AB}$. It will be at trace distance $\epsilon(\ge 0)$ from a target state $\theta^{(d)}_{AB}$, i.e., $\|\sigma_{AB}-\theta^{(d)}_{AB}\|_1=\epsilon$, where $\|X\|_1=\text{Tr}(\sqrt{X^\dagger X})$. We say that a protocol is a $P_{\{n_e\}_{e\in E},\epsilon}$ adaptive protocol if the average number of uses of channel $\mathcal N^e$ is upper bounded by $n_e$ for all edges and the protocol produces a state at most at a distance $\epsilon$ from a target state $\theta_{AB}^{(d)}$, where $d(\ge 1)$ can depend on the outcome of the protocol. 

The figure of merit of $P_{\{n_e\}_{e\in E},\epsilon}$ protocols is the average amount of the target entanglement produced, which is quantified by $\log_2 d$ for the states mentioned. From an operational point of view, a $d$-dimensional maximally entangled state or private state enables respectively the transmission of $\log_2 d$ qubits or the private communication of $\log_2 d$ bits. We denote the average entanglement produced---it might vary from round to round---by $\langle \log_2 d\rangle$.

We obtain the rate at which the protocol produces the entanglement, by dividing the average entanglement by the appropriate quantity of resources used. In contrast with the single channel case, one can consider several metrics: the number of channels used, the number of full uses of the network or the number of times a path of channels connecting $A$ with $B$ was used. These metrics could be related to time which is for engineering purposes a more convenient figure of merit (see \cite{azuma2017aggregating}, \cite{bauml2020linear}, or \cite{azuma2020tools} for detail).

The capacity of the quantum network is the optimal asymptotic rate for producing a target entangled state $\theta$ at which the error parameter $\epsilon$ can be made arbitrarily small. Following our previous discussion on the rate, each choice of rates gives rise to a different type of network capacity.

Let us denote by $n=\sum_e n_e$ an upper bound on the total number of channel uses and by $p_e=n_e/n$ the frequency that the protocol uses channel $\mathcal N_e$. Given a fixed set of frequencies, we define the capacity per channel use \cite{Azuma2016} as:
\begin{equation}\label{eq:capacitygivenfrequencies}
C_\text{c}^\theta (\mathcal G,\{p_e\}_{e\in E})=\lim_{\epsilon\rightarrow 0}\lim_{n\rightarrow\infty}\frac{1}{n}\sup_{P_{\{n_e\}_{e\in E},\epsilon} } \langle \log_2 d\rangle.
\end{equation}

Depending on the network scenario, the usage frequencies of the channels in the network can be free parameters. In this case, Eq.~\eqref{eq:capacitygivenfrequencies} can be maximized over the set $\{p_e\}_{e\in E}$ of frequencies to give a unique network capacity per channel use \cite{bauml2020linear}:
\begin{equation}
\label{eq:capacity}
C_\text{c}^\theta (\mathcal G)=\max_{p_e\geq 0,\sum_ep_e=1}C_\text{c}^\theta (\mathcal G,\{p_e\}_{e\in E}).
\end{equation}

To capture the capacity per network use, which we denote by $C_\text{n}^\theta (\mathcal G)$, we let all upper bounds on the average number of channel uses be equal $n_e=n_{e'}\ e,e'\in E$ and let $n$ denote the number of network uses, i.e., we let $n=n_e$, which then implies $p_e=1,\forall e\in E$. This quantity corresponds to the notion introduced by Pirandola in \cite{pirandola19} to capture the limits of so-called flooding protocols.
A third important scenario is the single-path per network use capacity \cite{pirandola19}, where the goal is to maximize the rate per use of a single path, though it is unclear if it can be expressed in a form similar to Eq.~\eqref{eq:capacitygivenfrequencies}.

If the target state $\theta$ is a maximally entangled state (see Eq.~\eqref{eq:ebit}), then these expressions represent a quantum capacity of the quantum network $\mathcal G$. If $\theta$ is a private state (see Eq.~\eqref{eq:pdit}) it represents a private capacity.

The distribution of entanglement between a single set of users is but one of many possible measures of usefulness of a quantum network. Networks typically serve many users and one might be interested in understanding the capacity of the network for distributing entanglement to multiple sets of users. Equation \eqref{eq:capacitygivenfrequencies} can be adapted to capture multiuser setups by modifying appropriately the figure of merit of the protocol $\langle \log_2 d\rangle$ and the definition of $P_{\{n_e\}_{e\in E},\epsilon}$ protocol \cite{bauml2020linear}. For instance, given $m$ sets of users and let $\langle\log_2 d^{(i)}\rangle$ be the average amount of entanglement that a $P_{\{n_e\}_{e\in E},\epsilon}$ protocol produces for set $i$ of users, then the maximization of $\min_{i=1}^m\langle \log_2 d^{(i)}\rangle$ leads to the maximum rate that can be guaranteed to all sets of users, called the worst-case network capacity, while the maximization of $\sum_{i=1}^m\langle \log_2 d^{(i)}\rangle$ leads to the maximum total rate, called the total network capacity.

\subsubsection{Entanglement based upper bounds}
While there is no known procedure for computing these capacities in general, there are several tools for bounding them both from above and from below leveraging the relation between the communication task and the distillation of the appropriate entangled state.

In the following, we present a formulation by \cite{rigovacca2018versatile} for abstract entanglement measures. This formulation generalizes earlier work by Pirandola \cite{pirandola19} for 
quantum networks composed of a specific type of channels (called teleportation simulable channels, explained later) with
the relative entropy of entanglement and by Azuma \textit{et al.} \cite{Azuma2016} for 
arbitrary quantum networks with 
the squashed entanglement. In particular, these two results build respectively on the PLOB \cite{Pirandola2015} and TGW \cite{Takeoka2014} bounds on the private capacity of an individual channel (see Sec.~\ref{sec:into} and Sec.~\ref{sec:IdealQR}).

In particular, let $\mathcal E$ be a measure of bipartite entanglement. That is, $\mathcal E$ is a function from the set of bipartite states into the positive real numbers that satisfy several requirements \cite{Horodecki2009}. In particular, it is not increasing on average under LOCC. We define the entanglement of channel $\mathcal N_{A\rightarrow B}$ as
\begin{equation}
\mathcal E(\mathcal N_{A\rightarrow B})\equiv \sup_{\rho_{AA'}}\mathcal E(\mathcal N_{A\rightarrow B}(\rho_{AA'}))\ ,
\end{equation}
where $\rho_{AA'}$ is a bipartite state with $A'$ isomorphic to $A$.

Now, let $\mathcal E$ be a bipartite entanglement measure that satisfies the following two inequalities:
\begin{enumerate}
\item[P1] (Continuity) If a bipartite state $\rho_{AB}$ is at epsilon distance from the target state $\theta_{AB}^{(d)}$, i.e., $\|\rho_{AB}-\theta_{AB}^{(d)} \|_1\leq\epsilon$, then $\mathcal E(\rho_{AB})\geq g(\epsilon)\log d-f(\epsilon)$ with $f$ and $g$ two real valued continuous functions that verify $\lim_{\epsilon\rightarrow 0}f(\epsilon)=0$ and $\lim_{\epsilon\rightarrow 0}g(\epsilon)=1$.
\item[P2] (Subadditivity) Given a bipartite state $\rho_{A_1A_2B_1}$, the entanglement in the $AB$-cut after sending the system $A_2$ through channel $\mathcal N_{A\rightarrow B}$ is not larger than the original entanglement in the $AB$-cut plus the entanglement of the channel: $\mathcal E(\sigma_{A_1B_2B_1})\leq \mathcal E(\rho_{A_1A_2B_1})+\mathcal E(\mathcal N_{A\rightarrow B})$, where $\sigma_{A_1B_2B_1}={\cal N}_{A_2\rightarrow B_2}(\rho_{A_1A_2B_1})$.
\end{enumerate}
Then, the capacity of the network for distributing some target state $\theta$ between two nodes $A$ and $B$ can be bounded from above by the following optimization formulae \cite{rigovacca2018versatile}:

\begin{align}
C_\text{c}^\theta(\mathcal G,\{p_e\}_{e\in E})&\leq \min_{\mathcal V\in V_{A;B}}\sum_{e\in \Delta(\mathcal V)}p_e\mathcal E(\mathcal N_e),\label{eq:qnetcapup}\\
C_\text{c}^\theta(\mathcal G)&\leq \max_{\substack{p_e\geq 0,\\ \sum_ep_e=1}}\min_{\mathcal V\in V_{A;B}}\sum_{e\in \Delta(\mathcal V)}p_e\mathcal E(\mathcal N_e),\label{eq:qnetcapupP}\\
C_\text{n}^\theta(\mathcal G)&\leq \min_{\mathcal V\in V_{A;B}}\sum_{e\in \Delta(\mathcal V)}\mathcal E(\mathcal N_e).
\label{eq:qnetcapupN}
\end{align}
Note that Eqs.~\eqref{eq:qnetcapup}-\eqref{eq:qnetcapupN} do not depend on any functional of more than one channel: equations \eqref{eq:qnetcapup} and \eqref{eq:qnetcapupP} depend only on the entanglement of each of the channels individually and the channel usage frequencies, while Eq.~\eqref{eq:qnetcapupN} depends only on the entanglement of the channels. The minimization is performed over $V_{A;B}$, the set of all cuts between $A$ and $B$. The intuition for this formula is that we could join all the nodes of the network into two virtual nodes, one including $A$ and one including $B$ (see Fig.~\ref{fig:qnetwork}~(b)). Distributing entanglement between these two virtual nodes is an easier task and can be done at a rate no larger than the entanglement of all the channels connecting the two virtual nodes. Since this argument provides a valid upper bound for any bipartition, the minimum provides the best upper bound of this form.

Fortunately, there are several entanglement measures that verify P1 and P2 for private states (and in consequence also for maximally entangled states). In particular the squashed entanglement \cite{takeoka2014squashed,Takeoka2014} and the max-relative entropy of entanglement \cite{christandl2017relative} satisfy both properties for arbitrary channels, while the relative entropy of entanglement is only known to satisfy both properties for a family of channels known as teleportation simulable, Choi simulable or stretchable channels \cite{Pirandola2015,Bennett1996,Gottesman1999,Horodecki1999,wolf2007quantum}. 

Leveraging an inequality from \cite{christandl2017relative}, Rigovacca \textit{et al.} \cite{rigovacca2018versatile} proved a hybrid relative entropy upper bound, where the entanglement measure in the upper bounds in Eqs.~\eqref{eq:qnetcapup}, \eqref{eq:qnetcapupP} and \eqref{eq:qnetcapupN} is the relative entropy of entanglement for teleportation simulable channels and the max-relative entropy of entanglement for the other channels. 
Therefore, the currently best option to give upper bounds in the form \eqref{eq:qnetcapup}, \eqref{eq:qnetcapupP} or \eqref{eq:qnetcapupN} to a given arbitrary quantum network is to use this hybrid relative-entropy bound or the squashed-entanglement bound. Many relevant channels such as the amplitude damping channel are not teleportation simulable. However, several channels of particular interest are teleportation simulable; this includes the depolarizing and dephasing channels, more generally mixed Pauli channels, the erasure channel and lossy bosonic channels. Remarkably for the lossy bosonic channels, which model optical fibers, the relative entropy of entanglement based upper bound is tight \cite{Pirandola2015}. In the following we define Choi-simulable channels and particularize the bounds for this case.

A channel $\mathcal N_{A\rightarrow B}$ is teleportation simulable if given a state $\rho_{A}$ that one wants to transmit through channel $\mathcal N_{A\rightarrow B}$ and the Choi state of the channel $\Gamma_{A'B}=\mathcal N_{A\rightarrow B}\left(\ket{\Phi_d}\bra{\Phi_d}_{A'A}\right)$,
there exists an LOCC protocol $\Lambda$ that simulates the action of the channel on any input state $\rho_{A}$:
\begin{equation}
\mathcal N_{A\rightarrow B}(\rho_{A})=\Lambda(\Gamma_{A'B}\otimes\rho_{A''})\ .
\end{equation}
To gain intuition on this equation one can think of the identity channel from $A$ to $B$. Then simulation can be obtained by teleportation, i.e., $\Lambda$ consists of a joint generalized Bell measurement on systems $A'A''$ and applying the appropriate correction to system $B$. More generally, this strategy works for any channel whose action commutes with the receiver's corrections of quantum teleportation \cite{Bennett93}, because, in this case, the correction to system $B$ can be regarded as correction for system $A$ before entering the channel ${\cal N}_{A\to B}$ and thus, this is merely a local teleportation to send a quantum state $\rho_{A''}$ to system $A$.

\subsubsection{Application of the upper bounds to linear networks}
\label{sub:linearnetwork}
In the following we focus on a particular use case: linear networks (see Fig. \ref{fig:qnetwork}). This use case of the upper bounds is of particular relevance to quantum repeater protocols. In this case, the cut-sets are the individual channels, highly simplifying the upper bounds.
The bounds on the capacities per channel \eqref{eq:qnetcapup}, \eqref{eq:qnetcapupP} and per network \eqref{eq:qnetcapupN} use take the form:
\begin{align}
C_{\text{c}}^\theta(\mathcal G,\{p_e\}_{e\in E}) &\leq \min_{e\in E}p_e\mathcal E(\mathcal N_e),\label{eq:uprepeaters}\\
C_{\text{c}}^\theta(\mathcal G) &\leq \frac{1}{\sum_{e \in E} \left[ \mathcal E(\mathcal{N}_e) \right]^{-1}} ,\label{eq:uprepeatersP}\\
C_{\text{n}}^\theta(\mathcal G) &\leq \min_{e\in E}\mathcal E(\mathcal N_e).
\label{eq:uprepeatersN}
\end{align}
The upper bound on the network capacity per channel use \eqref{eq:uprepeatersP} was derived in \cite{Azuma2016}.

As a first example, let us consider a linear network connected by lossy bosonic channels. For these channels, the choice of the relative entropy of entanglement (i.e., ${\cal E}=E_{\rm R}$) gives tight bounds. In particular, it was shown in \cite{Pirandola2015} that $E_{\rm R}(\mathcal N_e)=-\log_2(1-\eta_e)$, where $\eta_e$ is the transmittance of the lossy bosonic channel $\mathcal N_e$ of Eq.~(\ref{eq:lossch}). Then, if we insert this relation into Eqs.~\eqref{eq:uprepeaters}, \eqref{eq:uprepeatersP} and \eqref{eq:uprepeatersN}, we obtain the following expressions for the capacities of the network, including one derived in \cite{pirandola19}:
\begin{align}
C_{\text{c}}^\theta(\mathcal G,\{p_e\}_{e\in E}) &= \min_{e\in E}-p_e\log_2(1-\eta_e),
 \\
 C_{\text{c}}^\theta(\mathcal G) &= \frac{1}{\sum_{e\in E}(-\log_2(1-\eta_e))^{-1}},
 \\
C_{\text{n}}^\theta(\mathcal G) &= \min_{e\in E}-\log_2(1-\eta_e).
\label{eq:plobrepeaters}
\end{align}

\begin{figure}[t]
\centering
\includegraphics[width=\linewidth]{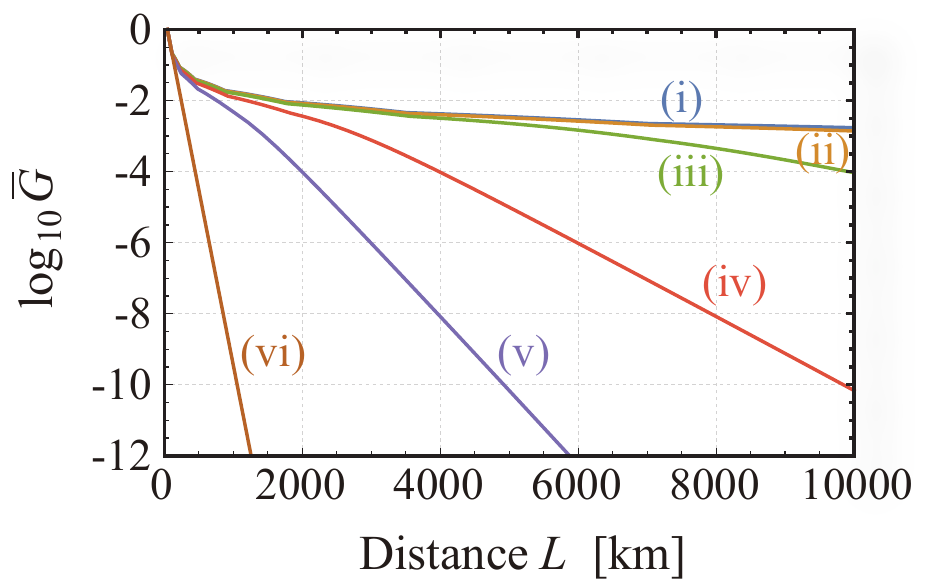}
\caption{Upper bound on the secret key rate achievable with a noisy linear network. In particular, the upper bound applies to a wide range of protocols including DLCZ \cite{Duan2001} and others \cite{Sangouard2011,kok2003construction,azuma2012quantum} when  implemented with matter quantum memories in the presence of dephasing noise. The linear network consists of a chain of repeaters equally separated and connected by an optical fiber with attenuation length 22\,km and spanning a total distance of $L$\,km. The curves labeled by (i-vi) correspond respectively with the following coherence times: $1.0 \times 10^{-2}$\,s, $5.0\times 10^{-3}$\,s, $2.5\times 10^{-3}$\,s, $1.0\times 10^{-3}$\,s, $5.0\times 10^{-4}$\,s, $1.0\times 10^{-4}$\,s. The upper bound in (vi) scales better than direct transmission, roughly     proportional to the square root of the PLOB bound, but equivalent to the intercity QKD protocols in Sec.~\ref{sec:milestones}. In consequence, with a coherence time of $1.0\times 10^{-4}$\,s there can be no advantage for a DLCZ-type repeater scheme compared with the simpler intercity QKD protocols. Figure from \cite{Azuma2016}. }
\label{fig:dlczupper}
\end{figure}

As a second example, we consider the performance of a DLCZ-type quantum repeater protocol \cite{Duan2001} (like one in Sec.~\ref{sec:IdealQR}) where the memory in the nodes is subject to decoherence and taking into account the time required to exchange classical communication between distant nodes.
Razavi \textit{et al.} \cite{Razavi2009} noticed that in contrast with the polynomial scaling with the total distance $L$ predicted by the DLCZ protocol, the performance with finite coherence times of quantum memories degrades exponentially with $\sqrt{L}$. Azuma \textit{et al.} \cite{Azuma2016} strengthened the results and showed that polynomial scalings for a large class of DLCZ-type protocols could be only possible above a threshold coherence time. 
In particular, see Fig.~\ref{fig:dlczupper}, the performance of any DLCZ-type repeater scheme with a memory coherence time below $1.0\times 10^{-4}$\,s is upper-bounded by an exponential on the square root of the total distance, and this kind of performance is achievable as described in Sec.~\ref{sec:milestones}. 
The key idea to apply upper bound~\eqref{eq:uprepeatersP} is that the memory noise can be modeled by a noisy quantum channel between the memory at the time when it stores a state and the memory at the moment that it releases the state. In consequence, the performance of any protocol using the noisy memory is bounded by the performance of an induced linear network (i.e., by using Eq.~\eqref{eq:uprepeatersP}).

\subsubsection{Capacity lower bounds via the aggregated repeater protocol}
Now let us look at a general lower bound on the capacity of quantum networks \cite{azuma2017aggregating}. This lower bound, based on aggregated quantum repeater protocol, matches the general upper bounds given in Eqs.~\eqref{eq:qnetcapup} and \eqref{eq:qnetcapupN} up to a prefactor. Moreover, the aggregation of even existing protocols \cite{Duan2001,Sangouard2011,Jiang2008,Mazurek2014Long-distanceMemory,li2013long} matches the lower bound on the capacity up to another prefactor for the case of optical quantum networks composed of lossy bosonic channels. This implies that the upper bounds have no scaling gap and yield good measures of the usefulness of a network. We note that, while we have exemplified the upper bounds with a linear network of repeaters in Sec.~\ref{sub:linearnetwork}, they can be applied to any quantum network with arbitrary topology, including distributed quantum computation setups.

In the following, we discuss the lower bound which corresponds with the achievable rate of the aggregated quantum repeater protocol introduced by Azuma \textit{et al.}~\cite{azuma2017aggregating}, see Fig. \ref{fig:aggrep}~(b). The goal of this protocol is to distribute entanglement between targeted nodes in the network which is later consumed to perform the appropriate communications task.

For each of the quantum channels in a given quantum network, let us consider a protocol that produces entangled states that are $\epsilon$-close to a maximally entangled state at a rate $R_e$ which can be different for each channel. This is possible for all channels provided that $R_e<Q(\mathcal N_e)$, i.e., provided that the rate is below the maximal rate of the channel for distributing maximally entangled states for a large enough number of channel uses (called the quantum capacity of the channel ${\cal N}_e$). Then, if each of the channels is used $n_e$ times, the whole network will be in a tensor product of entangled states and the whole network can be regarded as a multi-graph with $n_eR_e$ edges per edge in the original graph, where each edge in the multi-graph corresponds to a qubit maximally entangled state $\ket{\Phi_2}$.

We can use the resulting state to create maximally entangled states between Alice at node $A$ and Bob at node $B$. For each state it is necessary to perform entanglement swapping over a path of maximally entangled states connecting Alice with Bob. The number of maximally entangled states that can be distributed between Alice and Bob is then equivalent to the maximum number of edge disjoint paths connecting Alice with Bob in the multi-graph. This maximum number of paths is by Menger's theorem \cite{jungnickel2005graphs} equivalent to the value of the minimum cut of the graph:
\begin{equation}
M = \min_{\mathcal V\in V_{A;B}}\sum_{e\in\Delta(\mathcal V)}n_eR_e.
\label{eq:integermincut}
\end{equation}
This minimization can be solved in time proportional to a polynomial in the number of edges. However, since the number of edges grows with the number of uses, the full optimization is a priori intractable. Now, if we consider the achievable rate per channel use with the aggregated repeater protocol, Eq.~\eqref{eq:integermincut} becomes $\min_{\mathcal V\in V_{A;B}}\sum_{e\in\Delta(\mathcal V)}(n_e/n) R_e$. Moreover, for a number of uses $n$ large enough, any rate below the capacity of each channel is achievable. Consequently, the right-hand side of the following expression is achievable:
\begin{align}
C_{\text{c}}^\theta(\mathcal G,\{p_e\}_{e\in E})&\geq \min_{\mathcal V\in V_{A;B}}\sum_{e\in \Delta(\mathcal V)}p_e Q(\mathcal N_e),
\label{eq:rationalmincut}\\
C_\text{c}^\theta(\mathcal G)&\geq \max_{\substack{p_e\geq 0,\\ \sum_ep_e=1}}\min_{\mathcal V\in V_{A;B}}\sum_{e\in \Delta(\mathcal V)}p_e Q(\mathcal N_e) ,\label{eq:rationalmincutP}\\
C_{\text{n}}^\theta(\mathcal G)&\geq \min_{\mathcal V\in V_{A;B}}\sum_{e\in \Delta(\mathcal V)} Q(\mathcal N_e),
\label{eq:rationalmincutN}
\end{align}
where $p_e=n_e/n$. We note that the lower bounds are of the same form of the respective upper bounds in Eqs.~\eqref{eq:qnetcapup}-\eqref{eq:qnetcapupN} where the entanglement of the channel is replaced by the quantum capacity. Therefore, if ${\cal E}({\cal N}_e)=Q({\cal N}_e)$ holds for any $e$, these lower bounds \eqref{eq:rationalmincut}, \eqref{eq:rationalmincutP} and \eqref{eq:rationalmincutN} coincide with upper bounds \eqref{eq:qnetcapup}, \eqref{eq:qnetcapupP},  and \eqref{eq:qnetcapupN}. For example, this is indeed the case for quantum networks composed only of lossy bosonic channels.

The aggregation of quantum repeaters is also possible with minimizing cost \cite{azuma2023networking}. The cost here is a general
notion like a price to pay for presenting ebits between
two targeted nodes in a quantum network.

\subsubsection{Computability of the network capacities}
Let us now discuss how to compute both the lower and the upper bounds in Eqs.~\eqref{eq:qnetcapup}, \eqref{eq:qnetcapupN}, \eqref{eq:rationalmincut}, and \eqref{eq:rationalmincutN}.  This is indeed important in practice, for instance, to determine how a network provider should distribute entanglement to clients according to their requests.  All four equations depend only on the values of the entanglement of the individual channels. The four quantities are expressed as the solution of the minimum cut over an undirected graph. These optimization problems can be solved by a linear program in time polynomial in the number of nodes in the graph \cite{jungnickel2005graphs}. Similar arguments allow one to find efficiently lower and upper bounds not only on the capacities for two-party communication described above, but also on the worst-case and total quantum network capacities (see \ref{sec:qncap}) and for distributing GHZ states \cite{bauml2020linear}.

\section{Concluding remarks}
\label{sec:conclusion}
The quantum internet will have important applications in sensor networks, upscaling quantum computing and secure quantum communication \cite{VanDam2020}. To build the quantum internet, quantum repeaters have been proposed and studied extensively. This review has focused on the various generations of quantum repeaters as well as all-photonic quantum repeaters; we have seen that quantum repeaters are essential to realize an efficient quantum internet. Nonetheless, our discussion has been largely limited to a fiber-optical setting connecting two end nodes, Alice and Bob.

In this concluding section, we take a step back to think some more about how to build a quantum internet. We will discuss a few alternative designs and important issues facing the quantum internet---not only its efficiency, but also its cost and the uncertainty in the technology it would leverage. 

Cost can be a critical issue in realizing any technology. Although the conventional Internet is believed to contribute trillions of US dollars each year to the US economy, just upgrading the existing fiber optical network in the US to cover, say, 90 percent of households there would take an additional investment of over 100 billion US dollars
(see, e.g., \cite{cartesian2021}).
This figure is for a single country and for an upgrade to the existing, extensively developed, Internet. Therefore, it is not unreasonable to predict that the construction and operation of a global quantum internet would ultimately take decades and require investments of trillions of US dollars.
This is an astonishing number. Such an enormous investment would almost certainly come not only from governments, but also from for-profit commercial corporations.
For a comparison, the LIGO and LHC projects---endeavours admittedly more localized in scope---required only 1.1 billion and 4.75 billion dollars, respectively~\cite{Horgan2016, Roche2022}.
We have not even begun to estimate the cost of building various generations of quantum repeater structures on a global scale. Some detailed calculations, aided by a quantum network simulator, would be needed to address the cost issue more seriously.

On the other hand, as mentioned in the introductory section \ref{sec:into}, the Internet consumes a lot of energy through the transmission of optical signals. Furthermore, the sensing, monitoring and routing of the Internet require massive amounts of local computational power.
As the Internet grows, scalability becomes a challenge. A quantum internet could operate at single-photon level. It may well be interesting to explore whether a quantum internet could lead to huge savings in energy consumption. Similarly, it may be worthwhile to investigate whether quantum computing and quantum information processing could contribute to the management of the internet.

Next, let us imagine a world---sometime in the distant future---where quantum memories with long-term stability become widely available at low cost. In this case, to distribute entanglement, one could simply ship those stable quantum memories all over the world, physically, in the same way that we currently dispatch hard drives and mail (see, e.g., \cite{devitt2016high}). The apparent drawback would be latency, which means the delay before a transfer of data begins following an instruction for its transfer; however, this shipment could be done off-line, and entanglement swapping could be used to connect users via intermediate nodes {\it instantaneously} in the same way that a telephone network can connect the users. In this way, the latency issue could be alleviated.
With the physical shipment of quantum memory devices, the requirements of quantum repeaters could be reduced. This is just one way in which our design of the quantum internet is highly dependent on the available technology, in addition to cost.

Currently, quantum memories often operate at cryogenic temperatures and their lifetimes are often limited. If this is the case, quantum repeater nodes will need refrigerators. Notice that
all-photonic quantum repeaters may also require refrigerators (either in photonic graph state generation devices
or measurement devices).Suppose we wanted to connect someone in New York with another person in Tokyo---10,845\,km away---through undersea optical fibers. Then, optimistically, we would need to place a quantum repeater node every a few hundred kilometers under the sea. In this case, hundreds of repeater nodes would be needed. Placing cryogenic repeater nodes in undersea optical fibers, maintaining them, and providing the energy to operate them reliably are no easy feats, and would likely prove very costly.

As an alternative solution, ground-to-satellite quantum communication is a serious candidate. By preparing an entangled source of photons in a satellite, Charlie, and sending it to two ground stations, Alice and Bob, Charlie can act as an untrusted relay to connect two distant locations on the globe. Currently, line of sight is a serious restriction in ground-to-satellite communication. However, we can envision a future wherein space-grade long-lifetime quantum memories are available. By first sending one half of an entangled pair to Alice, storing the second half in the quantum memory on a rapidly moving quantum satellite, and later sending it to Bob, Charles can connect any two ground stations that have a line of sight to any point on the satellite's orbit. Besides this, a constellation of orbiting satellites could provide a continuous, on-demand entanglement distribution service to ground stations \cite{Khatri2021}. 
In principle, one could put quantum repeaters even on satellites to run a quantum repeater protocol \cite{liorni2021quantum}. However, this may also be challenging if the repeaters need cryogenic environment.

As mentioned earlier (see Sec.~\ref{sec:qrepeaters}), the probabilistic nature of a Bell state measurement in linear optics (for certain photonic encodings) is a key limiting factor in the design of both matter-based and all-photonic quantum repeaters. Indeed, without using additional ancillae or a different encoding, the success probability of a linear-optical Bell measurement is upper-bounded by 1/2. A game changer for the efficiency of quantum repeaters would therefore be a near-deterministic, high-fidelity entangling gate on photons. This could be based on, for example, an enhancement by quantum memories \cite{bhaskar2020experimental,Borregaard2019,Munro2012}.

For all-photonic repeaters in particular, a game-changer would be the deterministic generation of photonic graph states based on coupled quantum emitters such as quantum dots (see, e.g., \cite{li2021entangled}).
Alternatively, a hybrid approach with a single quantum emitter and subsequent fusions would also dramatically lower the resource requirements \cite{Hilaire2022}.
There exists another possibility of the development purely on all photonics: beginning with all-photonic intercity QKD (Sec.~\ref{sec:milestones}), proceeding to all-photonic quantum repeaters (Sec.~\ref{sec:memoryless}), and ending with linking fault-tolerant photonic quantum computers (e.g., \cite{knill2001scheme}). 
 
Another important area of research is the quantum interconnect (see, e.g., \cite{awschalom2021development}). Indeed, the ability to convert and transfer quantum information across different platforms will enhance the inter-operability of the future quantum internet.

In this review, we have focused on the distribution of bipartite entanglement. However, for many applications, including quantum sensing, it is often advantageous to use multipartite entangled states. Conceptually, we may build up multipartite states through successive teleportations. However, were we to do it with linear optics, the probabilistic nature of a Bell measurement would make the success probability of constructing an $n$-partite entangled state exponentially small. Therefore, there is value in further exploring the preparation and distribution of multipartite entanglement.

To conclude, we stress that a truly global quantum internet requires seamless operation across continents. As different countries are currently pursuing different approaches and strategies for the quantum internet, there will be a need for cooperation and standardization in the design, construction and operation of this major technology.

\acknowledgements{
We thank Stefan B\"{a}uml, Johannes Borregaard, Ronald Hanson, Tomoyuki Horikiri, Rikizo Ikuta, Jessica Illiano, Norbert L\"utkenhaus, Mattia Montagna, William J. Munro, Shoichi Murakami, Fatih Ozaydin, Stefano Pirandola, John Preskill, Mohsen Razavi, Tim Taminiau, Wolfgang Tittel, Takashi Yamamoto, Qiang Zhou, and Val Zwiller for their helpful
comments and suggestions on different versions of this manuscript.
K.A. is thankful for the support, in part, from CREST, JST JP-MJCR1671, from PREST, JST JP-MJPR1861, from Moonshot R\&D, JST JPMJMS2061, and from JSPS KAKENHI 21H05183 JP. S.E.E. was supported by the NSF (grant number 1741656), the EU Horizon 2020 programme (GA 862035 QLUSTER), and by ARO (MURI grant no. W911NF2120214).
D.E. was supported by the Netherlands Organization for Scientific Research (NWO/OCW), as part
of the Quantum Software Consortium program (project
number 024.003.037 / 3368). L.J. was supported by the ARO (W911NF-18-1-0020, W911NF-18-1-0212), ARO MURI (W911NF-16-1-0349), AFOSR MURI (FA9550-19-1-0399, FA9550-21-1-0209), DoE Q-NEXT, NSF (EFMA-1640959, OMA-1936118, EEC-1941583), NTT Research, and the Packard Foundation (2013-39273).
H.-K. Lo was supported by NSERC, Connaught Innovation, CFI, ORF, MITACS Accelerate,
 Huawei Canada, Royal Bank of Canada (RBC), the start-up grant by the University of Hong Kong, US Air Force, NRC-CSTIP program, and Innovative Solutions Canada program. I.T. was supported by the Ontario Graduate Scholarship.}
\bibliography{allreferences}

\end{document}